\documentclass[fleqn,usenatbib]{aa} 
\usepackage{float}
\usepackage{graphicx}	
\usepackage{amsmath}	
\usepackage{amssymb}

\usepackage{txfonts}
\usepackage{longtable}
\usepackage{graphicx}
\usepackage{supertabular,booktabs}
\usepackage[section]{placeins}

\makeatletter
\renewcommand*\aa@pageof{, page \thepage{} of \pageref*{LastPage}}
\makeatother
\DeclareRobustCommand{\VAN}[3]{#2}
\let\VANthebibliography\thebibliography
\def\thebibliography{\DeclareRobustCommand{\VAN}[3]{##3}\VANthebibliography}

\graphicspath{{./figures/}}

\usepackage[unicode=true,psdextra]{hyperref}
\begin{document} 

   \title{Core-collapse supernovae 2022prr, 2023ucy, and 2024ljc in the luminous infrared galaxy NGC 6745}
   \titlerunning{CCSNe 2022prr, 2023ucy, and 2024ljc in LIRG NGC 6745}

   \author{K.~K. Matilainen \inst{1}\fnmsep\thanks{E-mail: katja.matilainen@utu.fi}
        \and
            E. Kankare \inst{1}
        \and
            T. Nagao \inst{1}
        \and
            T. Reynolds \inst{1,2}
        \and
            A. Efstathiou \inst{3}
        \and
            S. Mattila \inst{1,3}
        \and
            C.~R. Angus \inst{4}
        \and
            A. Reguitti \inst{5,6}
        \and
            Y.-Z. Cai \inst{5,7}
        \and
            M. Fraser \inst{8}
        \and
            L. Galbany \inst{9,10}
        \and
            M. González-Bañuelos \inst{9,10}
        \and
            C.~P. Guti\'errez \inst{9,10}
        \and 
            T. Kangas \inst{11,1}
        \and 
            T.~L. Killestein \inst{12}
        \and
            T. Kravtsov \inst{1,11}
        \and
            P. Lundqvist \inst{13}
        \and
            S. Moran \inst{14}
        \and
            A. Pastorello \inst{5}
        \and
            A. Popowicz \inst{15}
        \and 
            I. Salmaso \inst{16}
        \and 
            M. Stritzinger \inst{17}
          }

   \institute{
        Department of Physics and Astronomy, University of Turku, 20014 Turku, Finland
    \and 
        Cosmic Dawn Center (DAWN), Niels Bohr Institute, University of Copenhagen, Jagtvej 128, 2200 København N, Denmark
    \and 
        School of Sciences, European University Cyprus, Diogenes street, Engomi, 1516 Nicosia, Cyprus
    \and  
        Astrophysics Research Centre, School of Mathematics and Physics, Queen’s University Belfast, Belfast BT7 1NN, UK
    \and  
        INAF – Osservatorio Astronomico di Padova, Vicolo dell'Osservatorio 5, I-35122 Padova, Italy
    \and  
        INAF – Osservatorio Astronomico di Brera, Via E. Bianchi 46, I-23807 Merate (LC), Italy
    \and  
        International Centre of Supernovae (ICESUN), Yunnan Key Laboratory of Supernova Research, Yunnan Observatories, Chinese Academy of Sciences (CAS), Kunming, 650216, China
    \and 
        School of Physics, University College Dublin, Belfield, Dublin 4, Ireland
    \and 
        Institute of Space Sciences (ICE-CSIC), Campus UAB, Carrer de Can Magrans, s/n, E-08193 Barcelona, Spain
    \and  
        Institut d'Estudis Espacials de Catalunya (IEEC), 08860 Castelldefels (Barcelona), Spain
    \and  
        Finnish Centre for Astronomy with ESO (FINCA), University of Turku, Vesilinnantie 5, Quantum, 20014 Turku, Finland
    \and  
        Department of Physics, University of Warwick, Gibbet Hill Road, Coventry CV4 7AL, UK
    \and 
        The Oskar Klein Centre, Department of Astronomy, Stockholm University, AlbaNova, SE-10691, Stockholm, Sweden
    \and 
        School of Physics and Astronomy, University of Leicester, University Road, Leicester LE1 7RH, UK
    \and 
        Silesian University of Technology, Akademicka 16, Gliwice, Poland.
    \and 
        INAF - Osservatorio Astronomico di Capodimonte, Salita Moiariello 16, 80131 Napoli, Italy
    \and 
        Department of Physics and Astronomy, Aarhus University, Ny Munkegade 120, DK-8000 Aarhus C,
Denmark
    }
   \date{Received 18.6.2026; accepted 17.9.2026}

 
  \abstract
   {}
   {The core-collapse supernovae (CCSNe) 2022prr, 2023ucy, and 2024ljc exploded within the central regions of the nearby luminous infrared galaxy (LIRG) NGC~6745.
   We present our follow-up data with the aim to characterise the spectrophotometric evolution of these CCSNe, determine their host-galaxy extinction, and to define their physical nature.
   }
    {The presented spectrophotometric data set covers the optical and near-infrared region for our three supernovae (SNe). The data are analysed and compared to other hydrogen-rich CCSNe, and a host-galaxy extinction value $A^\text{host}_V$ is derived based on broadband photometry for each of the three targets. 
    A spectral energy distribution (SED) for the host galaxy is constructed, and a combination of different galactic components are fitted to the SED to estimate the intrinsic CCSN rate and the age of the starburst of NGC~6745.}
  {SN~2022prr is a SN~2009ip-like event, whose light curve evolution  consists of two consecutive luminous events A and B with peak magnitudes of $M_r = -14.8$ and $M_r = -18.3$~mag, respectively. The high degree of $V$ and $R$-band polarisation ($P \sim 2.8$ \%) near maximum light suggests interaction with highly aspherical circumstellar material. The spectral evolution of the SN is dominated by narrow Balmer emission lines, which have a complex evolution over time. SN~2022prr joins the growing sample of SN~2009ip-like events that show early `flash ionisation' features of C~{\sc iii}, N~{\sc iii}, and He~{\sc ii}. 
  SN~2023ucy is found to be a normal Type~IIP SN. However, SN~2024ljc is a rapidly evolving Type~IIb SN with a post-maximum $r$-band decline of roughly 3.2~mag by +41~d whereas normal Type~IIb SNe decline typically by $1.5 \pm 0.3$~mag during the same timescale. 
  While the events exploded in a LIRG, the SNe are obscured by only moderate $V$-band host extinctions ranging from 0 to 1 mag. Based on the SED modelling we estimate an age of the host starburst to be $\sim$40 Myr, and the intrinsic CCSN rate to be $0.4 \pm 0.1$ SN per year.}
   {}


   \maketitle
%

\section{Introduction}

Due to their intrinsic brightness supernovae (SNe) can be observed in other galaxies up to cosmological distances. The most massive stars in our Universe ($\geqslant 8 M_\odot$) end their lives as core-collapse SNe \citep[CCSNe; ][]{Smartt2009}; these explosions terminally destroy the star leaving behind a compact remnant (i.e. a neutron star or a black hole). CCSNe are traditionally divided into two spectroscopically distinct categories: hydrogen-poor Type I SNe, and hydrogen-rich Type II SNe \citep{Minkowski1941}. Hydrogen-poor CCSNe can be further divided into Type Ib (helium-rich) and Type Ic (helium-poor) SNe. Subtypes of Type II SNe include Type IIn  identified by narrow emission lines of hydrogen, and Type IIb, which is an intermediate class between Type Ib and Type II SNe \citep{Filippenko1997}.

Luminous infrared galaxies (LIRGs) are by definition very bright galaxies at infrared (IR) wavelengths ($10^{11} L_\odot \leq L_\text{IR} < 10^{12} L_\odot$). LIRGs can have notably higher star formation rates compared to normal galaxies, which makes them the ideal environment for studying recent massive star formation \citep{Perez2021}. Massive stars that explode as CCSNe have relatively short life times in astronomical time scales of only a few million to a few tens of millions of years. Therefore, CCSNe trace the ongoing star formation. However, the number of discovered CCSNe in LIRGs is disproportionately low largely due to high host-galaxy extinction \citep{Kankare2008, Kankare2012b, Kankare2014, Kankare2021, Kool2018}. This effect is caused by large amounts of dust in these galaxies, which efficiently absorb and scatter visible light. Therefore, a large fraction of CCSNe in these galaxies can appear notably fainter or remain completely undetected at optical or IR wavelengths \citep{Mattila2012, Fox2021, Mantynen2025, Mantynen2026}.

LIRGs are relatively rare in the local Universe, but they dominate both star formation and CCSN rates in cosmological distances at the cosmic noon when the star formation rate in the Universe was at its peak \citep[e.g.][]{Magnelli2011, Madau2014}. 
LIRGs are often the result of multiple galaxies interacting, colliding, or merging with each other \citep{Sanders1988}. The star formation rate in these galaxies is strongest in their central regions. Recent findings suggest, that LIRGs have starburst episodes that make them produce predominantly certain types of CCSNe based on the age of this starburst phase \citep{Kankare2021}. 

In this article, we study the recent CCSNe SN~2022prr, SN~2023ucy, and SN~2024ljc in the LIRG NGC~6745. The host galaxy and the transients are described in Sect. \ref{sec:sample}. The spectrophotometric observations of the targets are presented in Sect.~\ref{sec:data}. The light curves and spectroscopic evolution are analysed in Sect. \ref{sec:analysis}, along with a model for the spectral energy distribution (SED) of the galaxy, and estimates for host-galaxy extinction derived from light curve comparisons. Discussions are provided in Sect. \ref{sec:discussion} and the final conclusions are presented in Sect. \ref{sec:conclusions}.

\section{Sample}\label{sec:sample}

\begin{figure}
\includegraphics[trim={2.5cm 12cm 2.5cm 2.5cm},clip,width=\linewidth]{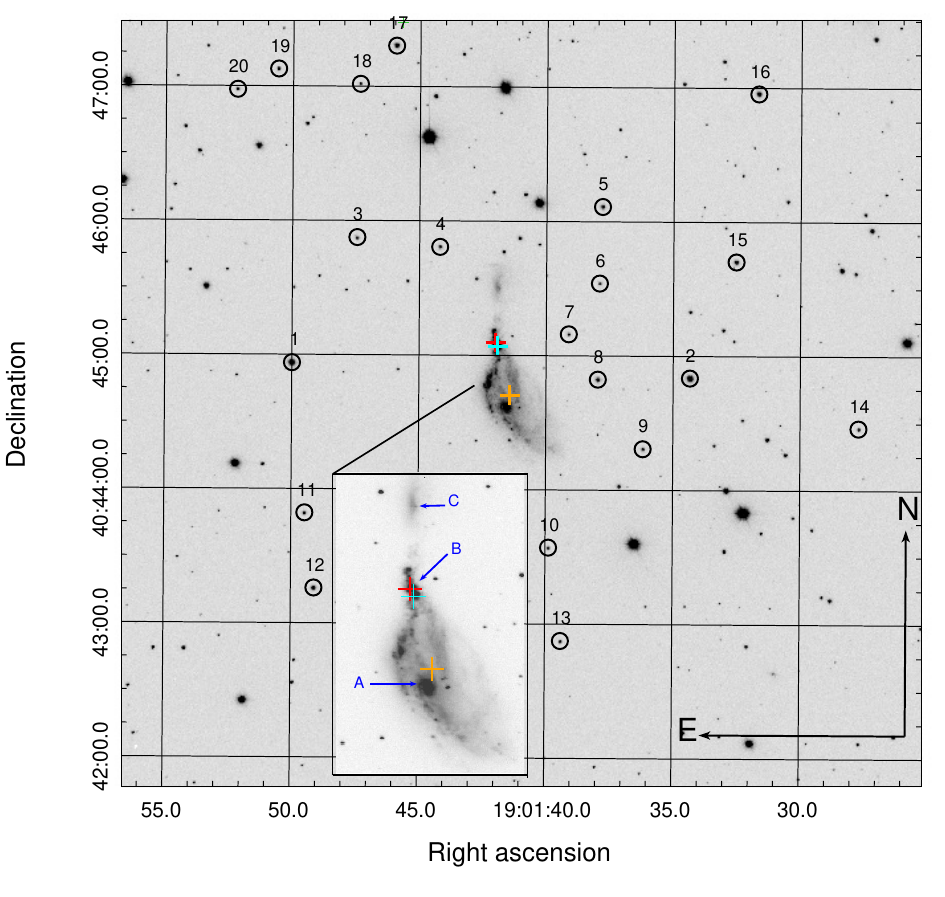}
\caption{Image of the field of NGC~6745 in $r$-band observed with ALFOSC at the Nordic Optical Telescope (MJD = 60525.0). Locations of the transients SN~2022prr, SN~2023ucy, and SN~2024ljc are marked in the image with cyan, red, and orange crosshairs, respectively. The field stars used to calibrate the photometry are circled in the image. The galaxy system components a, b, and c are marked in the subpanel image.}
\label{fig:FC}
\end{figure}

The LIRG NGC~6745 (IRAS 19000+4040) is an interacting galaxy system that is also known as the `Bird's Head galaxy', see Fig. \ref{fig:FC}. The system has been considered a triple system composed by the main galaxy component NGC~6745a, and two smaller components NGC 6745b and NGC~6745c \citep[for the nomenclature see][]{1978SvAL....4..261K, 1979SvAL....5..266V}. However, it appears that only the tidally disturbed large spiral galaxy NGC~6745a and the smaller northern early-type NGC~6745c companion are distinct galaxies and the NGC~6745b component is an interaction zone that forms a bridge between the two galaxies as a product of the past collision \citep[e.g.][]{Grijs2003}. The galaxy component NGC~6745c is absent of H~{\sc i} emission, whereas the interaction zone NGC~6745b and the eastern spiral arm of NGC~6745a are enhanced in star formation based on young and massive ($10^{6.5} M_{\odot} < M < 10^8 M_{\odot}$) super star clusters identified by \cite{Grijs2003}. Scaled with the adopted luminosity distance of 66.7 Mpc (see below), the infrared luminosity of NGC~6745 is $10^{11.0} L_{\odot}$, which makes the system a borderline LIRG at the low-luminosity end. 

Until recently, only one SN has previously been detected in the NGC~6745 system, the Type II SN~1999bx \citep{Jha1999, Green1999}. 
In archival pre-explosion images of the field of SN~1999bx from the \textit{Hubble} Space Telescope, \citet{VanDyk2003} identified four bright and blue sources consistent with the error region of the explosion site, which were likely clusters or luminous stellar objects. Recently, wide-field transient surveys have discovered three new SNe in this nearby galaxy in the time span of just two years: SN~2022prr, SN~2023ucy, and SN~2024ljc. These are all hydrogen-rich SNe akin to SN~1999bx. SN~2022prr is a SN~2009ip-like Type IIn event \citep{Jaeger2022}, SN~2023ucy is a Type IIP SN \citep{Taguchi2023}, and SN~2024ljc is a member of the Type IIb subclass \citep{Wise2024}.

We assume $H_0 = 73$ km s$^{-1}$ Mpc$^{-1}$, $\Omega_\text{M} = 0.27$, and $\Omega_\text{V} = 0.73$. 
The redshift $z = 0.0152$ and the Virgo infall corrected host luminosity distances of 66.7 Mpc were adopted via the NASA Extragalactic Database (NED)\footnote{\url{https://ned.ipac.caltech.edu}}, corresponding to a distance modulus of $\mu = 34.12$~mag. The Galactic dust map based Milky Way (MW) extinction of $A^\text{MW}_V = 0.373$~mag \citep{Schlafly2011} was retrieved from NED. For reddening, a standard extinction law by \cite{Cardelli1989} was adopted with $R_V = 3.1$.

\subsection{SN~2022prr}\label{sec:sample22prr}

\cite{Stanek2022} reported the discovery of the transient ASASSN-22jn (SN~2022prr) by the All Sky Automated Survey for SuperNovae (ASAS-SN) programme at a $g$-band magnitude of 17.3~mag on 2022 July 27 at 07:12:00 UT (MJD = 59787.30), with a $g$-band non-detection of >18.6~mag on 2022 July 23 at 11:31:12 UT (MJD = 59783.48). A Type IIn classification spectrum was reported by \cite{Jaeger2022} to the Transient Name Server (TNS) and obtained with the University of Hawai'i 88-inch (2.24~m) telescope on 2022 July 28 at 10:40:00 UT (MJD = 59788.44). 
SN~2022prr is located at $\alpha = 19^{\mathrm{h}}01^{\mathrm{m}}41\fs 90$, $\delta = +40\degr 45\arcmin 03\farcs 67$, which is $2\farcs 2$ from the centre of the component NGC~6745b and corresponds to a projected distance of 0.65 kpc. The projected distance to the centre of the main component NGC~6745a is $27\farcs 1$ (8.8~kpc).

\subsection{SN~2023ucy}\label{sec:sample23ucy}

\cite{De2023} reported the discovery of the transient ZTF23abhzxwm (SN~2023ucy) by the Zwicky Transient Facility (ZTF) survey at an $r$-band magnitude of 17.9~mag on 2023 October 05 at 03:38:05 UT (MJD = 60222.15), with a $g$-band non-detection of >19.9~mag on 2023 October 3 at 04:45:35 UT (MJD = 60220.20). A Type II classification spectrum was reported to the TNS by \cite{Taguchi2023} and obtained with the Okayama Observatory of Kyoto University 3.8~m Seimei telescope on 2023 October 16 at 11:42:51 UT (MJD = 60233.49). Furthermore, another public spectrum of SN~2023ucy was reported to the TNS by \cite{Teja2023}, obtained with the 2 m Himalayan Chandra Telescope (HCT-2m) with the Himalaya Faint Object Spectrograph And Camera (HFOSC) on 2023 October 20 at 15:26:41 UT (MJD = 60237.64). SN~2023ucy is located at $\alpha = 19^{\mathrm{h}}01^{\mathrm{m}}41\fs 99$, $\delta = +40\degr 45\arcmin 05\farcs 80$, which is consistent with the centre of the NGC~6745b component. The projected distance of SN~2023ucy to the centre of the main component NGC~6745a is $29\farcs 3$ (9.5~kpc).

\subsection{SN~2024ljc}\label{sec:sample24ljc}

\cite{Perez2024} reported the discovery of the transient ZTF24aasdtkt (SN~2024ljc) by the ZTF sky survey at an $r$-band magnitude of 18.7~mag on 2024 June 15 at 08:15:05 UT (JD = 2460476.84), with an $r$-band non-detection of >20.5~mag on 2024 June 12 at 10:12:50 UT (MJD = 60473.43). A  SN classification spectrum was reported to the TNS by \cite{Perez2024b}, obtained with the Liverpool Telescope on 2024 June 17 at 02:57:11 UT (MJD = 60478.10). They remarked, that the spectrum had a red continuum, likely due to obscuration and reddening in the host galaxy, and shows emission lines from the host galaxy. A later re-classification as a Type IIb was reported by \cite{Wise2024} on 2024 June 27 at 11:14:37 UT (MJD = 60488.47) and \cite{Gomez2024} on 2024 July 12 at 13:41:55 UT (MJD = 60503.57). SN~2024ljc is located at $\alpha = 19^{\mathrm{h}}01^{\mathrm{m}}41\fs 42$, $\delta = +40\degr 44\arcmin 42\farcs 24$, which is $5\farcs 7$ from the centre of the main galaxy NGC 6745a and corresponds to a projected distance of 1.8~kpc.


\section{Data}\label{sec:data}

\subsection{Optical and NIR photometry}\label{sec:photometry}

Our photometric follow-up of SN~2022prr, SN~2023ucy, and SN~2024ljc was carried out with the Nordic Optical Telescope \citep[NOT; ][]{Djupvik2010} on La Palma, Spain, via the NOT Unbiased Transient Survey 2 (NUTS2) programme. 
The optical imaging was obtained with the Alhambra Faint Object Spectrograph and Camera (ALFOSC) instrument in $uBgVri$ bands, and the near-IR (NIR) imaging was taken with the NOT near-infrared Camera and spectrograph (NOTCam). A few optical images of SN~2022prr and SN~2023ucy were obtained using the Copernico 1.82~m telescope with the Asiago Faint Object Spectrograph and Camera (AFOSC) and the 67/92~cm Schmidt telescope with the Moravian CCD camera in the Asiago Observatory, Italy. Additionally, public follow-up data from the ZTF survey \citep{Bellm2019} in $g$ and $r$ bands was made use of for all three targets. 

The ALFOSC imaging was reduced using the NUTS2 ALFOSCGUI pipeline\footnote{Foscgui is a graphic user interface aimed at extracting SN spectroscopy and photometry obtained with FOSC-like instruments. It was developed by E. Cappellaro. A package description can be found at \url{https://sngroup.oapd.inaf.it/foscgui.html}.}, which applies the standard reduction steps of bias substraction and flat-field correction to the images. 
The NOTCam imaging was reduced using a modified version of the NOTCAM 2.5 package\footnote{\url{https://www.not.iac.es/instruments/notcam/guide/observe.html}} within IRAF \citep{Tody1986, Tody1993}. The reductions steps included flat-field and distortion correction, sky subtraction, and stacking the individual exposures. The imaging from Asiago was reduced similarly with pipeline tools.

Template image subtraction was carried out with a slightly modified version of the ISIS 2.2 software package \citep{Alard1998, Alard2000} due to the strong host galaxy background emission. Selected ALFOSC images from our follow-up data sets were used as template images for the other SNe obtained before the SN explosion or after the SN had faded below a detection limit. 
The point spread function (PSF) photometry of SN~2022prr, SN~2023ucy, and SN~2024ljc was derived with the QUBA pipeline \citep{Valenti2011} based on standard IRAF tasks. The $ugri$ photometry was calibrated against the field star magnitudes in the Sloan Digital Sky Survey \citep[SDSS; ][]{York2000}, and the $BV$ magnitudes of the field stars were converted from the SDSS magnitudes using the transformations by \cite{Jester2005}. The $JHK$ band magnitudes of the field stars from the Two Micron All Sky Survey \citep[2MASS; ][]{Skrutskie2006} were used to calibrate the NIR magnitudes of the three SNe. The reported $BVJHK$ photometry is in the Vegamag system and the $ugri$ photometry in the AB system. The full photometry tables for the targets are found in the Appendix Tables \ref{tab:phot_SN2022prr}, \ref{tab:phot_SN2023ucy}, and \ref{tab:phot_SN2024ljc}.

\subsection{Spectroscopy}

The spectroscopic monitoring of SN~2022prr, SN~2023ucy, and SN~2024ljc was carried out with the NOT using the ALFOSC instrument via the NUTS2 programme. These observations were taken primarily with the grism Gr\#4 with the wavelength range of $3200 - 9600$~Å and a resolution of $\lambda/\Delta \lambda \sim 360$ or 280 with $1\farcs 0$ or $1\farcs 3$ slit, respectively. The ALFOSC spectra were reduced in a standard manner using the NUTS2 ALFOSCGUI pipeline. In this process the two-dimensional spectra were trimmed, bias and overscan subtracted, flat-field corrected using halogen lamp flats, and cosmic rays were removed from the data. The extracted one-dimensional spectra were wavelength calibrated using arc lamp spectra, and relative flux calibrated using observations of spectroscopic standard stars. Absolute flux calibration was carried out based on broadband photometry obtained close in time.

One additional late-time spectrum of SN~2022prr was obtained with the 2~m Faulkes Telescope North (FTN) using the Folded Low Order whYte-pupil Double-dispersed Spectrograph (FLOYDS) on Haleakala, Hawai'i, and one spectrum of SN~2023ucy was obtained with the 1.82~m Copernico telescope. The spectra were reduced based on standard methods using pipeline tools. The spectroscopic log of obsevations are reported in the Appendix Tables \ref{tab:spec22prr} to \ref{tab:spec24ljc}.

\subsection{Swift observations}

The High Energy Astrophysics Science Archive Research Center (HEASARC) database contains publicly available Level 2 pre-processed observations of SN~2022prr obtained with the Neil Gehrels Swift Observatory \citep{Gehrels2004} using the Ultra-violet Optical Telescope \citep[UVOT; ][]{Roming2005} and the X-Ray Telescope \citep[XRT; ][]{Burrows2005} instruments. The UVOT images were processed with the HEASARC High Energy Astrophysics software (HEAsoft) package. The HEAsoft uvotimsum task was used to combine the series of individual exposures to increase the S/N ratio. The uvotsource task in HEAsoft was used to carry out the aperture photometry of the transient with source and background aperture radii of 5\arcsec and 30\arcsec, respectively. The contribution of the strong host galaxy contamination at the location of the SN was measured from late-time UVOT images obtained on 2025 March 6 (MJD 60740.80) and subtracted accordingly. We report 8 epochs of UVOT Vegamag magnitudes from 2022 July 29 (MJD 59789.19) up to 2022 August 19 (MJD 59810.43) following which the contrast between the SN and the host galaxy was found to become generally too strong for robust template magnitude subtraction.

The sosta task in HEAsoft was used to calculate the count rates at the coordinates of SN~2022prr in the 0.2 $-$ 10 keV XRT images. This did not reveal a source with a S/N~$> 2$ detection. The weighted average Galactic H~{\sc I} column density for the coordinates of SN~2022prr is 8.35 $\times$ 10$^{20}$ cm$^{-2}$ based on the HEASARC nH tool\footnote{\url{https://heasarc.gsfc.nasa.gov/docs/tools.html}}. Based on this and assuming a power law photon index $\Gamma = 2$ the photon count upper limits were converted into unabsorbed flux upper limits using the HEASARC WebPIMMS tool and subsequently into luminosity upper limits taking into account the luminosity distance of SN~2022prr. The resulting upper limits are not very deep and would not exclude for example $\sim 10^{39}$ erg s$^{-1}$ X-ray emission detections for SN~2009ip near event B peak \citep{Ofek2013}. The photometric UVOT and XRT values are reported in Table \ref{tab:uv_22prr}.

\subsection{Polarimetry}

We carried out imaging polarimetry of SN~2022prr in the $V$ and $R$ bands using ALFOSC mounted on the NOT. Linear polarization measurements were obtained with a half-wave plate (HWP) set at four position angles ($0^{\circ}$, $22.5^{\circ}$, $45^{\circ}$, and $67.5^{\circ}$), in combination with a calcite plate. The data were reduced and analyzed following standard procedures \citep[e.g. ][]{Patat2017,Nagao2025} with the IRAF software package. After applying bias subtraction and flat-field corrections to all frames, aperture photometry was performed on both the ordinary and extraordinary beams of the source at each HWP angle. The aperture radius was set to 1.5 times the full width at half maximum (FWHM) of the PSF of the ordinary beam. The sky background was estimated from an annulus extending from four to five times the FWHM to minimize contamination from host-galaxy structure. Using these measurements, we derived the Stokes parameters, as well as the polarization degree and position angle. A correction for polarization bias was applied following the method by \citet[][]{Wang1997}.

\section{Analysis}\label{sec:analysis}

\subsection{Host-galaxy extinctions}\label{sec:extinction}

\begin{figure}
     \centering
     \includegraphics[width=\linewidth]{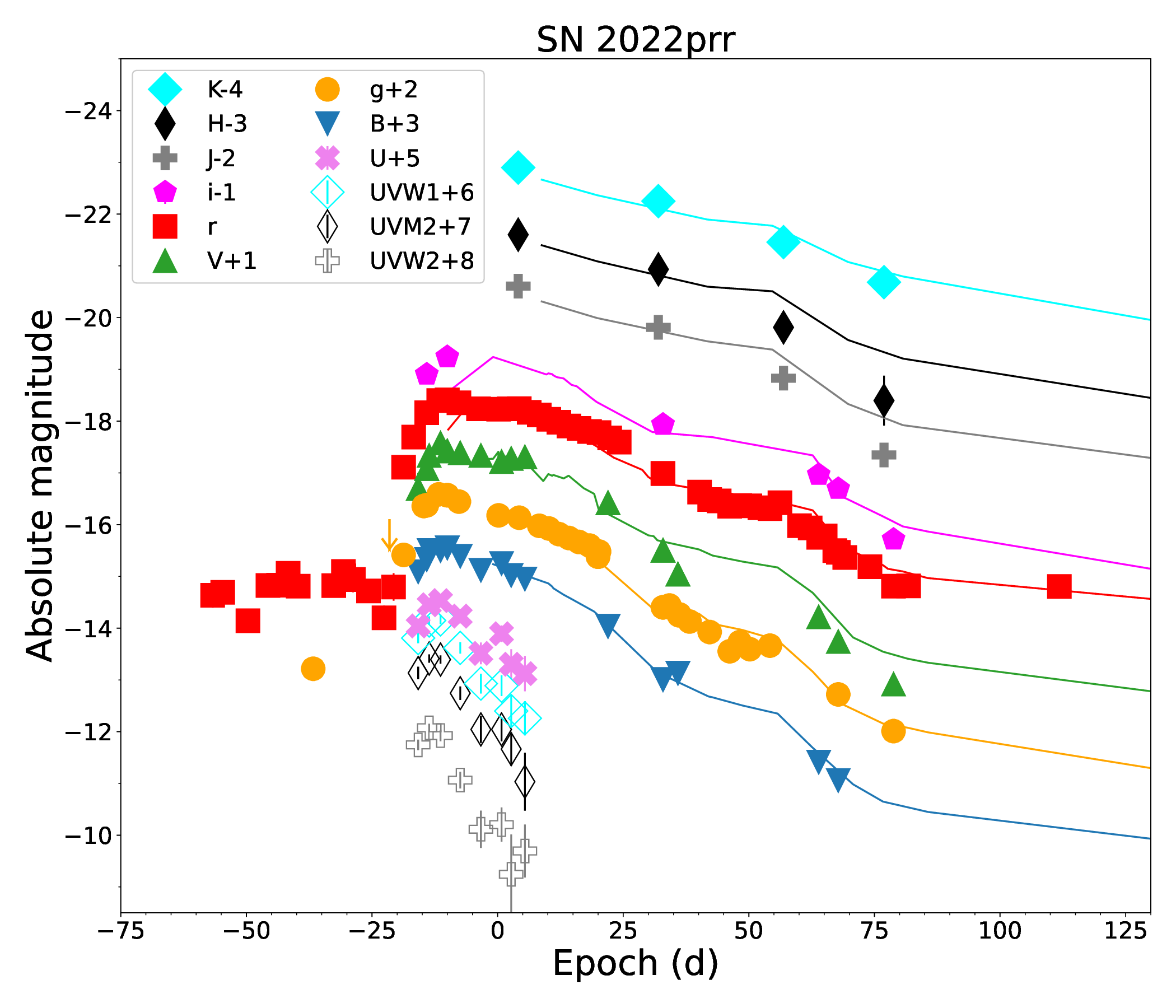}
     \includegraphics[width=\linewidth]{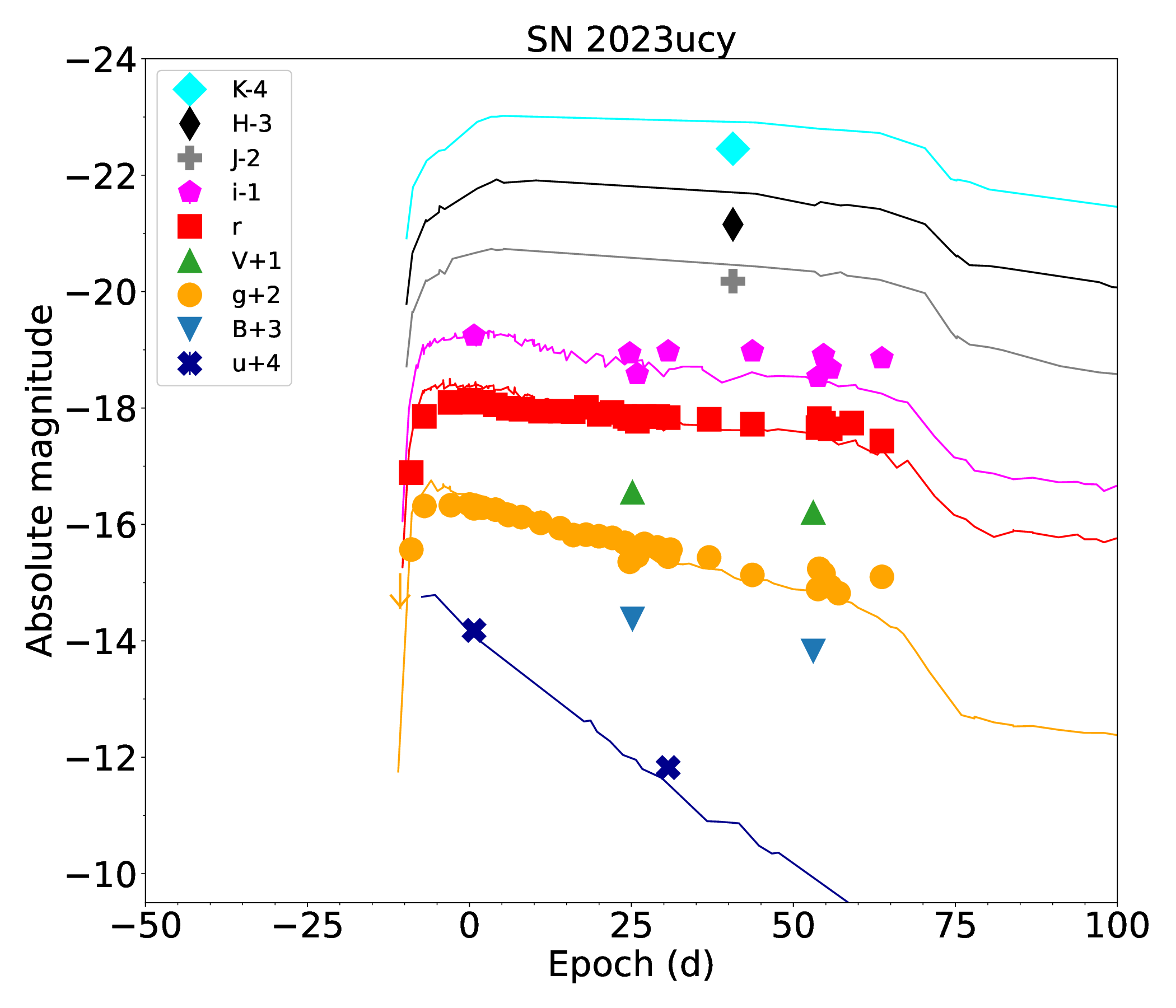}
     \includegraphics[width=\linewidth]{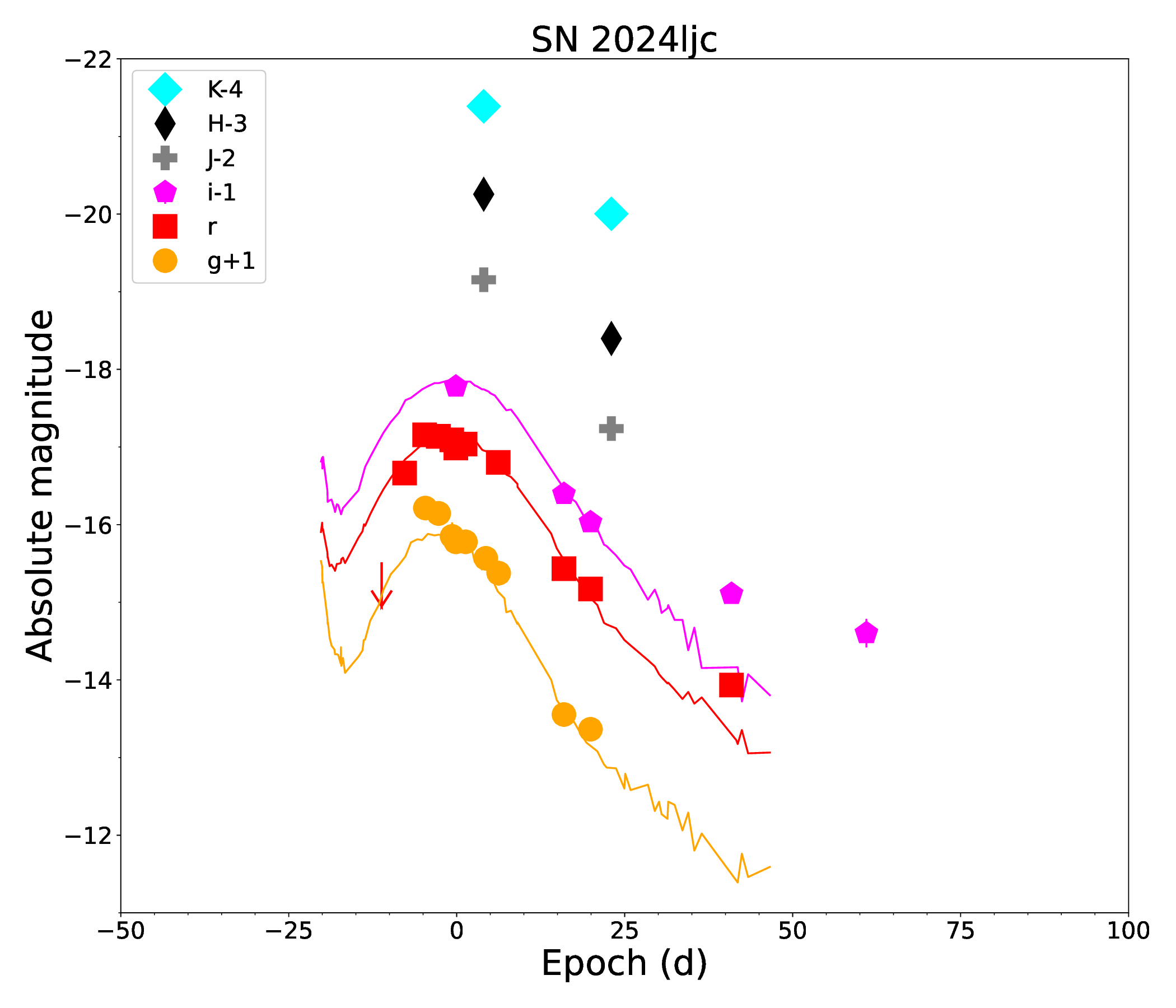}
     \caption{
     Light curves (points) of SN~2022prr (top), SN~2023ucy (middle), and SN~2024ljc (bottom), and the adopted templates (curves) of SN~2016bdu \citep{Pastorello2018}, SN~2023ixf \citep{Li2025, Jacobson2025}, and SN~2024aecx \citep{Xi2026}, respectively. The last non-detection of the transient by the survey that reported the discovery is noted with a downward pointing arrow. The light curves are corrected for the derived line-of-sight extinctions and the templates have been shifted in magnitude by a constant $C$ yielded by the fit.}
    \label{fig:extinction}
\end{figure}

The value of the host-galaxy extinction depends on the amount of foreground dust between the observer and the SN, and can vary greatly based on the exact location of the transient within the host galaxy. 
In particular, the extinction distribution in LIRGs can be very complex due to their often peculiar morphology, as is also the case with the multi-component system NGC~6745. For example, multiple CCSNe have been studied in the LIRG Arp 299 with host galaxy extinction estimates of $A_V$ ranging from 0.4 to 7~mag \citep{Kankare2021}. To estimate the host-galaxy extinction in the line-of-sight to SN~2022prr, SN~2023ucy, and SN~2024ljc, their broadband light curves were simultaneously compared with template SNe of the same subtype with similar temporal evolution and known extinction values, see Fig. \ref{fig:extinction}. The chosen templates were selected with the requirement of having light curves that are well sampled, include coverage of the rise phase, and extend up to $K$-band when necessary, and have an overall similar spectral and photometric evolution to our events. This did not result in many suitable options. The uncertainty between the intrinsic colour differences between the template and the SN is a caveat of this method. However, as reported below, the derived extinctions were relatively small, and it is unlikely that this method has heavily underestimated the extinctions. We estimated the $V$-band host-galaxy extinction, $A_V$, via minimising the $\chi^2$ value of the fit \citep{Kankare2014}. Other derived parameters were the epoch of the $r$-band light curve maximum, $t_0$, and a constant magnitude difference, $C$, applied in all the fitted bands between the studied SN and the comparison event. The equivalent width (EW) of the Na~{\sc i}~D interstellar absorption feature in the SN spectra is often used as an empirical method to estimate the host-galaxy line-of-sight extinction. However, the Na~{\sc i}~D EW is not a reliable proxy of host-galaxy reddening for SNe, in particular when low-resolution spectra are used \citep[e.g. ][]{Phillips2013}; therefore, we have opted not to use this method.

For SN~2022prr the most suitable comparison target was the SN~2009ip-like SN~2016bdu \citep{Pastorello2018}, which has a very similar light curve evolution during the event B of the transients. From this comparison we derived a small host-galaxy extinction value of $A_V = 0.1_{-0.1}^{+0.2}$~mag. The magnitude difference between the transients was $C = -0.3^{+0.1}_{-0.2}$~mag, which was applied to the light curve of SN~2016bdu in all bands. 
From this fit we also received an estimate of $\text{MJD} = 59805.0 \pm 1$ as the light curve peak epoch of SN~2022prr based on the light curve maximum of SN 2016bdu. However, the early evolution of the template SN is more rapid than that of SN~2022prr; therefore, we derived the maximum epoch (reference epoch $t=0$~d) via a polynomial fit instead, see Sect. \ref{sec:LC22prr}.

The host-galaxy extinction in the line-of-sight to SN~2023ucy was estimated by comparing the \textit{ugriJHK} light curve evolution of the SN to that of another Type IIP SN~2023ixf, which has similar absolute peak and plateau magnitudes \citep{Li2025, Jacobson2025}. Although SN~2023ixf has some departures from a prototypical SN~II, such as early excess in its light curve, and flash features in its early spectra, the intrinsic colours of Type IIP SNe are dominated by the temperature during the plateau phase, powered by hydrogen recombination. From this comparison fit an extinction value of $A_V = 0.4_{-0.1}^{+0.2}$~mag was retrieved, which is similar to that of SN~2022prr within errors. The magnitude difference between SN~2023ucy and SN~2023ixf was $C = -0.3_{-0.2}^{+0.1}$~mag, which was applied to the light curve of SN~2023ixf in all bands. From this fit we also estimate $\text{MJD} =60231 \pm 1$ as the light curve peak epoch of SN~2023ucy, which is adopted as the reference epoch ($t=0$~d) from here on.

The host-galaxy extinction of SN~2024ljc was estimated by comparing the \textit{gri} light curve evolution to that of another Type IIb SN~2024aecx, which has a similarly fast decline rate \citep{Xi2026}. This comparison yielded a host-galaxy extinction of $A_V = 1.1_{-0.1}^{+0.2}$~mag. The magnitude difference between the events was $C = 0.4_{-0.2}^{+0.1}$, which was added to the light curve of SN~2024aecx in all bands. From this fit we also estimate $\text{MJD} =60485 \pm 1$ as the light curve peak epoch of SN~2024ljc, which is adopted as the reference epoch ($t=0$~d) from here on.

The $g-r$ colour evolution of SN~2022prr, SN~2023ucy, and SN~2024ljc is very similar to their template SNe, and in the case of SN~2022prr and SN~2023ucy to multiple other event of the same subtype, see Fig. \ref{fig:colorevolution}. For all three targets, the $g-r$ colour was relatively close to zero near the light curve peak ($g-r = -0.09 \pm 0.01$, $-0.21 \pm 0.01$, and $0.21 \pm 0.01$~mag for SN~2022prr, SN~2023ucy, and SN~2024ljc, respectively).

\subsection{Light curve evolution}
\subsubsection{Light curve evolution of SN~2022prr}\label{sec:LC22prr}

\begin{figure}
    \centering
    \includegraphics[width=\linewidth]{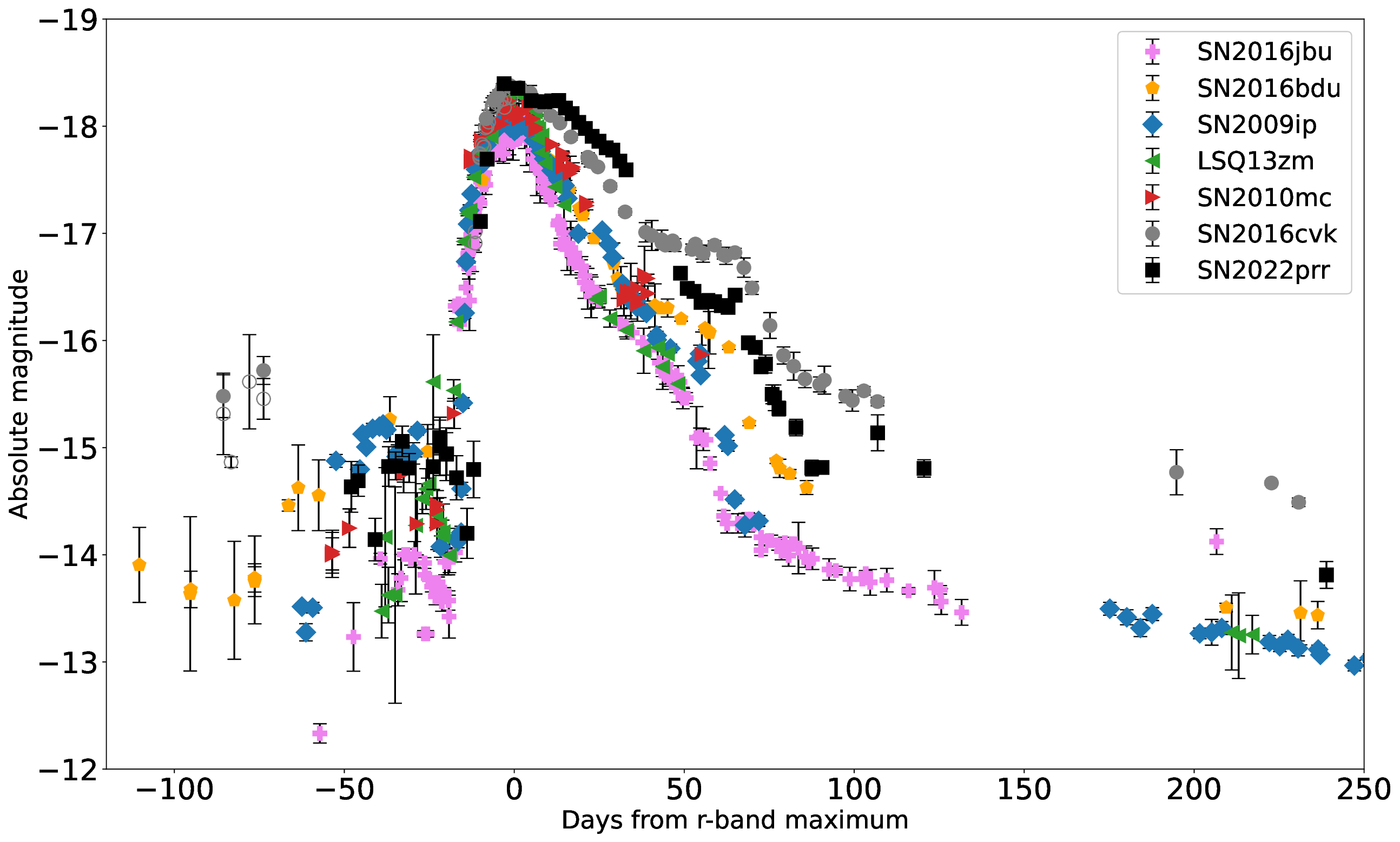}
    \caption{Extinction-corrected absolute $r$ (or $R_\text{AB}$) light curves for a selection of SN~2009ip-like events SN~2016jbu \citep{Brennan2022a}, SN~2016bdu \citep{Pastorello2018}, SN~2009ip \citep{Fraser2013,Fraser2015}, LSQ13zm \citep{Tartaglia2016}, SN~2010mc \citep{Ofek2013b}, and SN~2016cvk \citep{Matilainen2025}. Early $V$-band data are also included for SN~2016cvk.}
    \label{fig:22prr_LCcomp}
\end{figure}

\begin{figure}
\centering
\includegraphics[width=\linewidth]{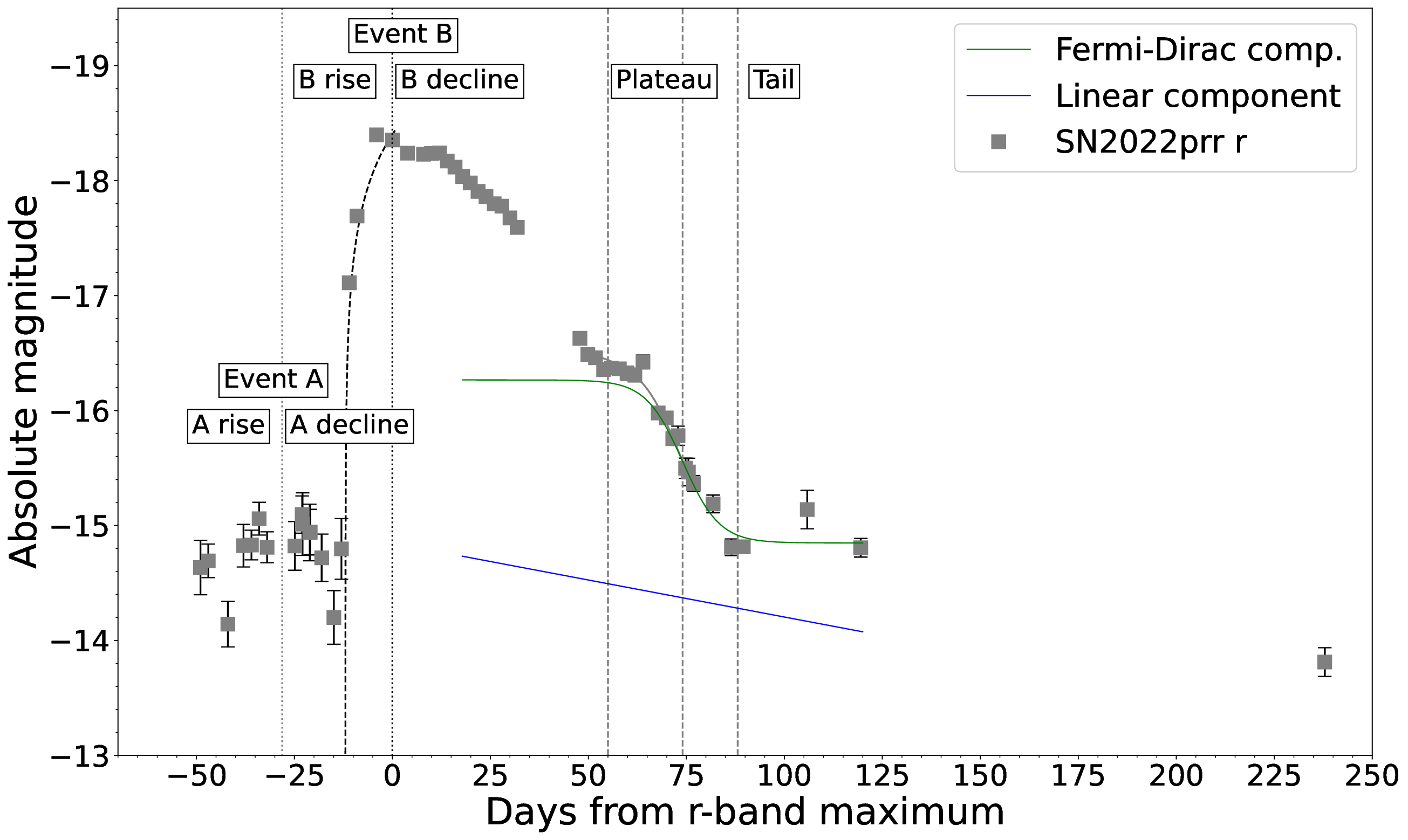}
\caption{Light curve phases of SN~2022prr. Fit results are shown for the event B rise phase and the plateau phase (grey dashed and solid curves, respectively), along with the two components of the plateau phase fit.}
\label{fig:LCphases22prr}
\end{figure}

SN~2022prr is a member of SN~2009ip-like events \citep[see e.g. ][]{Fraser2013, Ofek2013, Tartaglia2016, EliasRosa2016, Pastorello2018, Brennan2022a, Matilainen2025}. These transients have a very similar light curve evolution with two major events: a less bright precursor event A, followed by the main event B, see Fig. \ref{fig:22prr_LCcomp}. After a steep decline from the event B peak, their light curves flatten to an almost horizontal plateau phase, which turns to another sharp decline in the light curve, a so called `knee/ankle-stage' \citep[see e.g. ][]{Graham2014}. This drop is then followed by a late linear tail phase that declines very slowly. 

Following the steps taken in \cite{Matilainen2025} for the light curve of SN~2016cvk and other SN~2009ip-like events, selected functions were fitted to the different light curve phases of SN~2022prr. The events A and B were fitted with a second order polynomial via minimizing the $\chi^2$ value. The peak of the polynomial fit of event B is used as a reference epoch ($t=0$~d) throughout the light curve analysis, and the maximum of event A is similarly measured from the peak of its second-order polynomial fit. The plateau and the knee-ankle phase was fitted with a combination of a Fermi-Dirac function and a linear part, similar to the method used by \cite{Anderson2014} and \cite{Olivares2010} for the plateau phase of Type IIP SNe, see Fig. \ref{fig:LCphases22prr}. Finally, the tail was treated as a simple first-order polynomial. From these fits we determine $t_\text{peak,A}=-28$~d (measured from event B peak) as the peak epoch of event A, with a corresponding peak absolute magnitude of $M_\text{$r$,A}=-14.8 \pm 0.1$~mag in $r$-band. The event B peak absolute magnitude is estimated as $M_\text{$r$,B}=-18.3 \pm 0.1$~mag, at $\text{MJD} =59796.3$. From the Fermi-Dirac fits to the plateau phase we get $t_\text{mid, plateau}=63 \pm 1$~d as the middle epoch of the plateau, and the corresponding $r$-band absolute magnitude at this epoch is $M_\text{$r$, plateau}=-16.2 \pm 0.1$~mag. The duration of the plateau phase is $22 \pm 1$~d. We determine $t_\text{start, tail}= 88 \pm 1$~d as the beginning of the linear tail phase, and measure a decline rate of $\gamma_\text{tail}=0.6$~mag/100~d for the light curve in this phase.

The event A absolute peak magnitude of SN~2022prr is nearly identical to LSQ13zm and SN~2010mc \citep{Tartaglia2016, Ofek2013}. The event A shows a rise and decline phase, similar to SN~2009ip, SN~2016jbu, and LSQ13zm \citep{Fraser2013, Pastorello2013, Brennan2022a, Tartaglia2016}. The peak time of event A ($-28$ d) is similar to LSQ13zm, but somewhat late compared to other SN~2009ip-like transients, which vary between $-39$ and $-34$ d \citep{Matilainen2025}. The event B absolute peak magnitude is $-18.3$~mag, which is identical to SN~2016cvk, and in the brighter end of the narrow distribution of magnitudes between $-17.9$ and $-18.3$~mag of SN~2009ip-like transients \citep{Matilainen2025}. The plateau phase is also in line with other SN~2009ip-like transients. The plateau midpoint magnitude is very similar to that of SN~2016bdu \citep{Pastorello2018}. The midpoint of the plateau phase is rather late ($+63$ d), compared to the range between $+40$ and $+54$ d in the sample of \cite{Matilainen2025}. Similarly, the tail phase of SN~2022prr starts late ($+88$ d) compared to other SN~2009ip-like transients, for which the tail begins between $+67$ and $+82$~d. The tail of SN~2022prr declines slowly with a rate of $0.6 \pm 0.1$ mag in 100 days, similar for example within errors to that of SN~2016cvk \citep{Matilainen2025}. 

\subsubsection{Light curve evolution of SN~2023ucy}\label{sec:LC23ucy}

The light curve of SN~2023ucy follows a very typical evolution for a Type IIP SN. After a rapid rise to a peak $r$-band absolute magnitude of $M_\text{$r$,peak}=-18.0$~mag at $\text{MJD} =60231.6$, the light curve reaches a plateau from $+10$~d onward. This nearly horizontal phase lasts at least until $+64$~d, where our photometric coverage was interrupted due to solar conjunction. The average $r$-band absolute magnitude during the plateau is $M_\text{$r$,plateau}=-17.8$~mag. This makes the SN relatively luminous within the Type IIP subclass of CCSNe \citep[see e.g. ][]{Anderson2014}.

\subsubsection{Light curve evolution of SN~2024ljc}\label{sec:LC24ljc}

\begin{figure}
\centering
\includegraphics[width=\linewidth]{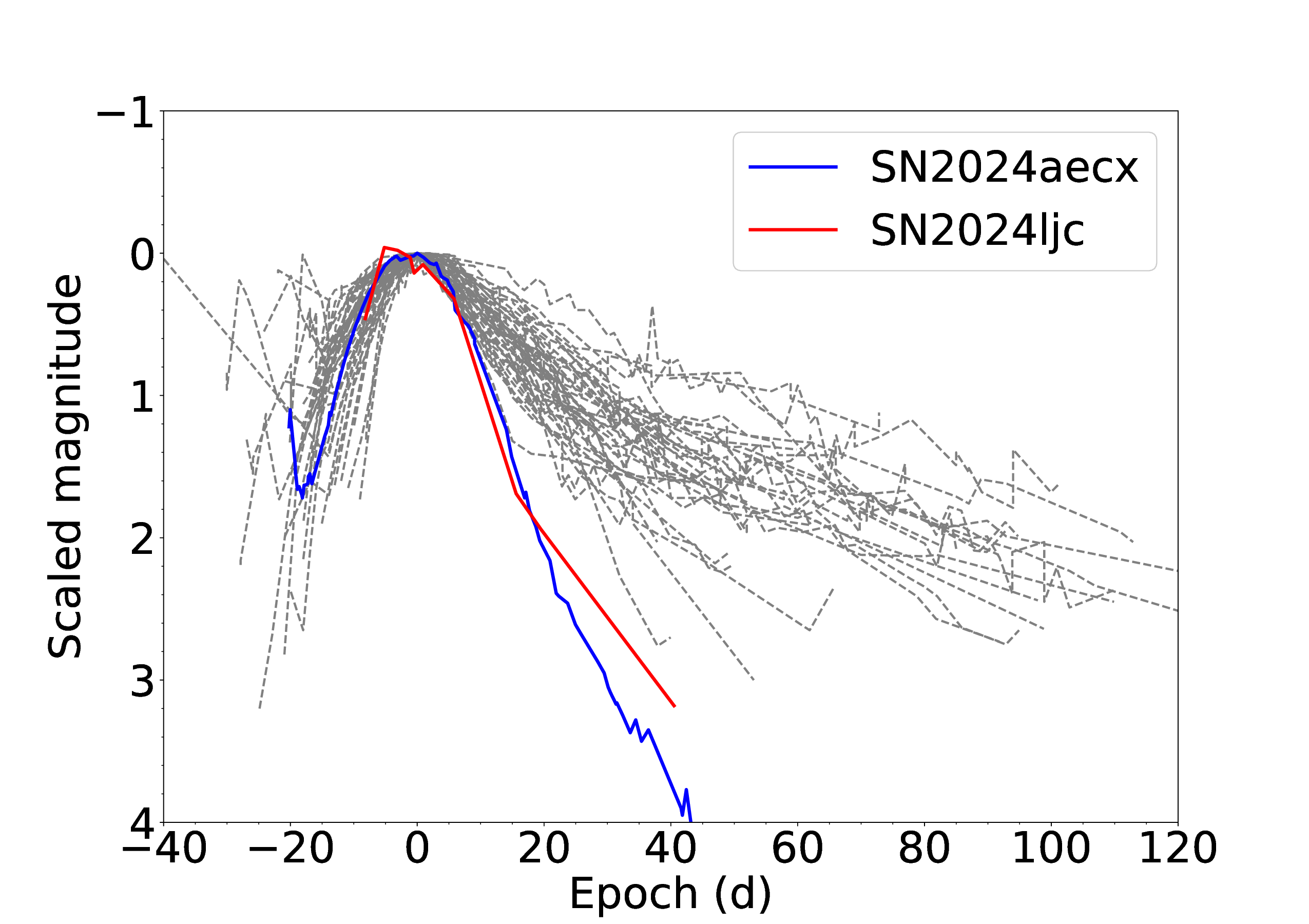}
\caption{Peak-normalised $r$-band light curves of SN~2024ljc (red) compared to those SN 2024aecx \citep[blue; ][]{Xi2026}, and a selection of other Type~IIb SNe from public ZTF data via ALeRCE (grey).}
\label{fig:comp_IIb}
\end{figure}

The light curve of SN~2024ljc has a peak $r$-band absolute magnitude of $M_\text{$r$,peak}=-17.1$~mag at $\text{MJD} =60483.4$. 
The rise to maximum light is not well documented in our data set; however, the 15 d decline rate after the peak is fast with post-maximum $\Delta m_{15}$ values of roughly 0.9, 1.3, 1.2, 1.0, 1.0, and 0.7~mag in $griJHK$, respectively.
In the $r$-band the light curve had declined by 3.2~mag in 41 days. This fast evolution is evident in a comparison to a sample of 43 transients classified as Type IIb SNe in the TNS database, which have publicly available ZTF $r$-band light curves via the Automatic Learning for the Rapid Classification of Events (ALeRCE) broker \citep{Forster2021}, and which have a well-sampled maximum light coverage, see Fig.~\ref{fig:comp_IIb}. For comparison, the 30 events in the abovementioned sample of Type~IIb SNe with light curves that extend beyond +41~d have typically declined by $1.5 \pm 0.3$~mag by that epoch. However, a notable exception to this is SN~2024aecx, which is a double-peaked and rapidly evolving stripped-envelope SN. Both \cite{Zou2026} and \cite{Xi2026} found weak signatures of H$\alpha$ in the early spectra of SN~2024aecx and considered the event to be a Type IIb SN; however, \cite{Tinyanont2026} has suggested the event is more akin to a Type Ic SN.

\subsection{Blackbody evolution of SN~2022prr}

To characterise the temperature and radial evolution of SN~2022prr, blackbody fits were computed to the spectral energy distributions (SEDs) constructed from available photometry between $-7.3$ and $+120.6$~d using the \textsc{SuperBol} package \citep{Nicholl2018}. 
The package generates an interpolated light curve for the bands used in the fits within the given time range, and creates a blackbody fit for each interpolated epoch. The photometry used in the fit includes optical data between $-6.0$ and $+120.6$ d, ultraviolet data between $-7.1$ and $+14.2$~d, and NIR data between $+13.4$ and $+86.2$ d (see Tables \ref{tab:phot_SN2022prr} and \ref{tab:uv_22prr}). The results for the blackbody temperature and radius of SN~2022prr are found in Fig. \ref{fig:bb_22prr}. 

The maximum blackbody temperature of SN~2022prr according to our fits is $T_\text{BB} = 16\,800 \pm 900$~K at $-7$~d from event B peak, which is consistent with other SN~2009ip-like transients. For comparison, the temperature of SN~2016cvk peaked at 14\,000~K around $-6$~d from its $r$-band peak \citep{Matilainen2025}, and SN~2016jbu reached its maximum temperature of $15\,000$~K at $-10$~d from peak \citep{Brennan2022b}. By the beginning of the plateau phase, the estimated $T_\text{BB}$ drops to around $5400 \pm 300$~K at $+51$~d and stays constant within errors for the duration of the plateau and the beginning of the tail phase. This is similar to that of other SN 2009ip-like transients, which stay between $5000 - 6000$~K during the plateau phase \citep{Matilainen2025, Brennan2022b, Fraser2013, Ofek2013, Pastorello2018, Thone2017, Tartaglia2016}. In these events the plateau can be interpreted as a signature of hydrogen recombination, supported by the observed temperatures and the presence of broad P Cygni features of the Balmer series and metal lines \citep[e.g. ][]{Salmaso2026}.

The evolution of the blackbody radius of SN~2022prr is also analogous to other SN~2009ip-like transients. At $+15$~d the radius reaches its largest value of $R_\text{BB} = (2.1 \pm 0.3) \times 10^{15}$~cm, which is close to the maximum radius of SN~2016cvk ($1.6 \times 10^{15}$~cm at $+24$~d), and nearly double the maximum radius found for SN~2016jbu ($1.2 \times 10^{15}$~cm at $+7$~d). The radius decreases from its peak value, to $(12.9 \pm 1.4) \times 10^{14}$~cm at the beginning of the plateau phase around $+51$~d, and to $(4.9 \pm 0.8) \times 10^{14}$~cm at the beginning of the tail phase around $+88$~d. SN~2016cvk had a roughly similar blackbody radius at the beginning of its tail phase ($8.5 \times 10^{14}$~cm at $+82$~d).

\subsection{Polarimetry of SN~2022prr}

\begin{figure}
     \centering
     \includegraphics[width=0.95\linewidth]{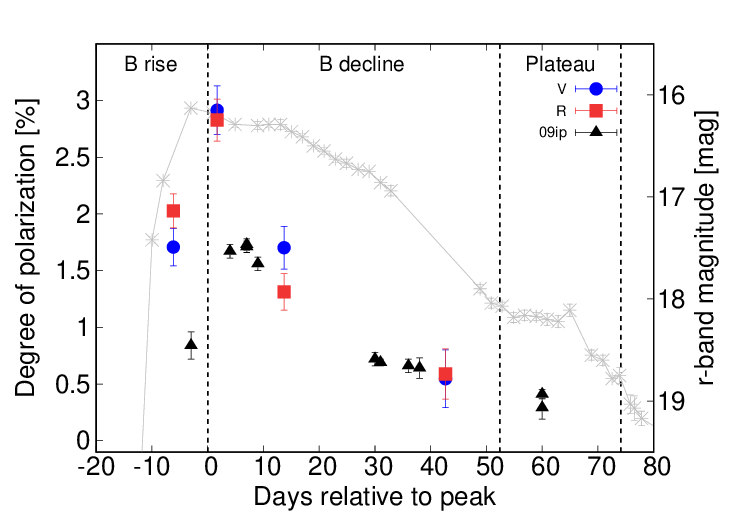}
     \includegraphics[width=0.95\linewidth]{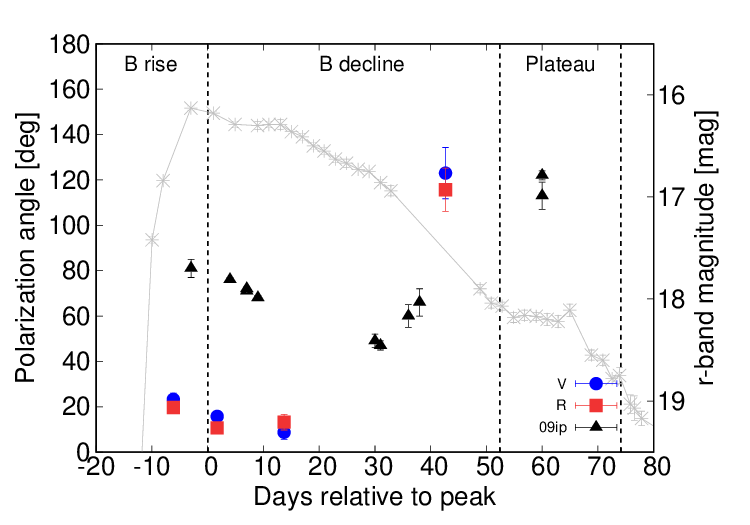}
     \includegraphics[width=0.95\linewidth]{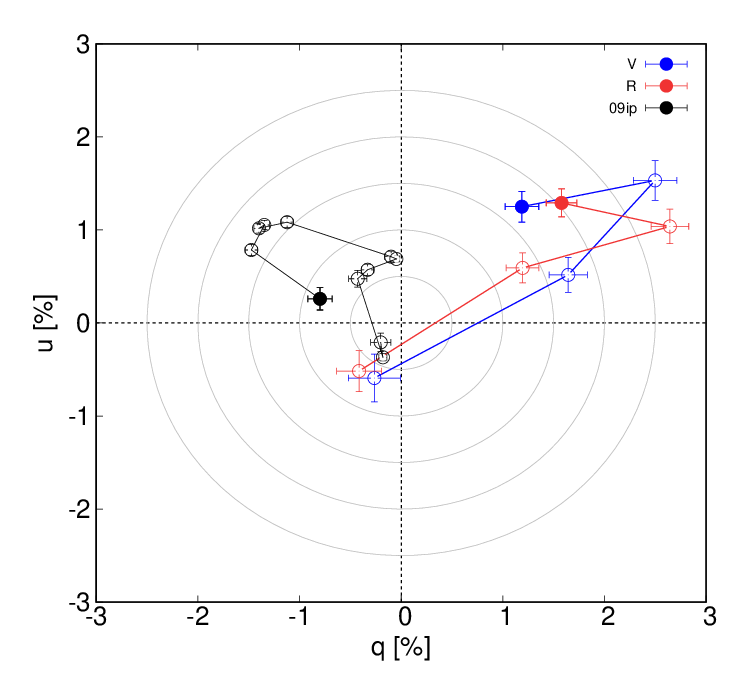}
     \caption{Time evolution of the $V$ and $R$-band polarization during the event B, including the polarization degree (top), polarization angle (middle), and Stokes $q$ and $u$ parameters (bottom). The comparison data of SN~2009ip were obtained from \cite{Mauerhan2014}. In the $q$-$u$ plane, the first epoch is marked with filled symbols.
     }
    \label{fig:pol}
\end{figure}

Figure~\ref{fig:pol} shows the time evolution of the degree and angle of $V$ and $R$-band polarization during event B. The $V$ and $R$-band measurements are broadly consistent across all phases, suggesting that the observed polarization is dominated by continuum emission. The polarization degree increases during the rising phase of the light curve, reaching a peak of $P \sim 2.8$ \% near maximum light, and then gradually declines at later times. The polarization angle remains relatively constant, except at the final polarimetric epoch when the optical light curve is simultaneously in transition to the beginning of the short plateau phase. Based on the empirical relation between interstellar extinction and polarization from \citet[][$P_{\rm max} \lesssim 9E(B-V)$ \%]{Serkowski1975}, the expected interstellar polarization (ISP) is likely limited to $P_{\rm max} \lesssim 1.4$ \%. This implies that the polarization near the peak of event B is predominantly intrinsic to the SN. The late-time change in the polarization angle is observed in both the $V$ and $R$ bands. Such behavior is more consistent with ISP than with line polarization, although a small intrinsic contribution cannot be entirely excluded. Similar polarimetric behavior was reported for SN~2009ip, which exhibited a peak polarization of $\sim 1.7$ \% and was interpreted as evidence for interaction with an aspherical circumstellar material (CSM) and an expanding ejecta \citep{Mauerhan2014}.

\subsection{Spectral energy distribution models for NGC 6745}\label{sec:SED}

\begin{figure}
    \centering
    \includegraphics[trim={0cm 0cm 0cm 0cm},clip,width=\linewidth]{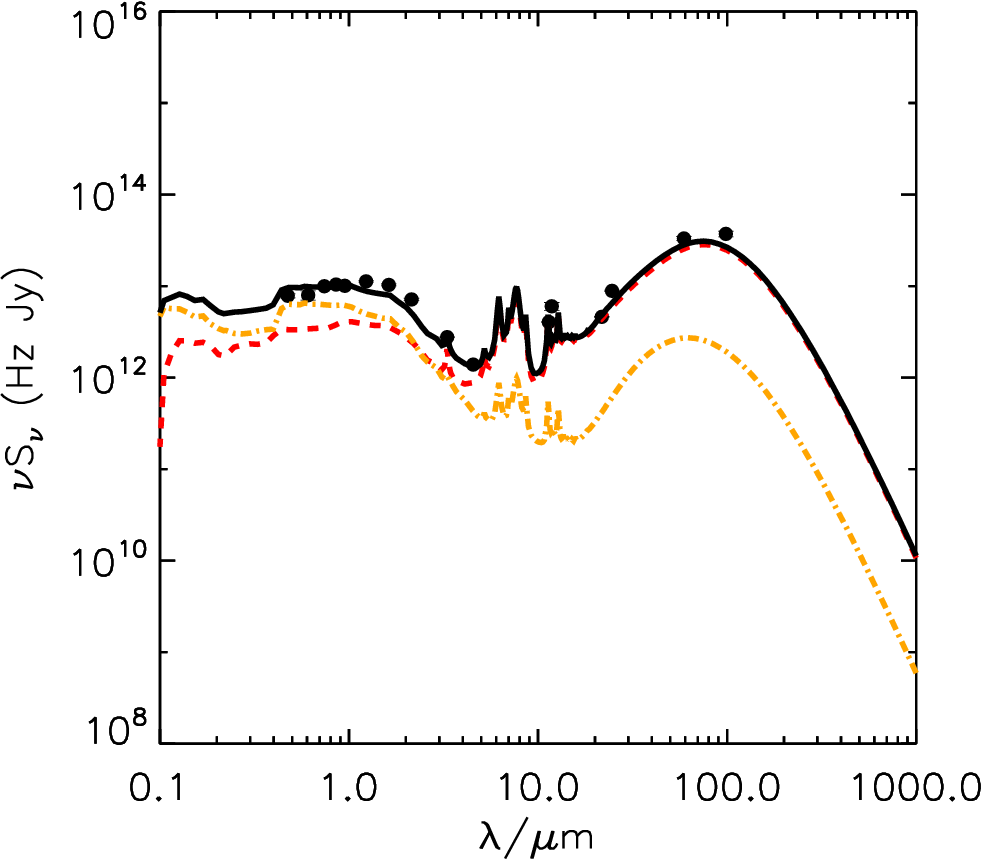}
    \caption{Model fit (solid black curve) to the SED data of the galaxy NGC 6745 (black points). The fit consists of a spheroidal galaxy (orange dot-dashed curve), and a starburst (dashed red) component.}
    \label{fig:SED}
\end{figure}

High-quality photometric data of the galaxy system has been reported by the Panoramic Survey Telescope and Rapid Response System (Pan-STARRS), 2MASS, the Widefield Infrared Survey Explorer (WISE), and the Infrared Astronomical Satellite (IRAS). The Pan-STARRS data was obtained via their Data Release 2 archive, and 2MASS, WISE, and IRAS data via the NED portal. The SED of the host galaxy NGC~6745 was fitted using the SED Analysis Through Markov Chains (SATMC) Monte Carlo code \citep{Johnson2013}, making use of different radiative transfer models for a starburst \citep{Efstathiou2000, Efstathiou2009}, active galactic nucleus \citep[AGN; ][]{Efstathiou1995, Efstathiou2013}, and a host galaxy \citep{Efstathiou2021, Efstathiou2022}. In the Monte Carlo fit a plethora of parameter combinations were tested, similar to the approach taken in \cite{Kankare2021} and \cite{Mantynen2025, Mantynen2026}. For the morphology of the host galaxy both a spheroidal and a disc model were considered, the possible starburst age was varied within a broad range (5 to 50~Myr), and the $e$-folding time of the starburst ranged within 1 to 40~Myr, and the fit was run both with and without an AGN component. The models assume a solar metallicity and a Salpeter initial mass function (IMF), and use the stellar population synthesis model from \cite{Bruzual1993} and \cite{Bruzual2003}.

The best fit for the SED of the galaxy, determined by the lowest $\chi^2$ value of the results, is shown in Fig. \ref{fig:SED}. In this model the AGN component is absent, and the fit consists of a starburst and a spheroidal galaxy component. The estimated rate of CCSNe per year in this model is $0.4 \pm 0.1$~yr$^{-1}$ from the starburst component. 
Based on the model, the estimated starburst age of the LIRG is $40 \pm 1$~Myr. One of the free parameters of the model is the optical depth of the spheroidal host galaxy. This is a global value for the whole host galaxy and the model fit suggests a relatively low value of $\tau \approx 0.1$. This is consistent with the relatively low extinction estimates of the studied CCSNe, which could have exploded in regions with somewhat higher local extinctions.

\begin{figure}
\centering
\includegraphics[trim={0cm 0cm 0cm 0cm},clip,width=\linewidth]{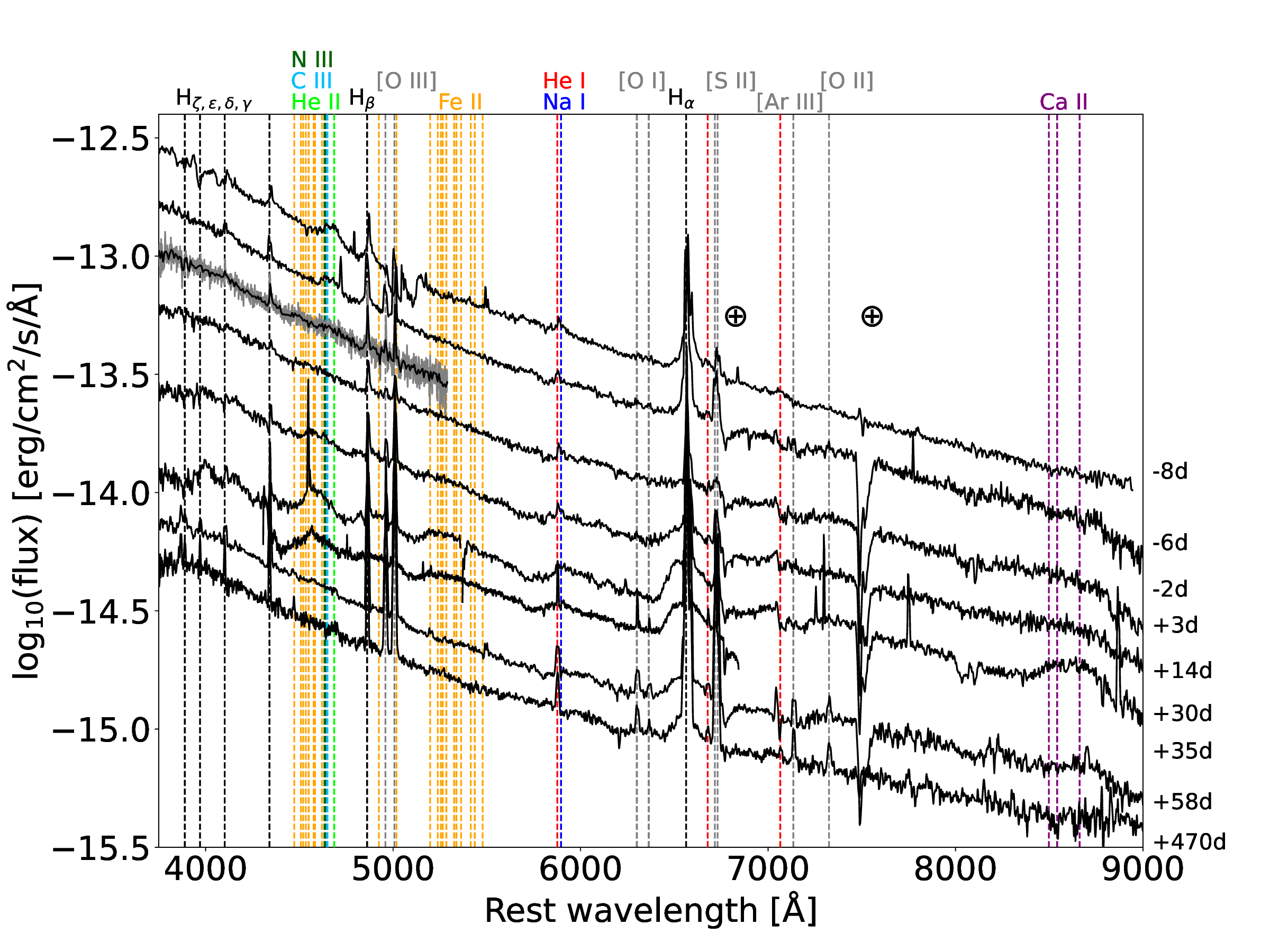}
\caption{Spectral time series of SN~2022prr with the epochs from the $r$-band maximum. The spectra have been dereddened and corrected to the wavelength rest frame. For the $-2$~d epoch both the original (dark grey) and binned (black) spectrum is presented. The wavelengths of the most prominent spectral lines are indicated with dashed vertical lines (host galaxy dominated lines in light grey) and telluric features with a $\oplus$ symbol. Logarithmic scale is used for flux, and the spectra have been vertically shifted for clarity. The $-8$~d spectrum by \cite{Jaeger2022} was obtained via the TNS.}\label{fig:22prr_spec}
\end{figure}

\subsection{Spectroscopic evolution}\label{sec:specevol}

\subsubsection{Spectroscopic evolution of SN~2022prr}\label{sec:specevol_22prr}

The most prominent features in the spectra of SN~2022prr are the Balmer emission lines, especially H$_\alpha$ and H$_\beta$, see Fig. \ref{fig:22prr_spec}. The narrow hydrogen emission features of the SN are blended with the extremely narrow hydrogen lines arising from the host galaxy, are unresolved, and cannot be disentangled. However, a broad H$_{\alpha}$ emission component appears in the $+14$~d spectrum of SN~2022prr for which we estimated a FWHM velocity of roughly 9000~km~s$^{-1}$. The width of the component declines slightly to around 8700~km~s$^{-1}$ in the $+30$ and $+35$~d spectra, and to 8000~km~s$^{-1}$ in the +58 d spectrum. In the late-time $+470$~d spectrum the broad H$_{\alpha}$ features suggests a velocity of $\sim$5000~km~s$^{-1}$. The FWHM velocities of the broad H$\alpha$ emission component during the event B decline and plateau phases are for example similar to those of SN 2009ip \citep[$\sim$~10000~km~s$^{-1}$; ][]{Margutti2014}. Contrary to most other SN~2009ip-like transients \citep{Fraser2013, Brennan2022a, Matilainen2025}, we do not observe a clear P~Cygni absorption component at the blue side of the Balmer emission lines. Unlike the profile of SN~2009ip, SN~2015bh, and SN~2016jbu \citep[][respectively]{Fraser2013,EliasRosa2016,Brennan2022a}, we also do not detect a clear double-peaked structure in the H$_\alpha$ or H$_\beta$ lines, see Fig. \ref{fig:09ip_spec}. This structure could either be unresolved due to the limited resolution in our spectra, or absent due to a different viewing geometry or CSM structure of the event. Furthermore, the H$_\alpha$ line is blended with the [N~{\sc ii}] doublet at 6548 and 6583~Å.

During the rise to event B peak, the H$_\alpha$ and H$_\beta$ profiles are narrow with broad electron scattering wings at the base of the spectral lines. During the event B decline phase a very broad emission component appears on the blue side of the emission lines. This is similar to the blue shoulder observed in SN~2016jbu at $+22$~d from event B maximum light \citep{Brennan2022a}. By the plateau phase this feature vanishes, and the H$_\alpha$ and H$_\beta$ lines become broader. The late-time spectrum at $+470$~d is dominated by the narrow host galaxy emission lines of the Balmer series, [O~{\sc iii}] at 4959 and 5007~Å, [O~{\sc i}] at 6300 and 6364~Å, [S~{\sc ii}] at 6716 and 6731~Å, [Ar~{\sc iii}] at 7135~Å, and [O~{\sc ii}] at 7325~Å, which are typical emission features in highly star-forming LIRGs \citep[e.g. ][]{Kim1995}. However, the H$_\alpha$ line has a broad base component, which likely arise from ongoing ejecta-CSM interaction.

Similar to the Balmer lines, narrow emission lines of He~{\sc i} are also visible throughout the evolution. The early spectra of SN~2022prr have a hot, blue continuum, and in the earliest epochs at $-8$ and $-6$~d a so called `flash ionisation feature' is also visible at around $\sim$4700~Å, see Sect. \ref{sec:flashfeature}. Other notable features are the tentative detections of the Ca~{\sc ii} NIR triplet and several Fe~{\sc ii} multiplets \citep[e.g. 37, 38, 41, 48, and 49; ][]{Moore1945} in the spectra from $+30$ d onward. However, we do not exclude contribution from other metal lines to the heavily blended regions. As is typical for SN~2009ip-like transients, clear and broad nebular features such as [O~{\sc i}] at 6300 and 6364~Å, and [Ca~{\sc ii}] at 7291 and 7323~Å, are absent in our spectra of SN~2022prr.

Recently, \cite{Medler2025} reported a compilation of NIR spectra of transients, which included three spectra of SN~2022prr observed with Keck-II using the Near-Infrared Echellette Spectrometer (NIRES) at epochs +15, +16, and +17~d from our estimated peak epoch (see their figure B5). Similar to SN 2009ip \citep{Fraser2013} or SN~2016cvk \citep{Matilainen2025} with NIR spectroscopic observations, the NIRES spectra of SN~2022prr reveal the presence of Paschen line series, He~{\sc i} at 10830 and 20587~Å, and a weak Brackett $\gamma$ line. We retrieved the spectra from the Weizmann Interactive Supernova Data Repository \citep[WISeREP; ][]{Yaron2012}. Both Paschen and He~{\sc i} lines show clear narrow P~Cygni profiles. We estimated a P~Cygni minima velocity of roughly $-$270~km~s$^{-1}$ compared to the line peak from the line profiles with the highest S/N ratio including the He~{\sc i} line at 10830~Å and the Pa$_{\beta}$, Pa$_{\gamma}$, and Pa$_{\delta}$ features.

\subsubsection{Flash feature of SN~2022prr}\label{sec:flashfeature}

\begin{figure}
\includegraphics[trim={0cm 0cm 0cm 0cm},clip,width=\linewidth]{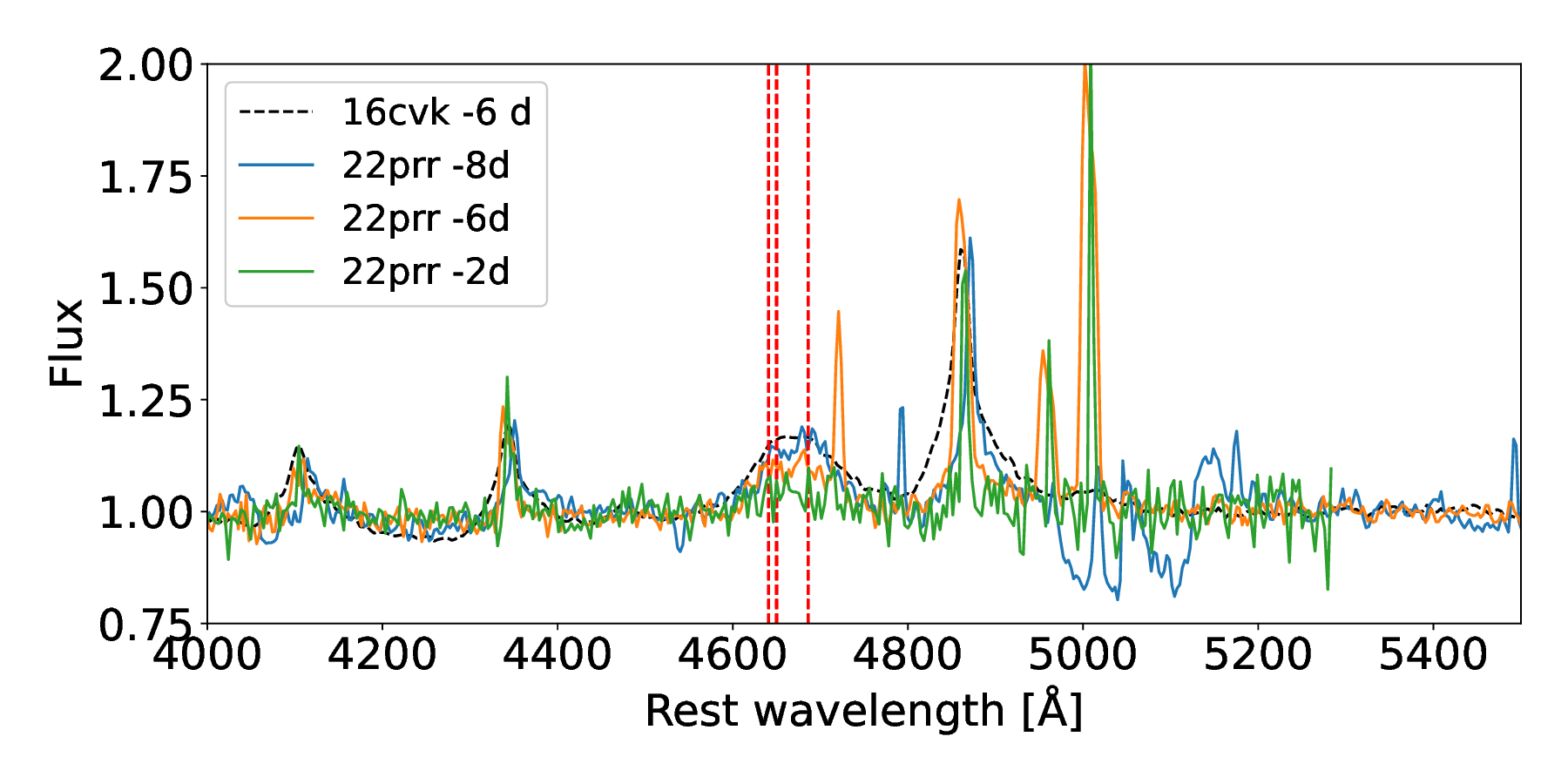}
\caption{A subsection of the spectra of SN~2022prr in three earliest epochs. The blended `flash ionization' feature is shown at $\protect\sim 4700$~Å. The locations of the N~{\sc iii} at 4641~Å, C~{\sc iii} at 4650~Å, and He~{\sc ii} at 4686~Å lines are marked with red dashed lines. The $-6$~d spectrum of SN~2016cvk is included for comparison, and the spectra have been scaled with the distance and normalised by dividing with the continuum.}
\label{fig:flashfeature}
\end{figure}

SNe surrounded by dense CSM show narrow, rapidly vanishing emission lines, so called `flash features', in their early spectra \citep[e.g. ][]{Kochanek2019}. These features originate from recombination of the CSM that was ionized either by the shock breakout UV flash of the SN \citep{Gal-Yam2014}, or by the ejecta-CSM interaction \citep{Jacobson2022}. 
Flash features are relatively common in Type IIn SNe in general \citep[e.g. ][]{Fassia2001, Khazov2016}, and have previously been reported in SN~2009ip-like transients in SN~2010mc \citep{Khazov2016}, SN~2015bh \citep{Thone2017}, as well as SN~2016jbu and SN~2016cvk \citep{Matilainen2025}.

In SN~2022prr such feature is present at $\sim$4700 Å in the $-8$ and $-6$ d spectra, see Fig. \ref{fig:flashfeature}, and has vanished at the next observed epoch at $-2$ d. This is likely a blended feature that includes the He~{\sc ii} line at 4686~Å, as well as emission from N~{\sc iii} (4634~Å and 4641~Å) and C~{\sc iii} (4648~Å and 4650~Å), similar to the feature found for example in SN~2016cvk \citep{Matilainen2025}, and overall the early spectra of SN~2022prr are analogous to those of SN~2016cvk. The early flash-feature spectrum of SN~2016cvk was interpreted to be consistent with models based on a solar-metallicity low-mass red supergiant (RSG), or a high-mass RSG, yellow supergiant (YSG) or blue supergiant (BSG) procursor with CNO-processed abundances, rather than a luminous blue variable (LBV) progenitor \citep{Matilainen2025}. This conclusion agrees also with the recently proposed relatively high rates and relatively low expected progenitor masses of SN 2009ip-like events by \cite{Salmaso2026}.

The duration of observed flash features varies between 5 and 10 days in normal Type II SNe, measured from their estimated date of explosion \citep{Bruch2023}, and is therefore powered by ejecta-CSM interaction due to the sufficiently long time scale ($>$ 1 d). In previously studied SN~2009ip-like events the duration of these features has been estimated as $14 \pm 1$~d for SN~2015bh, $12 \pm 3$~d for SN~2016jbu, and $16 \pm 5$ d for SN~2016cvk \citep{Matilainen2025}. To evaluate the duration of these features in the SN~2022prr spectra, we fit an exponential function $f(t) = a (t-t_\text{exp})^n$ to the measured flux during the event B rise phase \citep{Bruch2023}. The fit is shown in Fig. \ref{fig:LCphases22prr}. From this fit we estimate that $t_{\text{exp}}$, the onset of event B, occurred approximately 12~d before the event B peak; therefore, the flash feature at $\sim$4700~Å is visible for around $8 \pm 2$~d. This is slightly shorter than the flash feature time scales of other SN 2009ip-like events.

\subsubsection{Spectroscopic evolution of SN~2023ucy}\label{sec:specevol_23ucy}

\begin{figure}
\centering
\includegraphics[trim={0cm 0cm 1cm 1cm},clip,width=\linewidth]{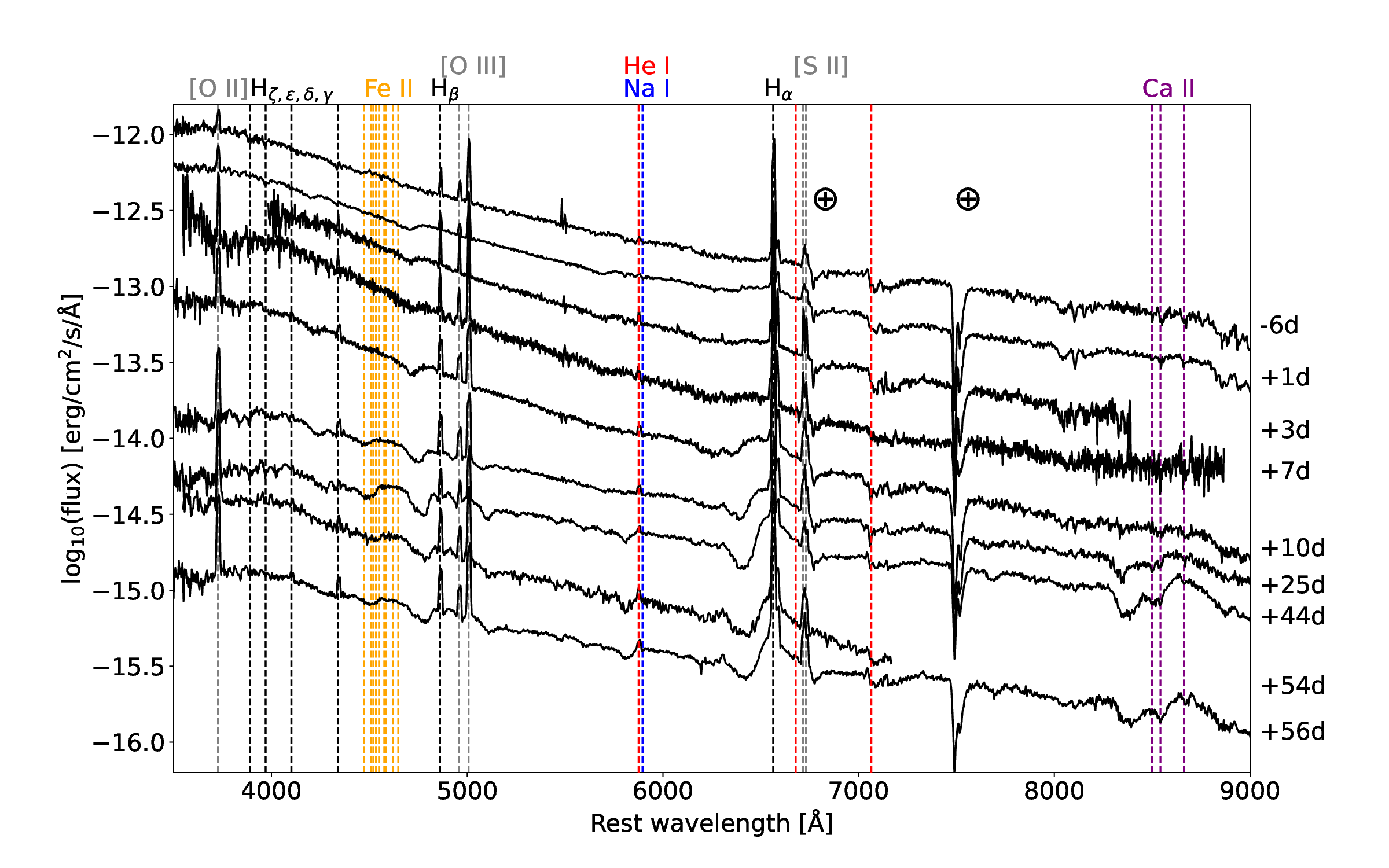}
\centering
 \caption{Spectral time series of SN~2023ucy with the epochs from the $r$-band maximum. The spectra have been dereddened and corrected to the rest frame wavelengths. The wavelengths of the most prominent spectral lines are indicated with dashed vertical lines (host galaxy dominated lines in light grey) and telluric features with a $\oplus$ symbol. Logarithmic scale is used for flux, and the spectra have been vertically shifted for clarity. The $+3$~d spectrum by \cite{Taguchi2023} and $+7$~d spectrum by \cite{Teja2023} were obtained via the TNS.}
\label{fig:23ucy_spec}
\end{figure}

\begin{figure}
\centering
\includegraphics[trim={0cm 0cm 1cm 1cm},clip,width=\linewidth]{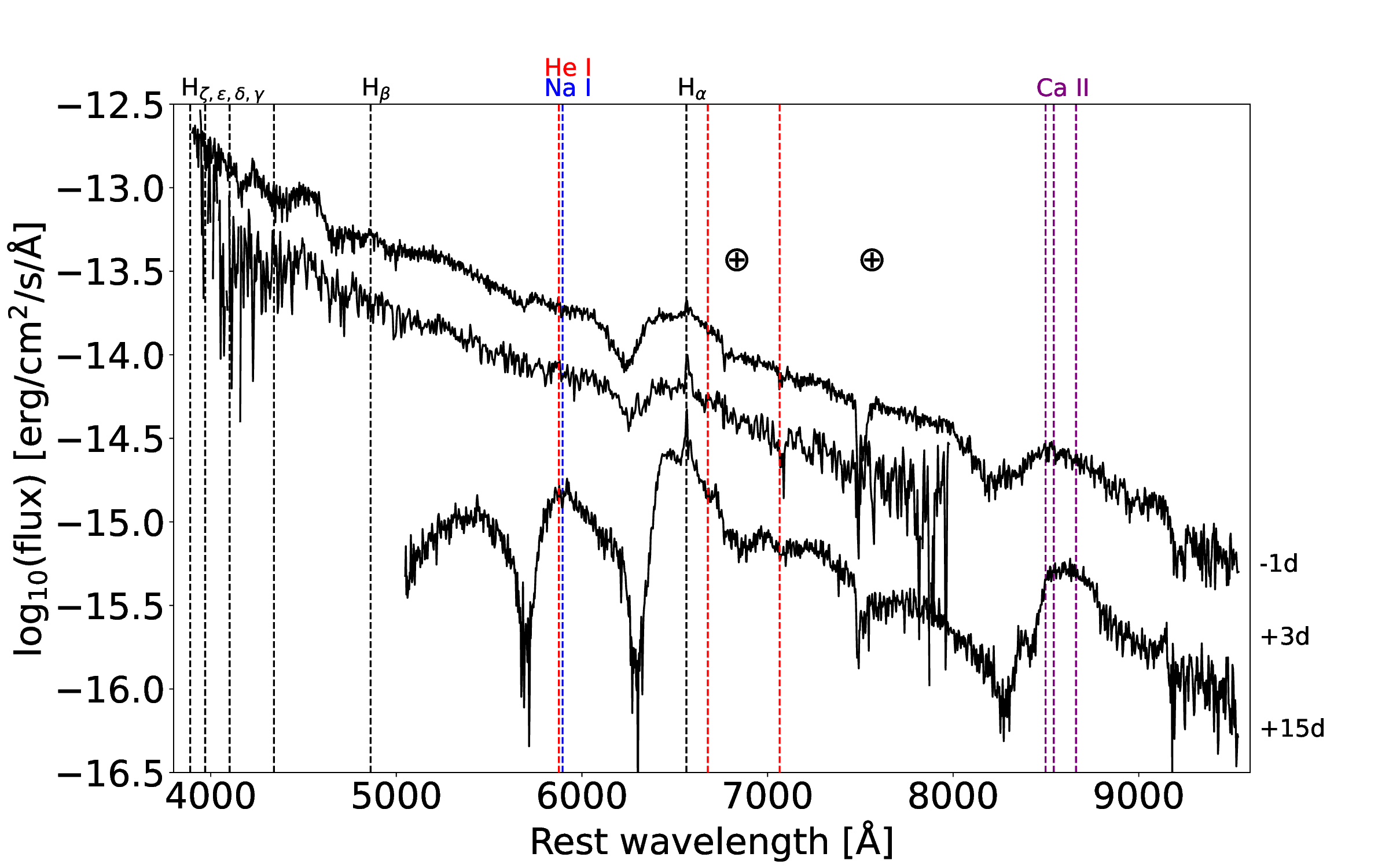}
\caption{Spectral time series of SN~2024ljc with the epochs from the $r$-band maximum. The spectra have been dereddened and corrected to the rest frame wavelengths. The wavelengths of the most prominent spectral lines are indicated with dashed vertical lines and telluric features with a $\oplus$ symbol. Logarithmic scale is used for flux, and the spectra have been vertically shifted for clarity. The $+3$~d spectrum by \cite{Wise2024} was obtained via the TNS.}
\label{fig:24ljc_spec_log}
\end{figure}

The earliest spectra of SN~2023ucy taken at $-6$~d and $+1$~d are dominated by a featureless blue continuum accompanied by extremely narrow host galaxy lines, such as H$_\alpha$, H$_\beta$, He~{\sc i} (5876~Å), [O~{\sc ii}] (3727~Å), [O~{\sc iii}] (4959, 5007~Å), and [S~{\sc ii}] (6716, 6731~Å), see Fig. \ref{fig:23ucy_spec}. At this stage it is difficult to distinguish the narrow lines originating from the CSM of the transient from the narrow host galaxy lines with an instrumental resolution of roughly 800~km~s$^{-1}$. We begin to see a P~Cygni profile in the H$_\beta$ and H$_\alpha$ line from $+1$ and $+10$~d onward, respectively.

From $+25$~d onward the spectrum of SN~2023ucy is very similar to other Type IIP SNe, such as SN~2023ixf and SN~2012ec, see Fig. \ref{fig:23ucy_like_spec}. Blended features of Fe~{\sc ii} at $\sim 4600$~Å and Ca~{\sc ii} NIR triplet at $\sim 8600$~Å emerge. At this stage the spectra are dominated by broad P~Cygni profiles seen in the H$_\alpha$ and H$_\beta$ lines, as well as in the Ca~{\sc ii} NIR triplet.

To get a better view of the H$_\alpha$ line evolution, we subtracted the local continuum around H$_\alpha$ and removed the narrow host lines from the spectrum. We fixed the wavelength of H$_\alpha$ at $v=0$~km~s$^{-1}$ for each epoch, and normalised the line flux between 0 and 1, see Fig. \ref{fig:23ucy_pcyg}. From a Gaussian fit to the data of SN~2023ucy we get $10\,200 \pm 200$~km~s$^{-1}$ as the velocity of the blue absorption minimum at $+10$~d, measured from the rest wavelength of H$_\alpha$. At $+25$~d the velocity declines sharply to $8100 \pm 300$~km~s$^{-1}$, and more gradually to $7000 \pm 200$~km~s$^{-1}$ at $+56$~d, see Fig. \ref{fig:linevel_23ucy}. For example, \cite{Zheng2025} reported similar velocities for the H$_\alpha$ line of SN~2023ixf, decreasing from $\sim 10\,000$~km~s$^{-1}$ at $+23$~d to $7000$~km~s$^{-1}$ at $+50$~d. \cite{Barbarino2015} notes a similar velocity trend for the absorption minimum of the H$_\alpha$ line of SN~2012ec, evolving from $11\,800$~km~s$^{-1}$ at $+9$~d to $5900$~km~s$^{-1}$ at $+56$~d, time measured from the estimated epoch of explosion.

Following similar steps of continuum-subtraction, normalisation, and curve fitting for the H$_\alpha$ line, a Gaussian was fit to the absorption component of the Ca~{\sc ii} NIR triplet of SN~2023ucy. We measure $6300 \pm 100$~km~s$^{-1}$ as the P~Cygni velocity at $+25$~d from the 8520~Å average of the two blended lines of the triplet at the shorter wavelengths (see Fig. \ref{fig:CaNIR_23ucy}), similar to the approach taken by \cite{Barbarino2015}. The Ca~{\sc ii} feature (and the Fe~{\sc ii} lines) become stronger in the $+44$~d spectrum, and begin to diminish in the last spectrum taken at $+56$~d. The velocity of the Ca~{\sc ii} NIR absorption feature decreases to $5400 \pm 100$~km~s$^{-1}$ at $+44$~d, and $4900 \pm 100$~km~s$^{-1}$ at $+56$~d. A potential emission line of O~{\sc i} at 8446~Å is also detected in the two latest $+44$ and $+56$~d spectra. In the data of SN~2012ec the Ca~{\sc ii} NIR triplet appears from $+49$~d onward, with a velocity of 5600~km~s$^{-1}$. The velocity is similar to the trend seen in SN~2023ucy in which the velocity had decreased to 4900~km~s$^{-1}$ at $+56$~d.

Based on the spectroscopic evolution of the emerging lines and their velocities during the plateau phase, SN~2023ucy is a quite normal Type IIP SN. Unfortunately, we do not have any late-time spectra of the nebular phase of SN~2023ucy.

\subsubsection{Spectroscopic evolution of SN~2024ljc}\label{sec:specevol_24ljc}

As is typical for Type IIb SNe, the early-phase spectrum of SN~2024ljc at $-1$~d from peak is blue with superposed low-contrast P~Cygni lines, see Fig. \ref{fig:24ljc_spec_log}. The most notable feature at this point is a rather faint, broad emission line of H$_\alpha$. From the Gaussian fit to the blue absorption component minimum of this feature we measure a velocity of $16\,400 \pm 100$ km~s$^{-1}$, see Fig. \ref{fig:ha_24ljc}. At $+15$~d from peak, the velocity of the absorption component decreases to $13\,100 \pm 200$~km~s$^{-1}$, while the emission component becomes more prominent. We found the rapid light curve evolution of SN~2024ljc to be similar to that of the Type IIb SN~2024aecx (see Sect. \ref{sec:extinction} and Sect. \ref{sec:LC24ljc}). However, \cite{Zou2026} determined a somewhat lower H$_\alpha$ velocity of 10\,000~km~s$^{-1}$ for SN~2024aecx near maximum light, decreasing to 7000~km~s$^{-1}$ at $+8$~d from peak. On the other hand, \cite{Tinyanont2026} suggested that the feature in SN~2024aecx arises from Si~{\sc ii} rather than H$_\alpha$. Conversely, the prototypical Type IIb SN~2011dh shares a similar H$_\alpha$ velocity trend with SN~2024ljc (see Fig. \ref{fig:linevel_24ljc}), decreasing from 18\,000~km~s$^{-1}$ near peak to 12\,000~km~s$^{-1}$ at $+20$~d \citep{Ergon2014}, despite its clearly slower photometric evolution. Within a comparison of the spectra of SN~2024ljc to those of a selection of classical Type IIb SNe, the strongest similarity is found with SN~2003bg \citep{Hamuy2003}, see Fig. \ref{fig:24ljc_comp_spec}. The peak epoch of SN~2003bg is uncertain; however, if we adopt the average estimate (MJD = 52713) from \cite{Mazzali2009} and the velocities from \cite{Hamuy2009} a similar H$_\alpha$ and He~{\sc i} velocity evolution is seen.

We identify a broad blended Ca~{\sc ii} NIR triplet at $\sim$~8700~Å in the earliest spectrum at $-1$~d. At $+15$~d from peak the Ca~{\sc ii} triplet becomes stronger. From the the 8520~Å average of the two blended lines of the Ca~{\sc ii} triplet at the shorter wavelengths we determine a velocity of $11\,400 \pm 100$~km~s$^{-1}$ at $-1$~d, and $9300 \pm 100$~km~s$^{-1}$ at $+15$~d, see Fig. \ref{fig:CaNIR_24ljc}. Similar results were presented for the Ca~{\sc ii} NIR triplet of SN~2024aecx by \cite{Xi2026}, who measured 12\,000~km~s$^{-1}$ as the highest line velocity, and \cite{Zou2026}, who determined a velocity of $\sim 10\,000$~km~s$^{-1}$ around +30~d. A clear P~Cygni feature of He~{\sc i} at 5876~Å blended with Na~{\sc i}~D is also visible in the +15~d spectrum of SN~2024ljc, with a line velocity of $9800 \pm 100$~km~s$^{-1}$. Typical for SNe with hydrogen features in their ejecta the H$_\alpha$ line arises from the outer ejecta with higher velocities and the He~{\sc i} feature from inner ejecta with lower velocities more similar to that of the bulk of the ejecta.

\section{Discussion}\label{sec:discussion}

The spectral evolution of SN~2022prr is dominated by narrow emission features of hydrogen and helium. The blue P~Cygni absorption component present in the Balmer lines of most other SN~2009ip-like transients such as SN~2009ip \citep{Fraser2015}, SN~2016cvk \citep{Matilainen2025}, SN~2010mc \citep{Ofek2013}, SN~2016jbu \citep{Brennan2022a}, SN~2015bh \citep{Thone2017}, and LSQ13zm \citep{Tartaglia2016}, is absent in the spectra of SN~2022prr. The absence of this absorption feature in SN~2022prr could potentially be explained by contamination from the host galaxy. It could also mean that these features are very shallow compared to other SN~2009ip-like transients, suggesting a lower ejecta mass in the line-of-sight of the transient, which could also indicate a lower envelope mass for the progenitor. Furthermore, we do not detect any signs of double-peaked structure in the H$_\alpha$ or H$_\beta$ lines, similar to SN~2016bdu \citep{Pastorello2018}, SN~2009ip \citep{Fraser2013}, and SN~2016cvk \citep{Matilainen2025}. Models by \cite{Kurfurst2020} predict, that SNe with circumstellar discs exhibit symmetrical, double-peaked line structures when viewed from a polar line-of-sight. The double-peaked pattern is less pronounced when the viewing angle is between the equator and the polar regions, and vanishes when the disc is observed from an edge-on angle. The lack of a visible double-peaked structure in the Balmer lines of SN~2022prr could be explained by a more edge-on viewing geometry, or differences in the CSM structure compared to other SN~2009ip-like transients  \citep[see also discussion by ][]{Brennan2022a}. However, we cannot fully exclude contamination from the host galaxy obscuring a double-peaked transient line profile due to the limited resolution of our spectra.
On the other hand, our polarimetric observations do suggest a high degree of $V$ and $R$-band polarisation ($P \sim 2.8$~\%), consistent with asymmetric CSM that could be disc-like. A possible channel for a disc-like CSM formation could be strong binary interaction between a H-rich massive star and a more compact companion shortly before the explosive transient event \citep{Smith2011b, Mauerhan2014}. Furthermore, the polarization angle shows a dramatic change before the onset of the short plateau phase. Concurrently broad features of Ca~{\sc ii} NIR triplet and Na~{\sc i} doublet appear in the $+35$~d spectrum and the broad component of the Balmer lines has become more prominent. This could suggest that we see a revealed highly aspherical ejecta component that was previously hidden by the CSM.

In the earliest $-8$ and $-6$~d spectra of SN~2022prr, we detect a blended flash ionisation feature of He~{\sc ii}, N~{\sc iii}, and C~{\sc iii} around $\sim$4700~Å, lasting for  $8 \pm 2$~d. These features are typical in SNe surrounded by dense CSM. A similar feature has also been reported in SN~2016cvk and SN~2016jbu \citep{Matilainen2025}, in SN~2010mc \citep{Khazov2016}, and in SN~2015bh \citep{Thone2017}, but surprisingly not in SN~2009ip, which has a high-quality spectroscopic data set. \cite{Matilainen2025} found flash features of SN~2016cvk to be consistent with a relatively low-mass massive star progenitor, plausibly a RSG. The similar flash feature spectra of SN~2022prr to that of SN~2016cvk suggests also a similar progenitor for this event. This is consistent with the relatively old starburst age of $\sim$40 Myr, which we derived for the LIRG host. 

SN~2023ucy is a Type IIP SN with a very typical spectrophotometric evolution. Its peak magnitude of $M_\text{r,peak}=-18.0$~mag is on the brighter side of Type II SNe. However, the evolution of the light curve is very similar for example to SN~2023ixf \citep{Li2025, Jacobson2025} and SN~2012ec \citep{Barbarino2015}, and the light curve reaches a plateau of $M_{r,\text{peak}}=-17.3$~mag from $+10$~d onward. The plateau phase lasts at least until $+64$~d, where our coverage is interrupted. The actual length of the plateau phase is unknown, but the lower limit is in line with the typical $>90$~d plateau duration for IIP SNe. The progenitors of IIP SNe are typically RSGs with zero-age main sequence (ZAMS) masses between $8-20 M_\odot$ \citep[e.g. ][]{Fang2025}, and the estimated starburst age of the host galaxy is in line with the lifetime of a RSG progenitor \citep{Ekstrom2012}.

SN 2024ljc is a rapidly evolving Type~IIb SN. Both theoretical work and observational evidence of precursor systems of nearby events such as SNe 1993J and 2011dh has supported binary progenitors for Type~IIb SNe \citep[e.g. ][]{Aldering1994, Woosley1994, Maund2011, Bersten2012}. Mass transfer via Roche-lobe overflow in a close binary system is an efficient mechanism for the primary star to have lost its outer envelope almost entirely with only $\lesssim$0.1 to $\lesssim$1~M$_{\odot}$ of hydrogen left to explain the spectroscopic characteristics of these events. However, it cannot be excluded that some Type~IIb SNe could originate from more massive single (e.g. WN-type Wolf-Rayet) stars that have lost their hydrogen envelope via very strong stellar winds \citep[e.g. ][]{Smartt2009}. 
The Type~IIb SN~2024ljc has a notably faster light curve evolution than is typical for this subtype. This suggests a low ejecta mass for the event. We find the photometric evolution of SN~2024ljc similar to that of another rapidly-evolving Type~IIb SN~2024aecx. Assuming that the main peak of the SN~2024aecx is powered by radioactive decay, \cite{Zou2026} estimated, based on modelling of the bolometric light curve for the event, $M_\mathrm{ejecta} = 0.70^{+0.18}_{-0.16}$~M$_{\odot}$ and $M_{\mathrm{Ni}} = 0.15 \pm 0.06$~M$_{\odot}$. With similar approach and conclusions, \cite{Xi2026} estimated $M_\mathrm{ejecta} = 1.55^{+0.18}_{-0.14}$~M$_{\odot}$ and $M_{\mathrm{Ni}} = 0.09 \pm 0.01$~M$_{\odot}$. As discussed for example by \cite{Zou2026}, the ejecta mass of SN~2024aecx has to be relatively small to result in short diffusion time scales and thus fast evolution of the light curve. Similar conclusion applies to SN~2024aecx. Furthermore, in addition to a small envelope mass, a relatively high explosion energy is likely also important and potentially how clumps of radioactive material expand in the ejecta \citep{Ergon2024}. \cite{Xi2026} noted also that incomplete gamma-ray trapping alone is unlikely to explain the extremely rapid optical decline of SN~2024aecx $\gtrsim$20~d from the main maximum; potential discussed effects included incomplete positron trapping, asymmetric ejecta, or formation of new dust. Recently, \cite{Wei2026}  modelled the light curve evolution of SN~2024aecx and derived parameters of $M_\text{ejecta} = 2.14^{+0.21}_{-0.19} M_\odot$ and $M_\text{Ni} = 0.050 \pm 0.002 M_\odot$, and proposed a need to introduce an additional suppression factor for the optical luminosity to explain the rapid evolution that could potentially arise from non-local energy escape, redistribution of optical emission into longer wavelengths, dust related effects, ejecta geometry, opacity changes, or line-formation processes.  Interestingly, \cite{Tinyanont2026} reported a prominent NIR excess developing for SN~2024aecx by $+32$~d from the main event; however, this was associated with an IR echo from pre-existing dust, likely in a face-on disc-like geometrical configuration.

\section{Conclusions}\label{sec:conclusions}

SN~2022prr, SN~2023ucy, and SN~2024ljc are all transients that show hydrogen in their spectra. The starburst age of their host galaxy NGC~6745 derived in the SED modeling of the galaxy in this paper, $40 \pm 1$~Myr, is consistent with the findings of \cite{Kankare2021}, suggesting LIRGs hosting primarily H-rich SNe have older starburst ages of $\gtrsim$30~Myr. 

The host-galaxy extinctions derived via light curve comparisons of the transients yield a very small host-galaxy extinction value ($A^\text{host}_V = 0.1^{+0.2}_{-0.1}$ and $0.4^{+0.2}_{-0.1}$~mag in the line-of-sight direction to SN~2022prr and SN~2023ucy, respectively) within the interaction zone NGC~6745b of the LIRG. This signals that the amount of interstellar dust in these regions is low and/or the events are located at the foreground of this region. Near the centre of the main galaxy, in the line-of-sight direction to SN~2024ljc the situation is somewhat different, and our comparisons yield a higher host extinction value of $A^\text{host}_V = 1.1^{+0.2}_{-0.1}$~mag.

The SN~2009ip-like transient SN~2022prr has a very similar light curve evolution as other SN~2009ip-like transients, with a fainter $-14.8$~mag event A followed by a bright $-18.3$~mag main event B. 
The high degree of $V$ and $R$-band polarisation ($P \sim 2.8$ \%) of SN~2022prr near maximum light suggests a very aspherical, possibly disc-like CSM interaction similar to SN~2009ip \citep{Mauerhan2014}. 
The spectroscopic evolution of SN~2022prr is dominated by hydrogen and helium features, which are intertwined with the narrow host galaxy emission lines, effectively shrouding most of the complex line evolution over time. In previous literature it has been suggested that the CSM of SN~2009ip-like transients is dense and disc-like, leading to double-peaked emission line profiles in many of the events. Our observations of SN~2022prr, however, do not show such structure. This could be due to a more edge-on viewing angle of the disc, or the feature could simply be masked by strong host galaxy emission lines. 
In the earliest spectrum of SN~2022prr at $-6$~d, we detected a blended flash ionisation feature of He~{\sc ii}, N~{\sc iii}, and C~{\sc iii} around $\sim$4700~Å. We determined the onset of event B as $-12$~d from the light curve maximum, and estimated that the flash features are visible for $8 \pm 2$~d. The relatively long duration of the feature suggests, that it is likely powered by radiation originating from ejecta-CSM interaction. The similarity of the flash feature and the early spectra of SN~2022prr and SN~2016cvk at $-6$~d hint at a similar progenitor for these SNe, which would likely be a low-mass RSG with a solar metallicity, or a high-mass RSG, YSG or BSG with CNO-processed abundances \citep{Matilainen2025}, rather than an LBV progenitor.

SN~2023ucy is a bright IIP SN with a peak magnitude of $M_{r,\text{peak}}=-18$~mag, and a light curve evolution very typical for its subtype. We can estimate only a very conservative lower limit for the plateau extending to $+63$~d when our photometric coverage ends. The spectroscopic evolution of SN~2023ucy is dominated by emission lines of H, Fe~{\sc ii}, and Ca~{\sc ii}, with line velocities similar to those of other normal Type IIP SNe. 

SN~2024ljc is a Type IIb SN with a notably fast light curve evolution, similar to Type IIb SN~2024aecx \citep{Xi2026}. From its peak $r$-band magnitude of $-17.1$~mag, the light curve declines at a rate of around 7.1~mag per 100~d. The fast light curve evolution suggests a relatively low ejecta mass for the event compared to normal Type IIb SNe, with the outer envelope of the progenitor star likely stripped by a binary companion. The obtained spectra of SN~2024ljc from $-2$ to $+15$~d are blue; the dominating spectral lines are broad P~Cygni features of H$_\alpha$ and the Ca~{\sc ii} NIR triplet with a velocity evolution similar to that of normal Type IIb SNe.

A large fraction of CCSNe in LIRGs remain undetected by both optical and NIR surveys \citep{Mattila2012, Fox2021, Mantynen2025, Mantynen2026}. Empirical relations between the IR luminosity of the CCSN rate of starbursting galaxies has been suggested in the literature \citep[e.g. ][]{Mattila2001}; however, to have robust estimates on the intrinsic CCSN rates of LIRGs detailed SED modelling of the galaxy is required. The increasing number of follow-up studies of CCSNe in LIRGs will help to characterize the extinction distribution of events in these galaxies, as well as link the properties of the SNe with the starburst parameters of their hosts, as was shown in \cite{Kankare2021}. This directly contributes to constraining the fraction of CCSNe that will remain undetected in LIRGs due to the host-galaxy extinction, and helps understand the populations of SNe in the star forming environments of LIRGs. These estimates are vital to robustly constrain the intrinsic CCSN rates that are dominated by LIRGs at cosmological distances. 

\begin{acknowledgements}

We thank Qiang Xi and Ning-Chen Sun for providing the light curve data of SN~2024aecx. 

We thank Rishabh Singh Teja for confirming the setup information of the HCT spectrum of SN~2023ucy.

KM acknowledges financial support from the Viljo, Yrjö and Kalle Väisälä foundation.

KM and EK acknowledge financial support from the Emil Aaltonen foundation.

SM, EK and TR acknowledge financial support from the Research Council of Finland project 350458.

AR and Y-ZC acknowledges financial support from the SOXS project (PI S. Campana). 
AR and AP acknowledge financial support from the PRIN-INAF 2022 "Shedding light on the nature of gap transients: from the observations to the models".

Y-ZC is supported by the National Natural Science Foundation of China (No. 12303054), the Yunnan Fundamental Research Projects (Grant Nos. 202401AU070063, 202501AS070078), the National Key Research and Development Program of China (Grant No. 2024YFA1611603), and the International Centre of Supernovae, Yunnan Key Laboratory (No. 202302AN360001). 

LG acknowledges financial support from CSIC, MCIN and AEI 10.13039/501100011033 under projects PID2023-151307NB-I00, PIE 20215AT016, CEX2020-001058-M, and by the MaX-CSIC Excellence Award MaX4-SOMMA-ICE.

MG-B acknowledges financial support from the Spanish Ministerio de Ciencia e Innovación (MCIN) and the Agencia Estatal de Investigación (AEI) 10.13039/501100011033 under the PID2023-151307NB-I00 SNNEXT project, from Centro Superior de Investigaciones Científicas (CSIC) under projects PIE 20215AT016, ILINK23001, COOPB2304, and the program Unidad de Excelencia María de Maeztu CEX2020-001058-M, and from the Departament de Recerca i Universitats de la Generalitat de Catalunya through the 2021-SGR-01270 grant. MG-B's work has been carried out within the framework of the doctoral program in Physics of the Universitat Autònoma de Barcelona.

CPG acknowledges financial support from grant RYC2024-050959-I, funded by MICIU/AEI/10.13039/501100011033 and the FSE+, as well as from projects PID2023-151307NB-I00, PIE 20215AT016, and CEX2020-001058-M, and the MaX-CSIC Excellence Award MaX4-SOMMA-ICE.

TK acknowledges support from the Research Council of Finland project 360274.

TLK acknowledges support via a Warwick Astrophysics prize post-doctoral fellowship, made possible thanks to a generous philanthropic donation.

SMo is funded by Leverhulme Trust grant RPG-2023-240.

IS acknowledges financial support from the SOXS Science Consortium.

Based on observations made with the Nordic Optical Telescope, owned in collaboration by the University of Turku and Aarhus University, and operated jointly by Aarhus University, the University of Turku and the University of Oslo, representing Denmark, Finland and Norway, the University of Iceland and Stockholm University at the Observatorio del Roque de los Muchachos, La Palma, Spain, of the Instituto de Astrofisica de Canarias. The NOT data were obtained under program IDs 62-507, 64-507, 65-005, 66-506, 66-701, 68-505, and 68-702.	

The data presented here were obtained in part with ALFOSC, which is provided by the Instituto de Astrofisica de Andalucia (IAA) under a joint agreement with the University of Copenhagen and NOT.

Observations from the NOT were obtained primarily through the NUTS2 collaboration, which is supported in part by the Instrument Centre for Danish Astrophysics (IDA), and the Finnish Centre for Astronomy with ESO (FINCA) via Academy of Finland grant nr 306531.

One epoch of SN 2022prr imaging was observed as part of the Finnish science school in November 2022 and one epoch of SN 2023ucy spectroscopy was observed as part of the Finnish observational astronomy course in October 2023, both organised by the University of Turku.

We are grateful to Natalie Allen, Kate Gould, Meghana Killi and Dazhi Zhou, who observed SN~2022prr as part of the Danish Summer Course in Observational Astrophysics at the NOT.

Based on observations collected at Copernico and Schmidt telescopes (Asiago Mount Ekar, Italy) of the INAF -- Osservatorio Astronomico di Padova.

We have made use of the Weizmann interactive supernova data repository (\url{www.weizmann.ac.il/astrophysics/wiserep}).

This publication makes use of data products from the Two Micron All Sky Survey, which is a joint project of the University of Massachusetts and the Infrared Processing and Analysis Center/California Institute of Technology, funded by the National Aeronautics and Space Administration and the National Science Foundation.

Funding for the SDSS and SDSS-II has been provided by the Alfred P. Sloan Foundation, the Participating Institutions, the National Science Foundation, the U.S. Department of Energy, the National Aeronautics and Space Administration, the Japanese Monbukagakusho, the Max Planck Society, and the Higher Education Funding Council for England. The SDSS Web Site is \url{http://www.sdss.org/}.

The SDSS is managed by the Astrophysical Research Consortium for the Participating Institutions. The Participating Institutions are the American Museum of Natural History, Astrophysical Institute Potsdam, University of Basel, University of Cambridge, Case Western Reserve University, University of Chicago, Drexel University, Fermilab, the Institute for Advanced Study, the Japan Participation Group, Johns Hopkins University, the Joint Institute for Nuclear Astrophysics, the Kavli Institute for Particle Astrophysics and Cosmology, the Korean Scientist Group, the Chinese Academy of Sciences (LAMOST), Los Alamos National Laboratory, the Max-Planck-Institute for Astronomy (MPIA), the Max-Planck-Institute for Astrophysics (MPA), New Mexico State University, Ohio State University, University of Pittsburgh, University of Portsmouth, Princeton University, the United States Naval Observatory, and the University of Washington. 

We acknowledge the use of public data from the Swift data archive.

This research used ASTROPY, a community-developed core Python package for Astronomy \citep{astropy}.
    
\end{acknowledgements}

\section*{Data Availability}

\bibliographystyle{aa}
\bibliography{lirg,PapersLit}

\onecolumn
\begin{appendix}
\section{Photometry tables, additional LC and spectra fits}

\begin{small}
\setlength\tabcolsep{2.5pt}
\begin{longtable}{cccccccccc}\caption{Photometry table of SN~2022prr in optical and NIR bands.}\label{tab:phot_SN2022prr}\\
\hline 
\hline 

 MJD & $m_B$($m_{B,\text{err}}$) & $m_g$($m_{g,\text{err}}$) & $m_V$($m_{V,\text{err}}$) & $m_r$($m_{r,\text{err}}$) & $m_\text{i}$($m_{i,\text{err}}$) & $m_J$($m_{J,\text{err}}$) & $m_H$($m_{H,\text{err}}$) & $m_K$($m_{K,\text{err}}$) & Telescope \\
  & (mag)& (mag)& (mag)& (mag)& (mag)& (mag)& (mag)& (mag)& \\
 \hline

\hline
\endfirsthead

\multicolumn{10}{c}{{\bfseries \tablename\ \thetable{} -- continued from previous page}} \\
\hline 
 MJD & $m_B$($m_{B,\text{err}}$) & $m_g$($m_{g,\text{err}}$) & $m_V$($m_{V,\text{err}}$) & $m_r$($m_{r,\text{err}}$) & $m_\text{i}$($m_{i,\text{err}}$) & $m_J$($m_{J,\text{err}}$) & $m_H$($m_{H,\text{err}}$) & $m_K$($m_{K,\text{err}}$) & Telescope \\
  & (mag)& (mag)& (mag)& (mag)& (mag)& (mag)& (mag)& (mag)& \\
  \hline
\endhead

\hline \multicolumn{10}{|r|}{{Continued on next page}} \\ \hline
\endfoot

\hline \hline
\endlastfoot
58435.13 &  $\cdots$  & 18.50(0.13) & $\cdots$  & $\cdots$  & $\cdots$  & $\cdots$  & $\cdots$  & $\cdots$  & ZTF  \\ 
59748.34 &  $\cdots$  & $\cdots$  & $\cdots$  & 19.90(0.24) & $\cdots$  & $\cdots$  & $\cdots$  & $\cdots$  & ZTF  \\ 
59750.34 &  $\cdots$  & $\cdots$  & $\cdots$  & 19.84(0.15) & $\cdots$  & $\cdots$  & $\cdots$  & $\cdots$  & ZTF  \\ 
59755.34 &  $\cdots$  & $\cdots$  & $\cdots$  & 20.39(0.20) & $\cdots$  & $\cdots$  & $\cdots$  & $\cdots$  & ZTF  \\ 
59759.32 &  $\cdots$  & $\cdots$  & $\cdots$  & 19.71(0.18) & $\cdots$  & $\cdots$  & $\cdots$  & $\cdots$  & ZTF  \\ 
59761.36 &  $\cdots$  & $\cdots$  & $\cdots$  & 19.70(0.13) & $\cdots$  & $\cdots$  & $\cdots$  & $\cdots$  & ZTF  \\ 
59763.32 &  $\cdots$  & $\cdots$  & $\cdots$  & 19.47(0.14) & $\cdots$  & $\cdots$  & $\cdots$  & $\cdots$  & ZTF  \\ 
59765.34 &  $\cdots$  & $\cdots$  & $\cdots$  & 19.72(0.14) & $\cdots$  & $\cdots$  & $\cdots$  & $\cdots$  & ZTF  \\ 
59768.34 &  $\cdots$  & 19.47(0.13) & $\cdots$  & $\cdots$  & $\cdots$  & $\cdots$  & $\cdots$  & $\cdots$  & ZTF  \\ 
59772.43 &  $\cdots$  & $\cdots$  & $\cdots$  & 19.71(0.21) & $\cdots$  & $\cdots$  & $\cdots$  & $\cdots$  & ZTF  \\ 
59774.30 &  $\cdots$  & $\cdots$  & $\cdots$  & 19.44(0.16) & $\cdots$  & $\cdots$  & $\cdots$  & $\cdots$  & ZTF  \\ 
59774.38 &  $\cdots$  & $\cdots$  & $\cdots$  & 19.52(0.27) & $\cdots$  & $\cdots$  & $\cdots$  & $\cdots$  & ZTF  \\ 
59776.23 &  $\cdots$  & $\cdots$  & $\cdots$  & 19.59(0.25) & $\cdots$  & $\cdots$  & $\cdots$  & $\cdots$  & ZTF  \\ 
59776.34 &  $\cdots$  & $\cdots$  & $\cdots$  & 19.59(0.20) & $\cdots$  & $\cdots$  & $\cdots$  & $\cdots$  & ZTF  \\ 
59779.27 &  $\cdots$  & $\cdots$  & $\cdots$  & 19.81(0.21) & $\cdots$  & $\cdots$  & $\cdots$  & $\cdots$  & ZTF  \\ 
59782.37 &  $\cdots$  & $\cdots$  & $\cdots$  & 20.33(0.23) & $\cdots$  & $\cdots$  & $\cdots$  & $\cdots$  & ZTF  \\ 
59784.30 &  $\cdots$  & $\cdots$  & $\cdots$  & 19.74(0.26) & $\cdots$  & $\cdots$  & $\cdots$  & $\cdots$  & ZTF  \\ 
59786.32 &  $\cdots$  & 17.27(0.04) & $\cdots$  & 17.42(0.03) & $\cdots$  & $\cdots$  & $\cdots$  & $\cdots$  & ZTF  \\ 
59788.28 &  $\cdots$  & $\cdots$  & $\cdots$  & 16.84(0.02) & $\cdots$  & $\cdots$  & $\cdots$  & $\cdots$  & ZTF  \\ 
59790.30 &  $\cdots$  & 16.32(0.03) & $\cdots$  & $\cdots$  & $\cdots$  & $\cdots$  & $\cdots$  & $\cdots$  & ZTF  \\ 
59790.89 & 16.42(0.05) & 16.31(0.02) & 16.51(0.05) & 16.37(0.02) & 16.52(0.02) & $\cdots$  & $\cdots$  & $\cdots$  & Copernico+AFOSC\\
59793.27 &  $\cdots$  & 16.10(0.03) & $\cdots$  & 16.13(0.02) & $\cdots$  & $\cdots$  & $\cdots$  & $\cdots$  & ZTF  \\ 
59794.96 & 16.20(0.02) & 16.12(0.01) & 16.17(0.02) & 16.12(0.01) & 16.19(0.02) & $\cdots$  & $\cdots$  & $\cdots$  & 67/92-Schmidt+Moravian \\
59797.31 &  $\cdots$  & 16.24(0.03) & $\cdots$  & 16.18(0.02) & $\cdots$  & $\cdots$  & $\cdots$  & $\cdots$  & ZTF  \\ 
59801.21 &  $\cdots$  & $\cdots$  & $\cdots$  & 16.29(0.03) & $\cdots$  & $\cdots$  & $\cdots$  & $\cdots$  & ZTF  \\ 
59805.25 &  $\cdots$  & 16.50(0.03) & $\cdots$  & 16.30(0.04) & $\cdots$  & $\cdots$  & $\cdots$  & $\cdots$  & ZTF  \\ 
59807.29 &  $\cdots$  & $\cdots$  & $\cdots$  & 16.29(0.03) & $\cdots$  & $\cdots$  & $\cdots$  & $\cdots$  & ZTF  \\ 
59809.11 &  $\cdots$  & $\cdots$  & $\cdots$  & $\cdots$  & $\cdots$  & 15.65(0.01) & 15.61(0.02) & 15.28(0.02) & NOT+NOTCam  \\ 
59809.35 &  $\cdots$  & 16.55(0.03) & $\cdots$  & 16.29(0.05) & $\cdots$  & $\cdots$  & $\cdots$  & $\cdots$  & ZTF  \\ 
59811.32 &  $\cdots$  & $\cdots$  & $\cdots$  & 16.36(0.02) & $\cdots$  & $\cdots$  & $\cdots$  & $\cdots$  & ZTF  \\ 
59813.29 &  $\cdots$  & 16.71(0.04) & $\cdots$  & 16.41(0.03) & $\cdots$  & $\cdots$  & $\cdots$  & $\cdots$  & ZTF  \\ 
59815.24 &  $\cdots$  & 16.76(0.05) & $\cdots$  & 16.50(0.03) & $\cdots$  & $\cdots$  & $\cdots$  & $\cdots$  & ZTF  \\ 
59817.18 &  $\cdots$  & 16.87(0.05) & $\cdots$  & 16.55(0.03) & $\cdots$  & $\cdots$  & $\cdots$  & $\cdots$  & ZTF  \\ 
59819.18 &  $\cdots$  & 16.94(0.04) & $\cdots$  & 16.63(0.03) & $\cdots$  & $\cdots$  & $\cdots$  & $\cdots$  & ZTF  \\ 
59821.18 &  $\cdots$  & 17.02(0.04) & $\cdots$  & 16.67(0.04) & $\cdots$  & $\cdots$  & $\cdots$  & $\cdots$  & ZTF  \\ 
59823.30 &  $\cdots$  & 17.09(0.02) & $\cdots$  & 16.73(0.02) & $\cdots$  & $\cdots$  & $\cdots$  & $\cdots$  & ZTF  \\ 
59824.97 & $\cdots$ & 17.30(0.05) & $\cdots$ & $\cdots$ & $\cdots$ & $\cdots$  & $\cdots$  & $\cdots$  & 67/92-Schmidt+Moravian \\
59825.24 &  $\cdots$  & 17.20(0.06) & $\cdots$  & 16.75(0.02) & $\cdots$  & $\cdots$  & $\cdots$  & $\cdots$  & ZTF  \\ 
59826.95 & 17.71(0.10) & $\cdots$ & 17.18(0.03) & $\cdots$ & $\cdots$ & $\cdots$  & $\cdots$  & $\cdots$  & 67/92-Schmidt+Moravian \\
59827.31 &  $\cdots$  & $\cdots$  & $\cdots$  & 16.86(0.03) & $\cdots$  & $\cdots$  & $\cdots$  & $\cdots$  & ZTF  \\ 
59829.16 &  $\cdots$  & $\cdots$  & $\cdots$  & 16.94(0.05) & $\cdots$  & $\cdots$  & $\cdots$  & $\cdots$  & ZTF  \\ 
59836.96 &  $\cdots$  & $\cdots$  & $\cdots$  & $\cdots$  & $\cdots$  & 16.45(0.02) & 16.28(0.02) & 15.92(0.04) & NOT+NOTCam  \\ 
59837.93 & 18.74(0.18) & 18.28(0.08) & 18.09(0.11) & 17.54(0.0240 & 17.49(0.04) & $\cdots$  & $\cdots$  & $\cdots$  & 67/92-Schmidt+Moravian \\
59839.18 &  $\cdots$  & 18.25(0.04) & $\cdots$  & $\cdots$  & $\cdots$  & $\cdots$  & $\cdots$  & $\cdots$  & ZTF  \\ 
59839.28 &  $\cdots$  & 18.28(0.08) & $\cdots$  & $\cdots$  & $\cdots$  & $\cdots$  & $\cdots$  & $\cdots$  & ZTF  \\ 
59840.88 & 18.62(0.15) & $\cdots$ & 18.56(0.19) & $\cdots$ & $\cdots$ & $\cdots$  & $\cdots$  & $\cdots$  & 67/92-Schmidt+Moravian \\
59841.14 &  $\cdots$  & 18.44(0.08) & $\cdots$  & $\cdots$  & $\cdots$  & $\cdots$  & $\cdots$  & $\cdots$  & ZTF  \\ 
59841.19 &  $\cdots$  & 18.42(0.08) & $\cdots$  & $\cdots$  & $\cdots$  & $\cdots$  & $\cdots$  & $\cdots$  & ZTF  \\ 
59843.20 &  $\cdots$  & 18.56(0.10) & $\cdots$  & $\cdots$  & $\cdots$  & $\cdots$  & $\cdots$  & $\cdots$  & ZTF  \\ 
59845.16 &  $\cdots$  & $\cdots$  & $\cdots$  & 17.90(0.04) & $\cdots$  & $\cdots$  & $\cdots$  & $\cdots$  & ZTF  \\ 
59847.16 &  $\cdots$  & 18.76(0.13) & $\cdots$  & 18.04(0.05) & $\cdots$  & $\cdots$  & $\cdots$  & $\cdots$  & ZTF  \\ 
59849.13 &  $\cdots$  & $\cdots$  & $\cdots$  & 18.07(0.04) & $\cdots$  & $\cdots$  & $\cdots$  & $\cdots$  & ZTF  \\ 
59851.19 &  $\cdots$  & 19.13(0.10) & $\cdots$  & 18.18(0.05) & $\cdots$  & $\cdots$  & $\cdots$  & $\cdots$  & ZTF  \\ 
59853.15 &  $\cdots$  & 18.95(0.12) & $\cdots$  & 18.16(0.05) & $\cdots$  & $\cdots$  & $\cdots$  & $\cdots$  & ZTF  \\ 
59855.23 &  $\cdots$  & 19.09(0.14) & $\cdots$  & 18.17(0.04) & $\cdots$  & $\cdots$  & $\cdots$  & $\cdots$  & ZTF  \\ 
59857.18 &  $\cdots$  & $\cdots$  & $\cdots$  & 18.20(0.06) & $\cdots$  & $\cdots$  & $\cdots$  & $\cdots$  & ZTF  \\ 
59859.16 &  $\cdots$  & 19.02(0.13) & $\cdots$  & 18.22(0.06) & $\cdots$  & $\cdots$  & $\cdots$  & $\cdots$  & ZTF  \\ 
59861.25 &  $\cdots$  & $\cdots$  & $\cdots$  & 18.11(0.06) & $\cdots$  & $\cdots$  & $\cdots$  & $\cdots$  & ZTF  \\ 
59861.91 &  $\cdots$  & $\cdots$  & $\cdots$  & $\cdots$  & $\cdots$  & 17.43(0.04) & 17.4(0.08) & 16.72(0.04) & NOT+NOTCam  \\ 
59865.13 &  $\cdots$  & $\cdots$  & $\cdots$  & 18.55(0.05) & $\cdots$  & $\cdots$  & $\cdots$  & $\cdots$  & ZTF  \\ 
59867.18 &  $\cdots$  & $\cdots$  & $\cdots$  & 18.60(0.05) & $\cdots$  & $\cdots$  & $\cdots$  & $\cdots$  & ZTF  \\ 
59868.85 & 20.34(0.06) & $\cdots$  & 19.38(0.03) & 18.78(0.02) & 18.46(0.04) & $\cdots$  & $\cdots$  & $\cdots$  & NOT+ALFOSC  \\ 
59870.18 &  $\cdots$  & $\cdots$  & $\cdots$  & 18.75(0.08) & $\cdots$  & $\cdots$  & $\cdots$  & $\cdots$  & ZTF  \\ 
59872.12 & $\cdots$  & $\cdots$  & $\cdots$  & 19.03(0.09) & $\cdots$  & $\cdots$  & $\cdots$  & $\cdots$  & ZTF  \\ 
59872.84 & 20.69(0.09) & 19.96(0.09) & 19.86(0.04) & 19.07(0.12) & 18.73(0.04) & $\cdots$  & $\cdots$  & $\cdots$  & NOT+ALFOSC  \\ 
59874.13 &  $\cdots$  & $\cdots$  & $\cdots$  & 19.17(0.07) & $\cdots$  & $\cdots$  & $\cdots$  & $\cdots$  & ZTF  \\ 
59879.12 &  $\cdots$  & $\cdots$  & $\cdots$  & 19.34(0.08) & $\cdots$  & $\cdots$  & $\cdots$  & $\cdots$  & ZTF  \\ 
59881.91 &  $\cdots$  & $\cdots$  & $\cdots$  & $\cdots$  & $\cdots$  & 18.91(0.11) & 18.81(0.48) & 17.49(0.16) & NOT+NOTCam  \\
59883.81 &  $\cdots$  & 20.67(0.10) & 20.68(0.06) & 19.72(0.07) & 19.71(0.05) & $\cdots$  & $\cdots$  & $\cdots$  & NOT+ALFOSC \\
59886.82 &  $\cdots$  & $\cdots$  & $\cdots$  & 19.72(0.04) & $\cdots$  & $\cdots$  & $\cdots$  & $\cdots$  & NOT+ALFOSC  \\
59916.80 & $\cdots$  & $\cdots$  & $\cdots$  & 19.73(0.08) & $\cdots$  & $\cdots$  & $\cdots$  & $\cdots$  & NOT+ALFOSC  \\ 
60035.19 & $\cdots$  & 21.30(0.20) & $\cdots$  & 20.72(0.12) & $\cdots$  & $\cdots$  & $\cdots$  & $\cdots$  & NOT+ALFOSC  \\ 
\end{longtable}

\begin{longtable}{ccccccccc}\caption{Photometry table of SN~2022prr in Swift UV ($UVW2$, $UVM2$, $UVW1$, $U$, $B$, and $V$) and X-ray ($0.2 - 10$~keV) bands with errors given in brackets.}\label{tab:uv_22prr}\\
\hline
\hline

        MJD	& Epoch & $m_{UVW2}$  &$m_{UVM2}$  &$m_{UVW1}$  &$m_\text{U}$  &$m_\text{B}$  &$m_\text{V}$  &	$L_{0.2-10\text{ keV}}$ \\
        	& (d) & (mag) & (mag) & (mag) & (mag) & (mag) &	(mag)& $(10^{41}$ erg~$s^{-1})$ \\
            
\hline
\endfirsthead
\endhead
\hline
\endfoot
\hline \hline
\endlastfoot
    
        59789.19 & $-7.1$ & 15.72(0.10) & 15.47(0.12) & 15.43(0.10) & 15.82(0.14) & 16.66(0.14) & 16.90(0.30) &	$<$4.9 \\
        59791.38 & $-4.9$ & 15.39(0.08) & 15.19(0.08) & 15.09(0.08) & 15.41(0.12) & 16.25(0.12) & 16.26(0.18) &	$<$3.6 \\
        59793.64 &$-2.6$& 15.54(0.11) &	15.21(0.08) &	15.07(0.09) &	15.33(0.12) &	16.25(0.12) &	16.02(0.18) &	$<$11.3 \\
        59797.56  &$1.3$& 16.40(0.16) & 15.86(0.13) & 15.62(0.11) & 15.63(0.14) & 16.36(0.13) & 16.21(0.18) &	$<$3.7 \\
        59801.67  &$5.4$& 17.35(0.36) & 16.56(0.26) & 16.31(0.19) & 16.35(0.21) & 16.63(0.13) & 16.26(0.18) &	$<$2.1 \\
        59805.78 &$9.5$& 17.26(0.33) & 16.56(0.23) & 16.35(0.20) & 15.97(0.16) & 16.50(0.12) & 16.38(0.20) &	$<$3.3 \\
        59807.70  &$11.4$& 18.22(0.77) & 16.94(0.32) & 16.84(0.30) & 16.56(0.29) & 16.73(0.14) & 16.32(0.19) &	$<$1.8 \\
        59810.43 &$14.2$& 17.77(0.51) & 17.57(0.56) & 16.98(0.33) & 16.74(0.34) & 16.80(0.15) & 16.29(0.19) & $<$3.2 \\      
    \hline
 \end{longtable}

\begin{longtable}{ccccccccccc}\caption{Photometry table of SN~2023ucy in optical and NIR bands.}\label{tab:phot_SN2023ucy}\\
\hline 
\hline 

 MJD & $m_u$($m_\text{u,err}$)& $m_B$($m_{B,\text{err}}$) & $m_g$($m_{g,\text{err}}$) & $m_V$($m_{V,\text{err}}$) & $m_r$($m_{r,\text{err}}$) & $m_\text{i}$($m_{i,\text{err}}$) & $m_J$($m_{J,\text{err}}$) & $m_H$($m_{H,\text{err}}$) & $m_K$($m_{K,\text{err}}$) & Telescope \\
  & (mag)& (mag)& (mag)& (mag)& (mag)& (mag)& (mag)& (mag)& (mag)& \\
 \hline

\hline
\endfirsthead

\multicolumn{11}{c}{{\bfseries \tablename\ \thetable{} -- continued from previous page}} \\
\hline 
 MJD & $m_u$($m_\text{u,err}$)& $m_B$($m_{B,\text{err}}$) & $m_g$($m_{g,\text{err}}$) & $m_V$($m_{V,\text{err}}$) & $m_r$($m_{r,\text{err}}$) & $m_\text{i}$($m_{i,\text{err}}$) & $m_J$($m_{J,\text{err}}$) & $m_H$($m_{H,\text{err}}$) & $m_K$($m_{K,\text{err}}$) & Telescope \\
 & (mag)& (mag)& (mag)& (mag)& (mag)& (mag)& (mag)& (mag)& (mag)&\\
 \hline
\endhead

\hline \multicolumn{11}{|r|}{{Continued on next page}} \\ \hline
\endfoot

\hline \hline
\endlastfoot

60222.65 &  $\cdots$  &  $\cdots$  & 17.47(0.06) &  $\cdots$  & 17.90(0.05) & $\cdots$  & $\cdots$  & $\cdots$  & $\cdots$  & ZTF  \\ 
60224.69 &  $\cdots$ &  $\cdots$   & 16.72(0.02) &  $\cdots$  & 16.93(0.03) & $\cdots$  & $\cdots$  & $\cdots$  & $\cdots$  & ZTF  \\ 
60228.66 &  $\cdots$  &  $\cdots$  & 16.71(0.02) &  $\cdots$  & 16.69(0.03) & $\cdots$  & $\cdots$  & $\cdots$  & $\cdots$  & ZTF  \\ 
60231.64 &  $\cdots$ &  $\cdots$   & 16.70(0.03)&  $\cdots$   & 16.66(0.03) & $\cdots$  & $\cdots$  & $\cdots$  & $\cdots$  & ZTF  \\ 
60231.85 & 17.16(0.01) &  $\cdots$  & 16.77(0.01) &  $\cdots$  & 16.69(0.01) & 16.38(0.04) & $\cdots$  & $\cdots$  & $\cdots$  & NOT+ALFOSC  \\ 
60233.64 &  $\cdots$ &  $\cdots$   & 16.75(0.04) &  $\cdots$  & 16.69(0.02) & $\cdots$  & $\cdots$  & $\cdots$  & $\cdots$  & ZTF  \\ 
60235.63 &  $\cdots$ &  $\cdots$   & 16.79(0.02) &  $\cdots$  & 16.74(0.02) & $\cdots$  & $\cdots$  & $\cdots$  & $\cdots$  & ZTF  \\ 
60237.66 &  $\cdots$  &  $\cdots$  & 16.88(0.02) &  $\cdots$   & 16.80(0.03) & $\cdots$  & $\cdots$  & $\cdots$  & $\cdots$  & ZTF  \\ 
60239.62 &  $\cdots$ &  $\cdots$   & 16.92(0.03) &  $\cdots$  & 16.81(0.03) & $\cdots$  & $\cdots$  & $\cdots$  & $\cdots$  & ZTF  \\ 
60242.62 &  $\cdots$  &  $\cdots$  & 17.02(0.04) &  $\cdots$  & 16.85(0.03) & $\cdots$  & $\cdots$  & $\cdots$  & $\cdots$  & ZTF  \\ 
60245.67 &  $\cdots$  &  $\cdots$  & 17.11(0.05) &  $\cdots$  & 16.85(0.04) & $\cdots$  & $\cdots$  & $\cdots$  & $\cdots$  & ZTF  \\ 
60247.66 &  $\cdots$  &  $\cdots$  & 17.22(0.08) &  $\cdots$  & 16.85(0.04) & $\cdots$  & $\cdots$  & $\cdots$  & $\cdots$  & ZTF  \\ 
60249.68 &  $\cdots$  &  $\cdots$  & 17.22(0.03) &  $\cdots$  & 16.78(0.04) & $\cdots$  & $\cdots$  & $\cdots$  & $\cdots$  & ZTF  \\ 
60251.66 &  $\cdots$  &  $\cdots$  & 17.24(0.03) &  $\cdots$  & 16.90(0.03) & $\cdots$  & $\cdots$  & $\cdots$  & $\cdots$  & ZTF  \\ 
60253.66 &  $\cdots$  &  $\cdots$  & 17.27(0.06) &  $\cdots$  & 16.87(0.04) & $\cdots$  & $\cdots$  & $\cdots$  & $\cdots$  & ZTF  \\ 
60255.64 &  $\cdots$ &  $\cdots$   & 17.36(0.03) &  $\cdots$  & 16.93(0.05) & $\cdots$  & $\cdots$  & $\cdots$  & $\cdots$  & ZTF  \\ 
60255.88 &  $\cdots$  &  $\cdots$  & 17.68(0.10) &  $\cdots$  & 16.97(0.02) & 16.68(0.01) & $\cdots$  & $\cdots$  & $\cdots$  & NOT+ALFOSC  \\ 
60256.77 & $\cdots$ & 17.77(0.06) & 17.58(0.03) & 17.35(0.09) & 17.02(0.02) & 17.04(0.01)& $\cdots$& $\cdots$& $\cdots$ & Copernico+AFOSC \\
60258.61 &  $\cdots$  &  $\cdots$  & 17.38(0.05) &  $\cdots$  & 16.94(0.04) & $\cdots$  & $\cdots$  & $\cdots$  & $\cdots$  & ZTF  \\ 
60260.71 &  $\cdots$  &  $\cdots$  & 17.44(0.04) &  $\cdots$  & 16.94(0.05) & $\cdots$  & $\cdots$  & $\cdots$  & $\cdots$  & ZTF  \\ 
60261.81 & 19.52(0.06) &  $\cdots$  & 17.60(0.01) &  $\cdots$  & 16.96(0.01) & 16.65(0.01) & $\cdots$  & $\cdots$  & $\cdots$  & NOT+ALFOSC  \\ 
60262.66 &  $\cdots$  &  $\cdots$  & 17.47(0.05) &  $\cdots$  & $\cdots$  & $\cdots$  & $\cdots$  & $\cdots$  & $\cdots$  & ZTF  \\ 
60268.64 &  $\cdots$  &  $\cdots$  & 17.61(0.07) &  $\cdots$  & 16.99(0.04) & $\cdots$  & $\cdots$  & $\cdots$  & $\cdots$  & ZTF  \\ 
60271.81 &  $\cdots$  &  $\cdots$  & $\cdots$  &  $\cdots$  & $\cdots$  & $\cdots$  & 16.17(0.02) & 16.11(0.03) & 15.75(0.02) & NOT+NOTCam  \\ 
60274.81 &  $\cdots$  &  $\cdots$  & 17.91(0.06) &  $\cdots$  & 17.07(0.01) & 16.64(0.01) & $\cdots$  & $\cdots$  & $\cdots$  & NOT+ALFOSC  \\ 
60284.71 & $\cdots$ & 18.32(0.07) & 18.15(0.09) & 17.69(0.14) & 17.13(0.02) & 17.08(0.01)& $\cdots$& $\cdots$& $\cdots$ & Copernico+AFOSC \\
60285.64 &  $\cdots$  &  $\cdots$  & 17.80(0.07) &  $\cdots$  & 16.97(0.05) & $\cdots$  & $\cdots$  & $\cdots$  & $\cdots$  & ZTF  \\ 
60285.79 &  $\cdots$  &  $\cdots$  & 17.89(0.06) &  $\cdots$  & 17.05(0.02) & 16.71(0.03) & $\cdots$  & $\cdots$  & $\cdots$  & NOT+ALFOSC  \\ 
60286.82 &  $\cdots$  &  $\cdots$  & 18.11(0.04) &  $\cdots$  & 17.15(0.02) & 16.94(0.09) & $\cdots$  & $\cdots$  & $\cdots$  & NOT+ALFOSC  \\ 
60288.62 &  $\cdots$  &  $\cdots$  & 18.22(0.14) &  $\cdots$  & $\cdots$  & $\cdots$  & $\cdots$  & $\cdots$  & $\cdots$  & ZTF  \\ 
60290.64 &  $\cdots$ &  $\cdots$   & $\cdots$  &  $\cdots$  & 17.05(0.05) & $\cdots$  & $\cdots$  & $\cdots$  & $\cdots$  & ZTF  \\ 
60294.80 &  $\cdots$ &  $\cdots$   & 17.94(0.06) &  $\cdots$  & 17.36(0.04) & 16.76(0.04) & $\cdots$  & $\cdots$  & $\cdots$  & NOT+ALFOSC  \\ 
60400.16 &  $\cdots$  &  $\cdots$  & 19.73(0.17) &  $\cdots$  & $\cdots$  & $\cdots$  & $\cdots$  & $\cdots$  & $\cdots$  & NOT+ALFOSC  \\ 
\hline
\end{longtable}

\begin{table*}
\centering
\caption{Photometry table of SN~2024ljc in optical and NIR bands.}
\label{tab:phot_SN2024ljc}
\begin{tabular}{cccccccc}
\hline
\hline
 MJD & $m_g$($m_{g,\text{err}}$) & $m_r$($m_{r,\text{err}}$) & $m_\text{i}$($m_{i,\text{err}}$) & $m_J$($m_{J,\text{err}}$) & $m_H$($m_{H,\text{err}}$) & $m_K$($m_{K,\text{err}}$) & Telescope \\
 & (mag)& (mag)& (mag)& (mag)& (mag)& (mag)&\\
 \hline
60476.84 &  $\cdots$  & 18.74(0.05) & $\cdots$  & $\cdots$  & $\cdots$  & $\cdots$  & ZTF  \\ 
60479.93 & 18.66(0.07) & 18.24(0.07) & $\cdots$  & $\cdots$  & $\cdots$  & $\cdots$  & ZTF  \\ 
60481.91 & 18.73(0.08) & 18.26(0.06) & $\cdots$  & $\cdots$  & $\cdots$  & $\cdots$  & ZTF  \\ 
60483.91 & 19.03(0.18) & 18.31(0.06) & $\cdots$  & $\cdots$  & $\cdots$  & $\cdots$  & ZTF  \\ 
60483.97 & 19.10(0.01) & 18.41(0.01) & 18.29(0.01) & $\cdots$  & $\cdots$  & $\cdots$  & NOT+ALFOSC  \\ 
60485.91 & 19.10(0.13) & 18.36(0.06) & $\cdots$  & $\cdots$  & $\cdots$  & $\cdots$  & ZTF  \\ 
60488.12 &  $\cdots$  & $\cdots$  & $\cdots$  & 17.40(0.03) & 17.14(0.02) & 16.90(0.06) & NOT+NOTCam  \\ 
60488.93 & 19.31(0.10) & $\cdots$  & $\cdots$  & $\cdots$  & $\cdots$  & $\cdots$  & ZTF  \\ 
60490.85 & 19.50(0.16) & 18.60(0.05) & $\cdots$  & $\cdots$  & $\cdots$  & $\cdots$  & ZTF  \\ 
60500.16 & 21.32(0.03) & 19.97(0.02) & 19.68(0.02) & $\cdots$  & $\cdots$  & $\cdots$  & NOT+ALFOSC  \\ 
60504.01 & 21.51(0.09) & 20.23(0.04) & 20.04(0.04) & $\cdots$  & $\cdots$  & $\cdots$  & NOT+ALFOSC  \\ 
60507.12 &  $\cdots$  & $\cdots$  & $\cdots$  & 19.32(0.12) & 19.00(0.10) & 18.29(0.21) & NOT+NOTCam  \\ 
60525.00 &  $\cdots$  & 21.47(0.05) & 20.96(0.04) & $\cdots$  & $\cdots$  & $\cdots$  & NOT+ALFOSC  \\ 
60545.08 &  $\cdots$  & $\cdots$  & 21.47(0.18) & $\cdots$  & $\cdots$  & $\cdots$  & NOT+ALFOSC  \\ 
\hline
\hline
\end{tabular}
\end{table*}

\end{small}

\begin{table*}
\caption{Spectroscopic log of observations for SN~2022prr.}
\centering
\begin{tabular}{cccccccc}
\hline
\hline
MJD & Epoch & Grism & Slit & $\lambda / \Delta \lambda$ & $\lambda$ & $t_\text{exp}$ & Telescope + instrument \\
 & (d) & & (") & & (\AA) & (s) & \\
\hline
59788.4 & $-8$ & $-$ & $-$ & $1200$ & 3400 $-$ 10000 & 1800 & UH88+SNIFS \\
59790.1 & $-6$ & Gr\#4 & 1.3 & 280 & 3200 $-$ 9600 & 600 & NOT+ALFOSC \\
59794.1 & $-2$ & Gr\#18 & 1.0 & 1000 & 3450 $-$ 5350 & 150 & NOT+ALFOSC \\
59799.0 & $+3$ & Gr\#4 & 1.0 & 360 & 3200 $-$ 9600 & 300 & NOT+ALFOSC \\
59810.0 & $+14$ & Gr\#4 & 1.0 & 360 & 3200 $-$ 9600 & 300 & NOT+ALFOSC \\
59825.9 & $+30$ & Gr\#4 & 1.3 & 280 & 3200 $-$ 9600 & 600 & NOT+ALFOSC \\
59830.9 & $+35$ & Gr\#19 & 1.3 & 750 & 4400 $-$ 6950 & 2400 & NOT+ALFOSC \\
59853.9 & $+58$ & Gr\#4 & 1.3 & 280 & 3200 $-$ 9600 & 1200 & NOT+ALFOSC \\
60266.2 & $+470$ & Red + Blue & 2.0 & 400 + 700 & 5400 $-$ 10000 & 2700 & FTN+FLOYDS\\
\hline
\hline
\end{tabular}
\label{tab:spec22prr}
\end{table*}

\begin{table*}
\caption{Spectroscopic log of observations for SN~2023ucy.}
\centering
\begin{tabular}{cccccccc}
\hline
\hline
MJD & Epoch & Grism & Slit & $\lambda / \Delta \lambda$ & $\lambda$ & $t_\text{exp}$ & Telescope + instrument \\
 & (d) & & (") & & (\AA) & (s) & \\
\hline
60225.0 & $-6$ & Gr\#4 & 1.0 & 360 & 3200 $-$ 9600 & 900 & NOT+ALFOSC \\
60231.9 & $+1$ & Gr\#4 & 1.0 & 360 & 3200 $-$ 9600 & 1200 & NOT+ALFOSC \\
60233.5 & $+3$ &  $-$  & $-$ & 500 & 4100 $-$ 8500 & 1200 & Seimei+KOOLS-IFU \\
60237.6 & $+7$ & Gr\#7 + Gr\#8 & $1.92$ & 1000 + 1200 & 3800 $-$ 9200 & 1500 & HCT-2m+HFOSC \\ 
60240.9 & $+10$ & Gr\#4 & 1.3 & 280 & 3200 $-$ 9600 & 900 & NOT+ALFOSC \\ 
60255.9 & $+25$ & Gr\#4 & 1.0 & 360 & 3200 $-$ 9600 & 1200 & NOT+ALFOSC \\ 
60274.8 & $+44$ & Gr\#4 & 1.3 & 280 & 3200 $-$ 9600 & 1200 & NOT+ALFOSC \\ 
60284.7 & $+54$ & VPH7 & 1.69 & $-$ & 3550 $-$ 7170 & 1800 & Copernico+AFOSC \\ 
60286.8 & $+56$ & Gr\#4 & 1.3 & 280 & 3200 $-$ 9600 & 1200 & NOT+ALFOSC \\
\hline 
\hline
\end{tabular}
\label{tab:spec23ucy}
\end{table*}

\begin{table*}
\caption{Spectroscopic log of observations for SN~2024ljc.}
\centering
\begin{tabular}{cccccccc}
\hline
\hline
MJD & Epoch & Grism / Grating & Slit & $\lambda / \Delta \lambda$ & $\lambda$ & $t_\text{exp}$ & Telescope + instrument \\
 & (d) & & (") & & (\AA) & (s) & \\
\hline
60484.0 & $-1$ & Gr\#4 & 1.0 & 360 & 3200 $-$ 9600 & 2400 & NOT+ALFOSC \\ 
60488.0 & $+3$ & VPH & $1.8$ &$350$ & 4020 $-$ 8100 & 1500 & LT+SPRAT \\ 
60500.1 & $+15$ & Gr\#4 & 1.0 & 360 & 3200 $-$ 9600 & 2400 & NOT+ALFOSC \\ 
\hline
\hline
\end{tabular}

\label{tab:spec24ljc}
\end{table*}

\begin{figure}
\centering
\includegraphics[trim={0cm 0cm 0cm 0cm},clip,width=0.3\linewidth]{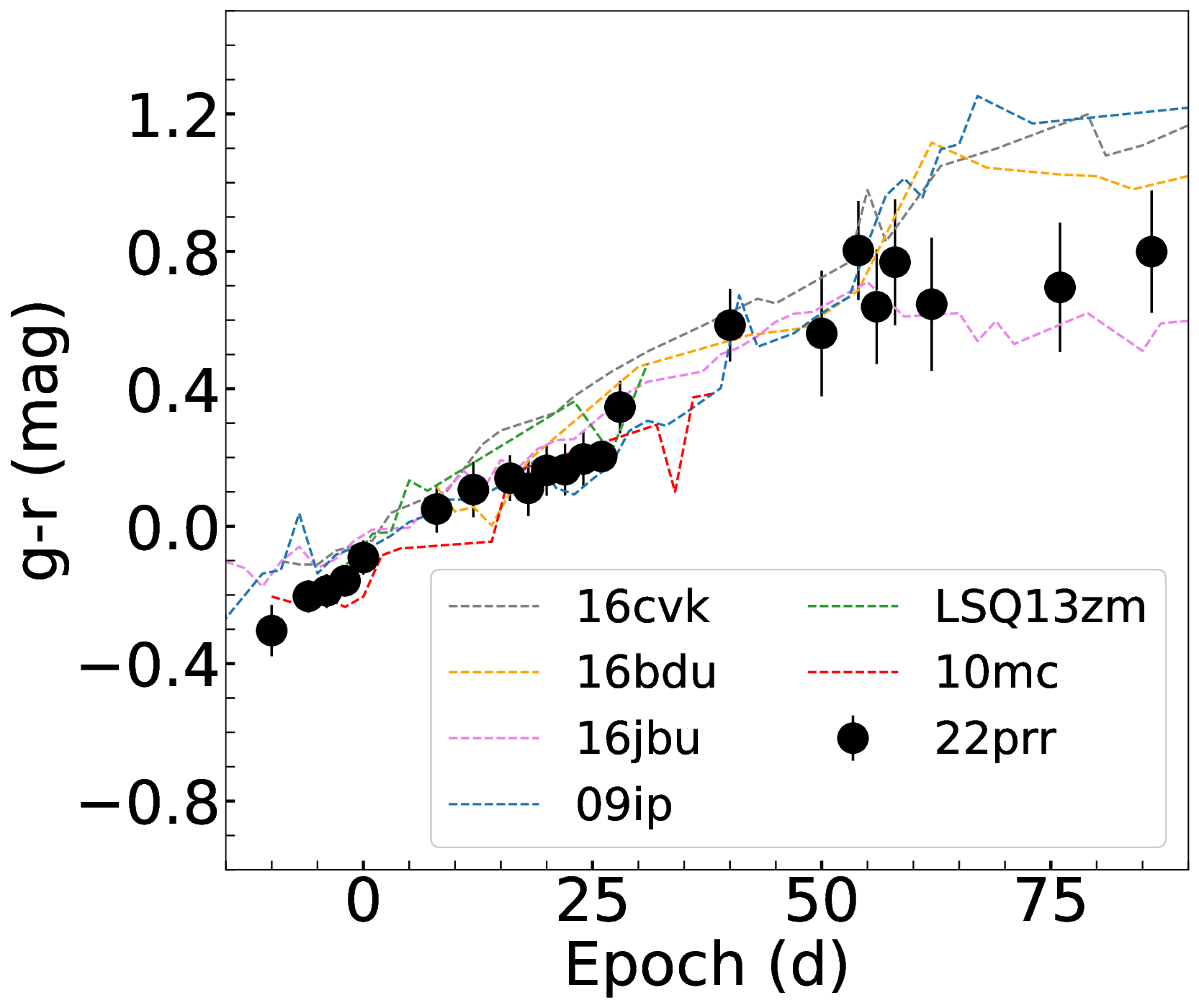}
\includegraphics[trim={0cm 0cm 0cm 0cm},clip,width=0.3\linewidth]{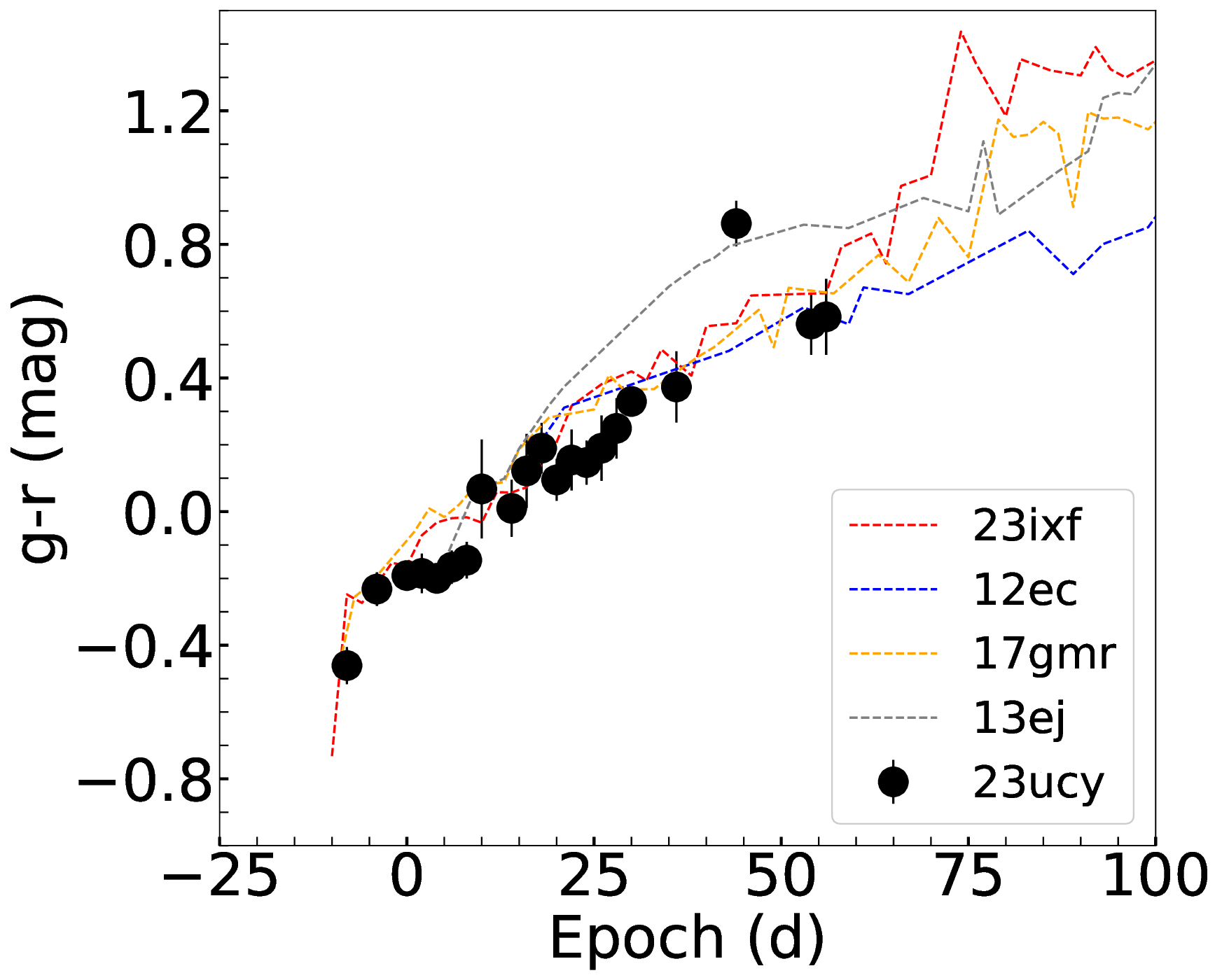}
\includegraphics[trim={0cm 0cm 0cm 0cm},clip,width=0.3\linewidth]{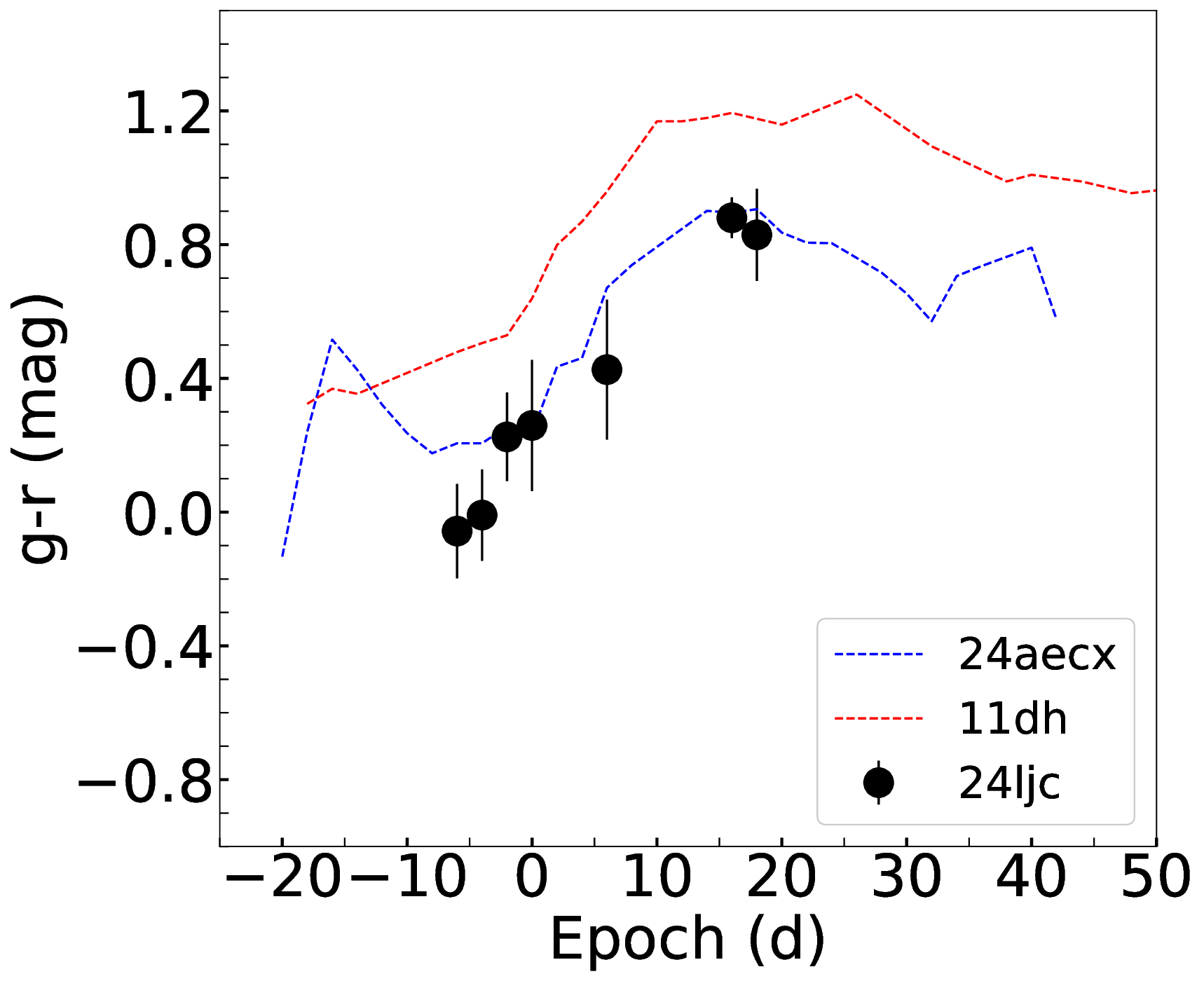}

\caption{Left: colour evolution $(g-r)$ of SN~2022prr, compared to events SN~2016cvk \citep{Matilainen2025}, SN~2016bdu \citep{Pastorello2018}, SN~2016jbu \citep{Brennan2022a}, SN~2009ip \citep{Graham2017}, LSQ13zm \citep{Tartaglia2016}, and SN~2010mc \citep{Ofek2013b}. Middle: color evolution $(g-r)$ of SN~2023ucy, compared to events SN~2012ec \citep{Barbarino2015}, SN~2017gmr \citep{Andrews2019}, and SN~2013ej \citep{Dhungana2016}. Right: color evolution $(g-r)$ of SN~2024ljc, compared to events SN~2011dh \citep{Ergon2014} and SN~2024aecx \citep{Xi2026}. Epoch is measured from the $r$-band light curve peak of each event, and the magnitudes have been corrected for extinction.}
\label{fig:colorevolution}
\end{figure}

\begin{figure}
\centering
\includegraphics[width=\linewidth]{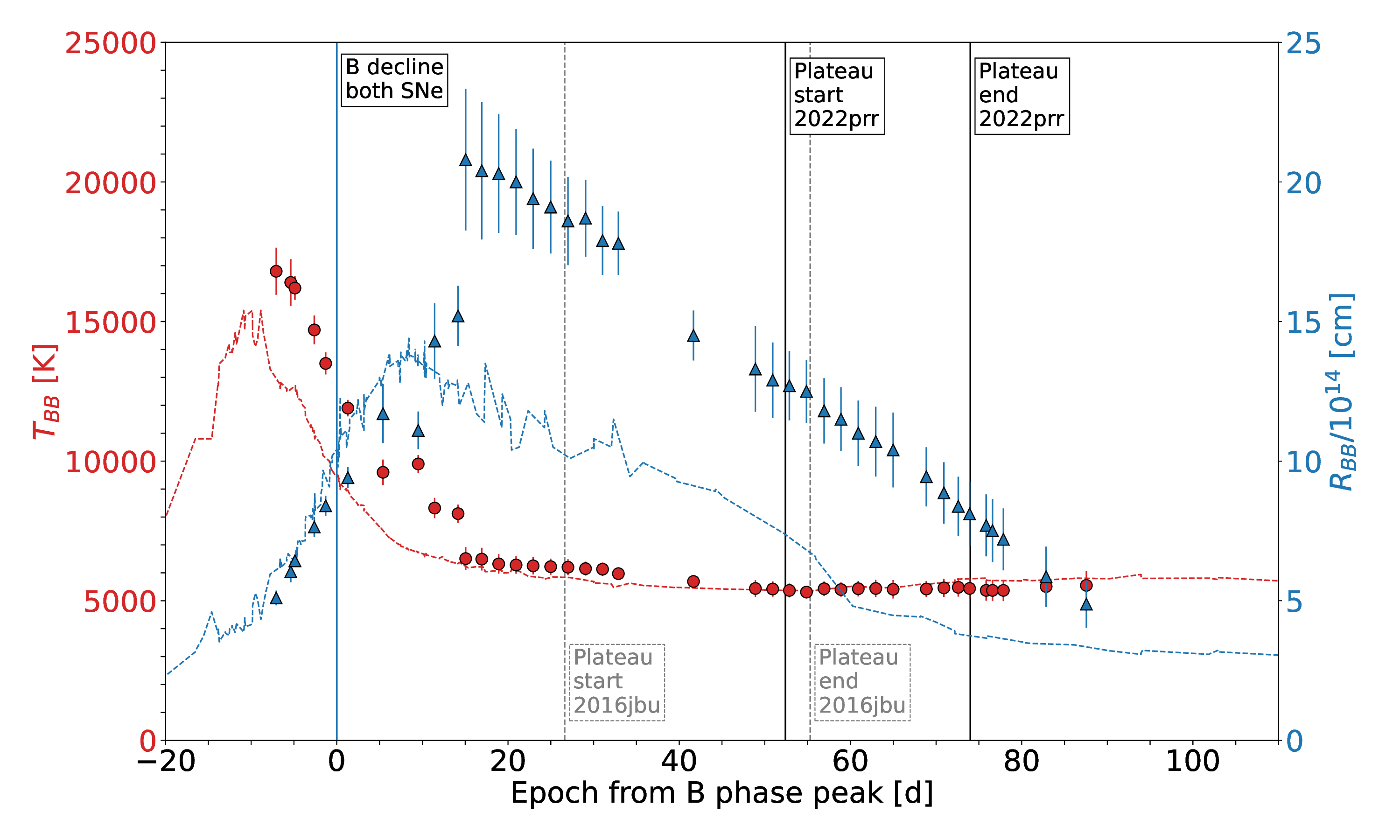}

\caption{Evolution of the blackbody temperature, $T_\text{BB}$ (red circles), and radius, $R_\text{BB}$ (blue triangles), of SN~2022prr. Values reported for SN~2016jbu \citep{Brennan2022b} are shown for comparison with dashed curves.}
\label{fig:bb_22prr}
\end{figure}

\begin{figure*}
    \centering
    \begin{minipage}{0.5\linewidth}
    \includegraphics[width=\linewidth]{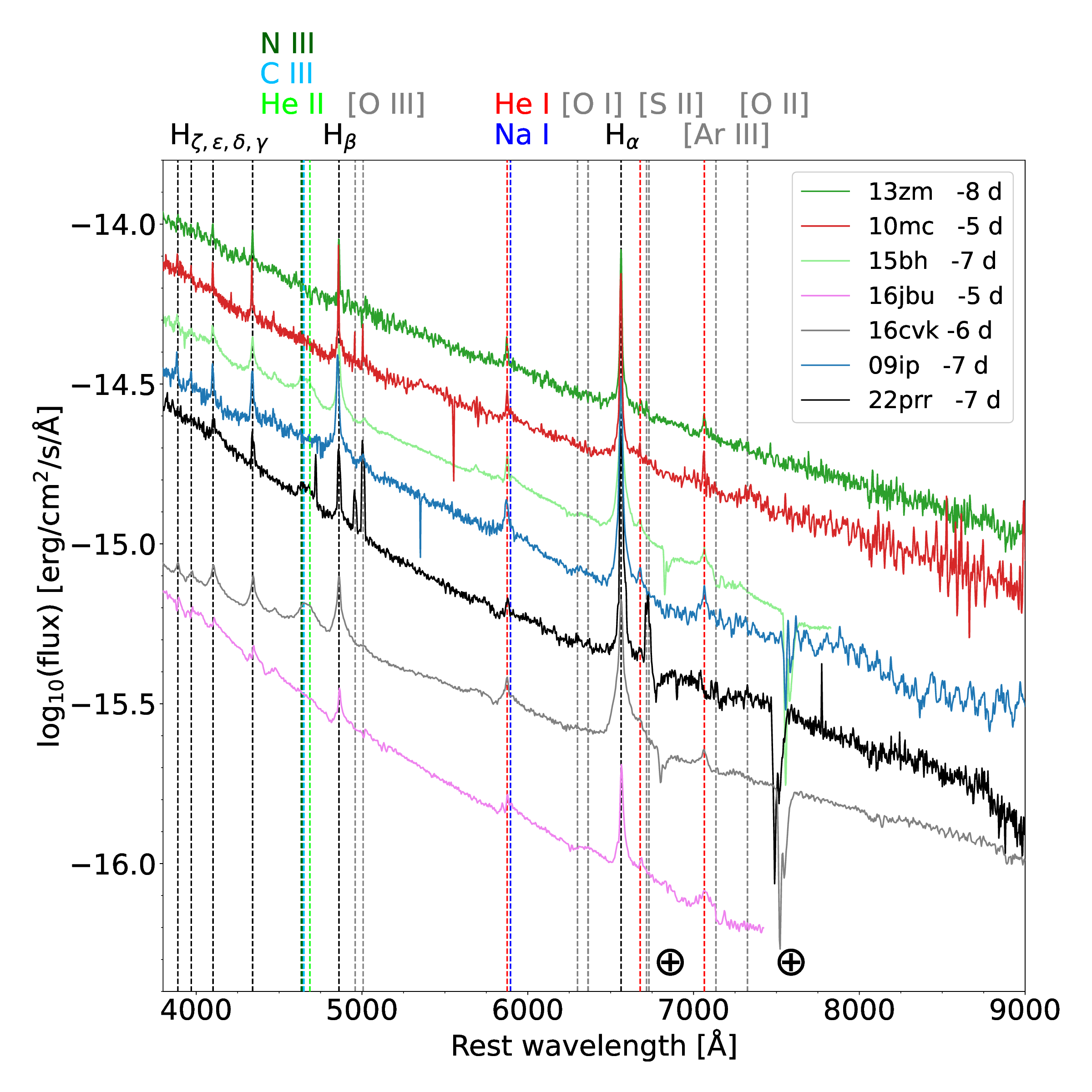}
    \includegraphics[width=\linewidth]{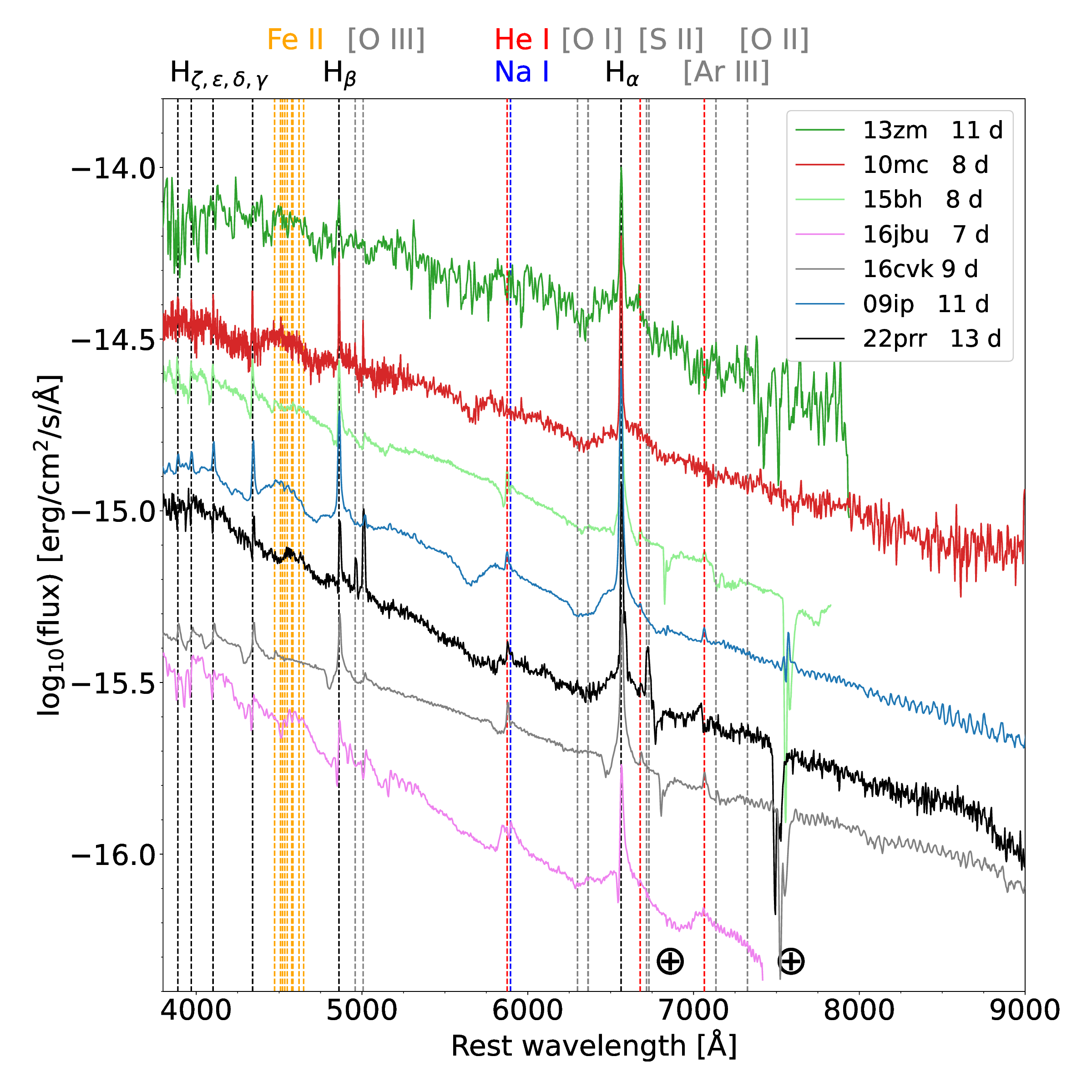}
    \end{minipage}\begin{minipage}{0.5\linewidth}
    \includegraphics[width=\linewidth]{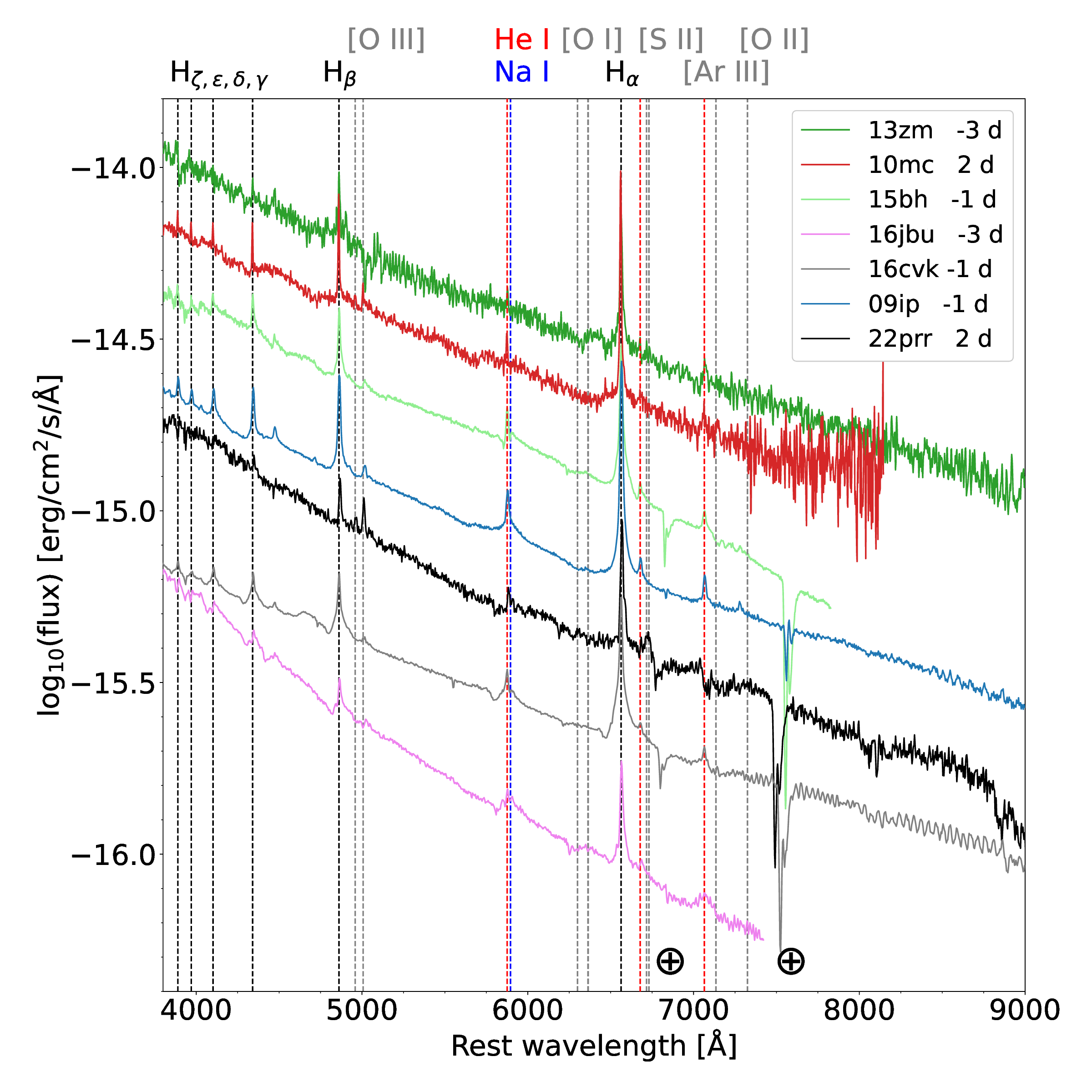}
    \includegraphics[width=\linewidth]{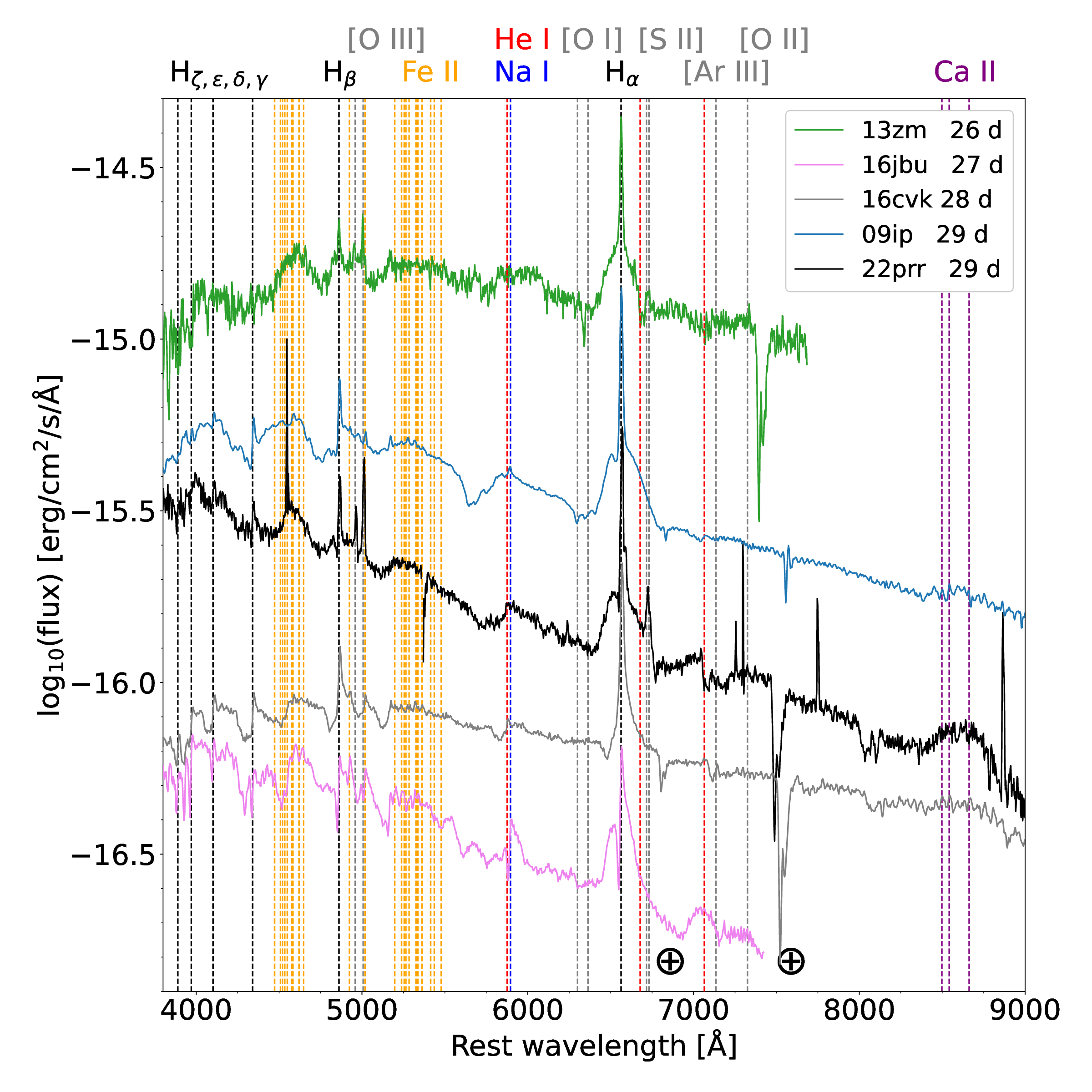}
    \end{minipage}
    \caption{Spectra of SN~2022prr and other SN~2009ip-like transients LSQ13zm \citep{Tartaglia2016}, SN~2010mc \citep{Ofek2013b}, SN~2015bh \citep{Thone2017}, SN~2016jbu \citep{Brennan2022a}, SN~2016cvk \citep{Matilainen2025}, and SN~2009ip \citep{Fraser2013}. The spectra have been dereddened and corrected to the wavelength rest frame. The wavelengths of the most prominent spectral lines are indicated with dashed vertical lines (host galaxy dominated lines in light grey) and telluric features with a $\oplus$ symbol. Logarithmic scale is used for flux, and the spectra have been vertically shifted for clarity.}
    \label{fig:09ip_spec}
\end{figure*}

\begin{figure*}
    \centering
    \begin{minipage}{0.5\linewidth}
    \includegraphics[width=\linewidth]{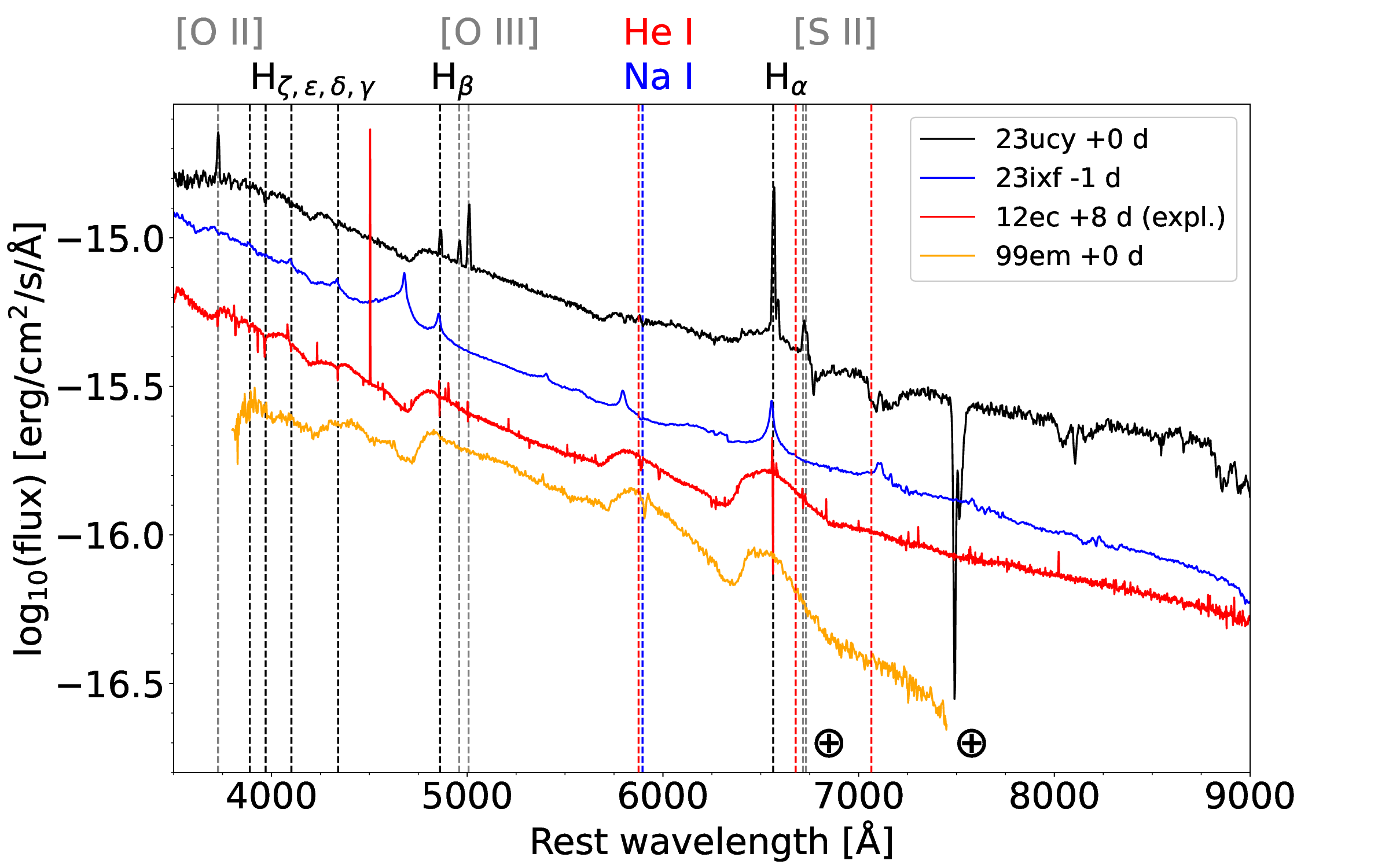}
    \includegraphics[width=\linewidth]{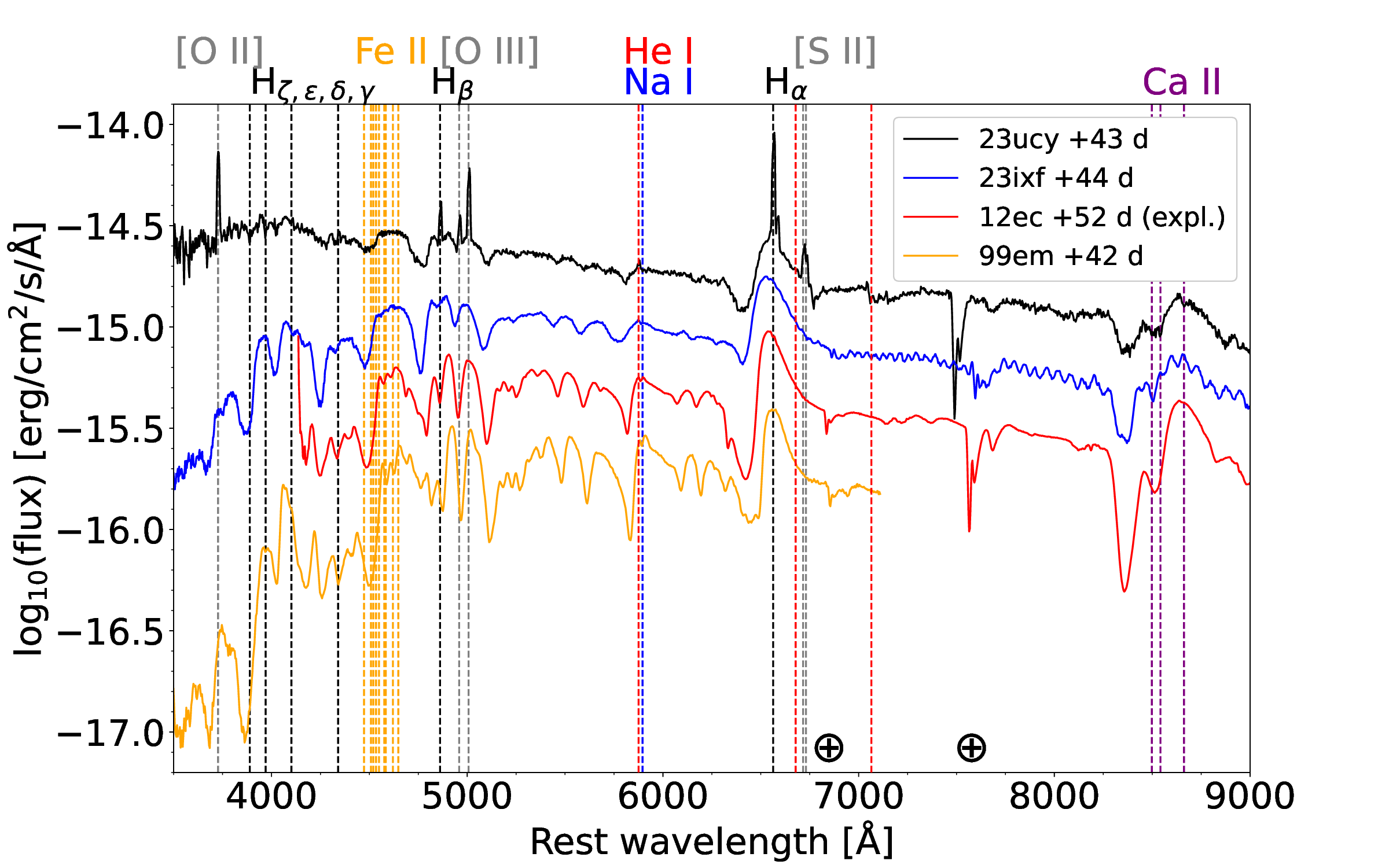}
     \end{minipage}\begin{minipage}{0.5\linewidth}
    \includegraphics[width=\linewidth]{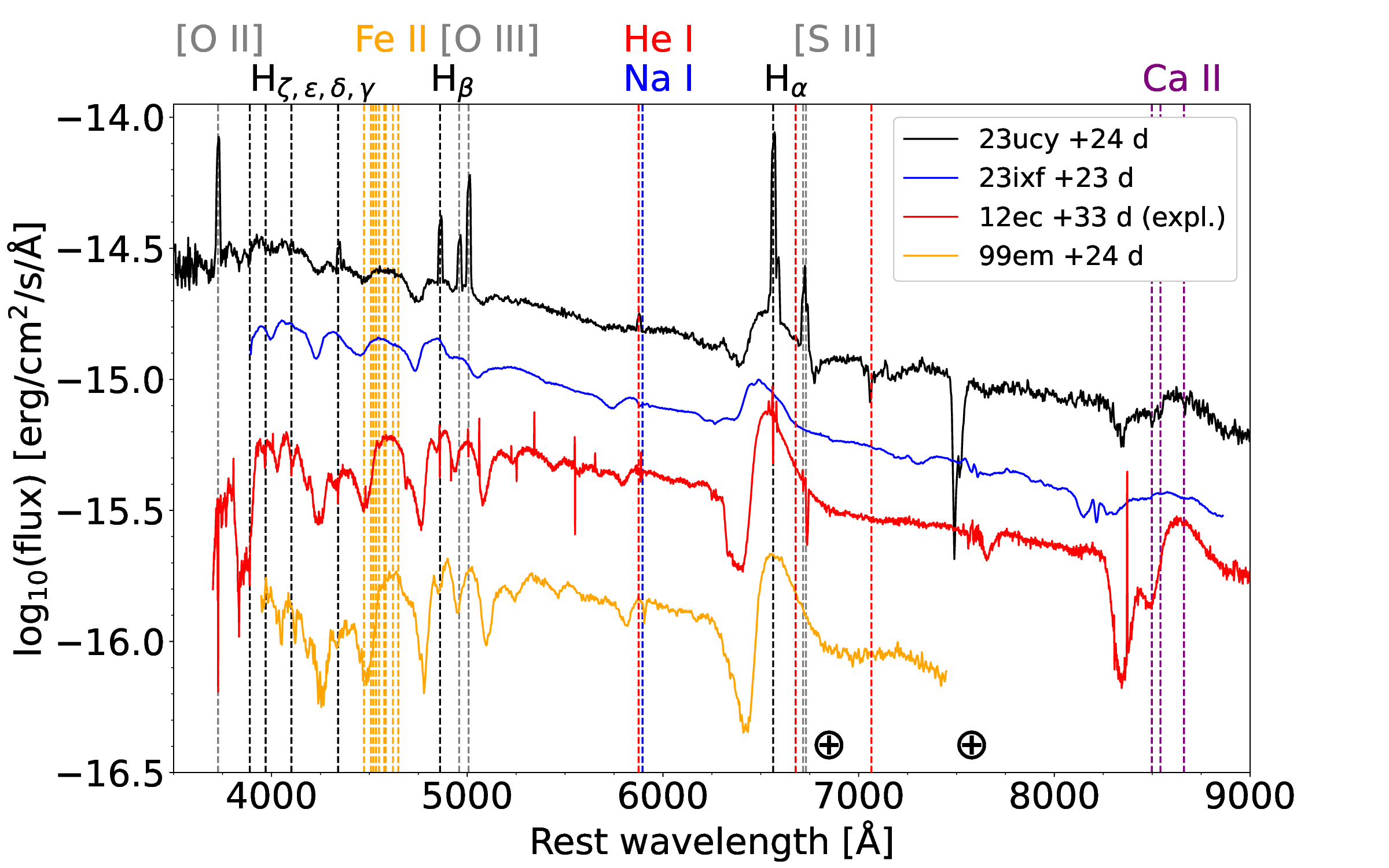}
    \includegraphics[width=\linewidth]{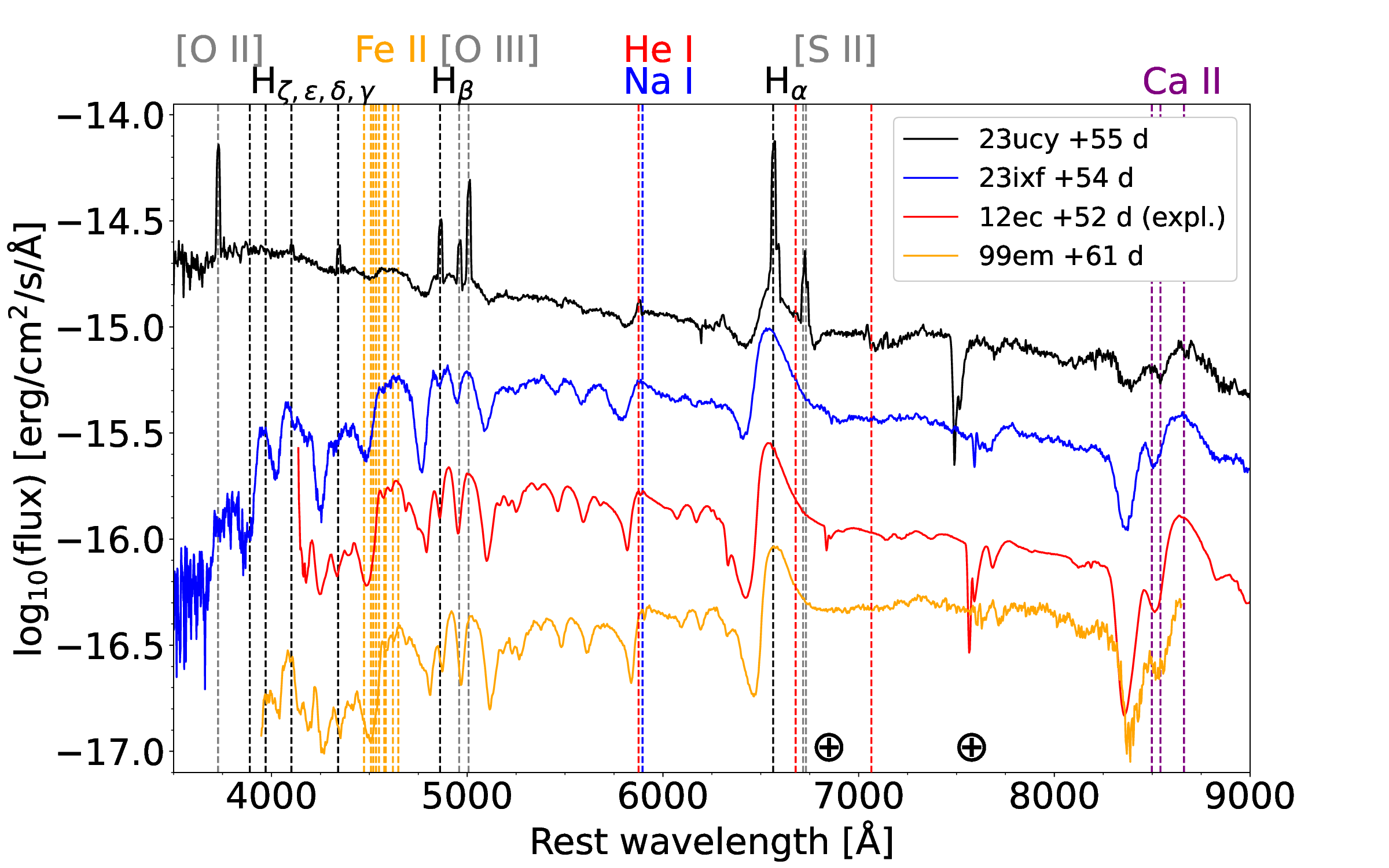}
    \end{minipage}
    \caption{Spectra of SN~2023ucy compared to those of Type IIP SN~1999em \citep{Leonard2002}, SN~2012ec \citep{Maund2013, Childress2016, Nagao2023}, and SN~2023ixf \citep{Zhang2023, Bostroem2023}. The spectra have been dereddened and corrected to the wavelength rest frame, and the epochs are measured from $r$-band maximum if available, or from the estimated time of explosion. The wavelengths of the most prominent spectral lines are indicated with dashed vertical lines (host galaxy dominated lines in light grey) and telluric features with a $\oplus$ symbol. Logarithmic scale is used for flux, and the spectra have been vertically shifted for clarity.}
    \label{fig:23ucy_like_spec}
\end{figure*}

\begin{figure*}
    \centering
    \begin{minipage}{0.5\linewidth}
    \includegraphics[width=\linewidth]{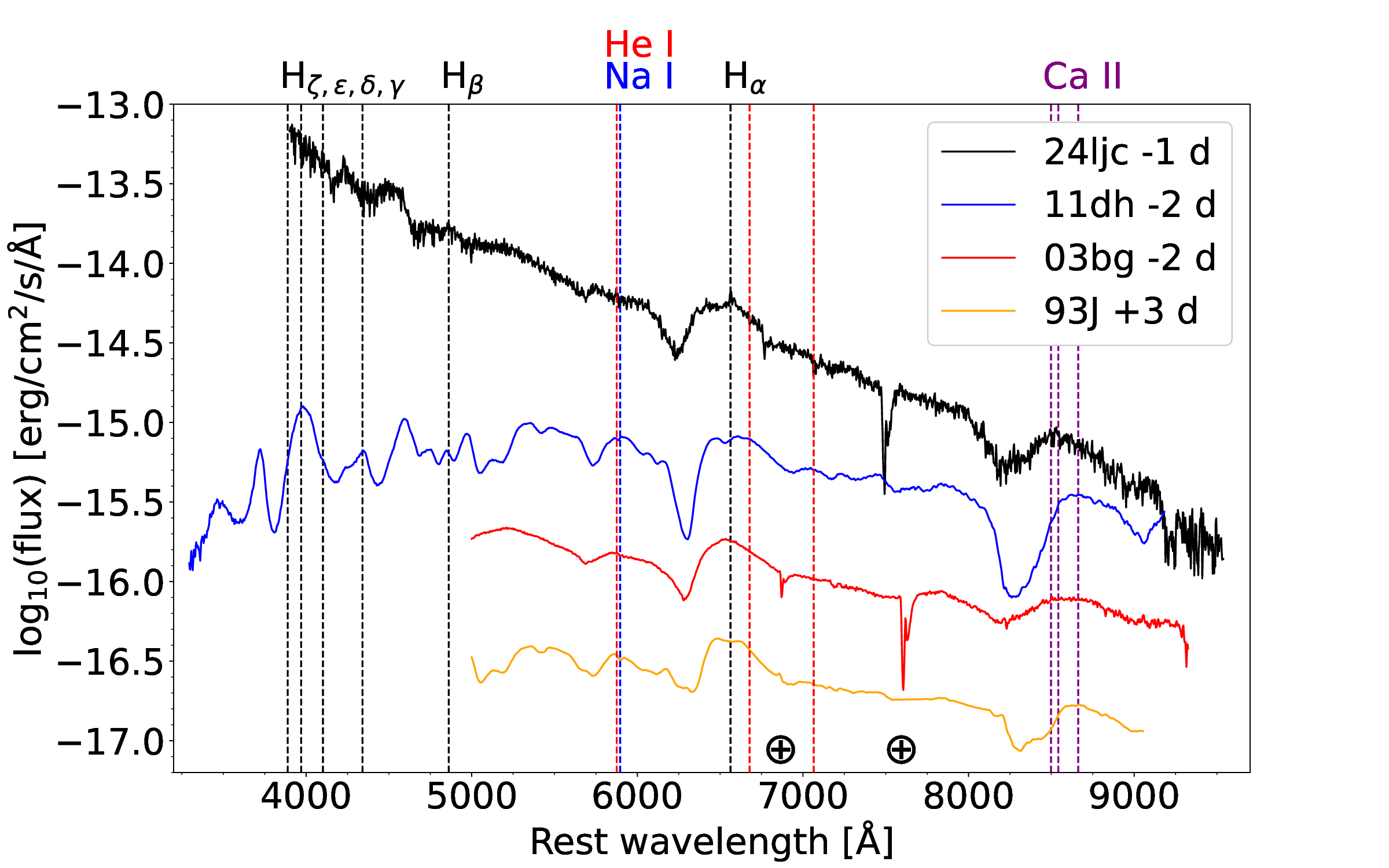}
    \end{minipage}\begin{minipage}{0.5\linewidth}
    \includegraphics[width=\linewidth]{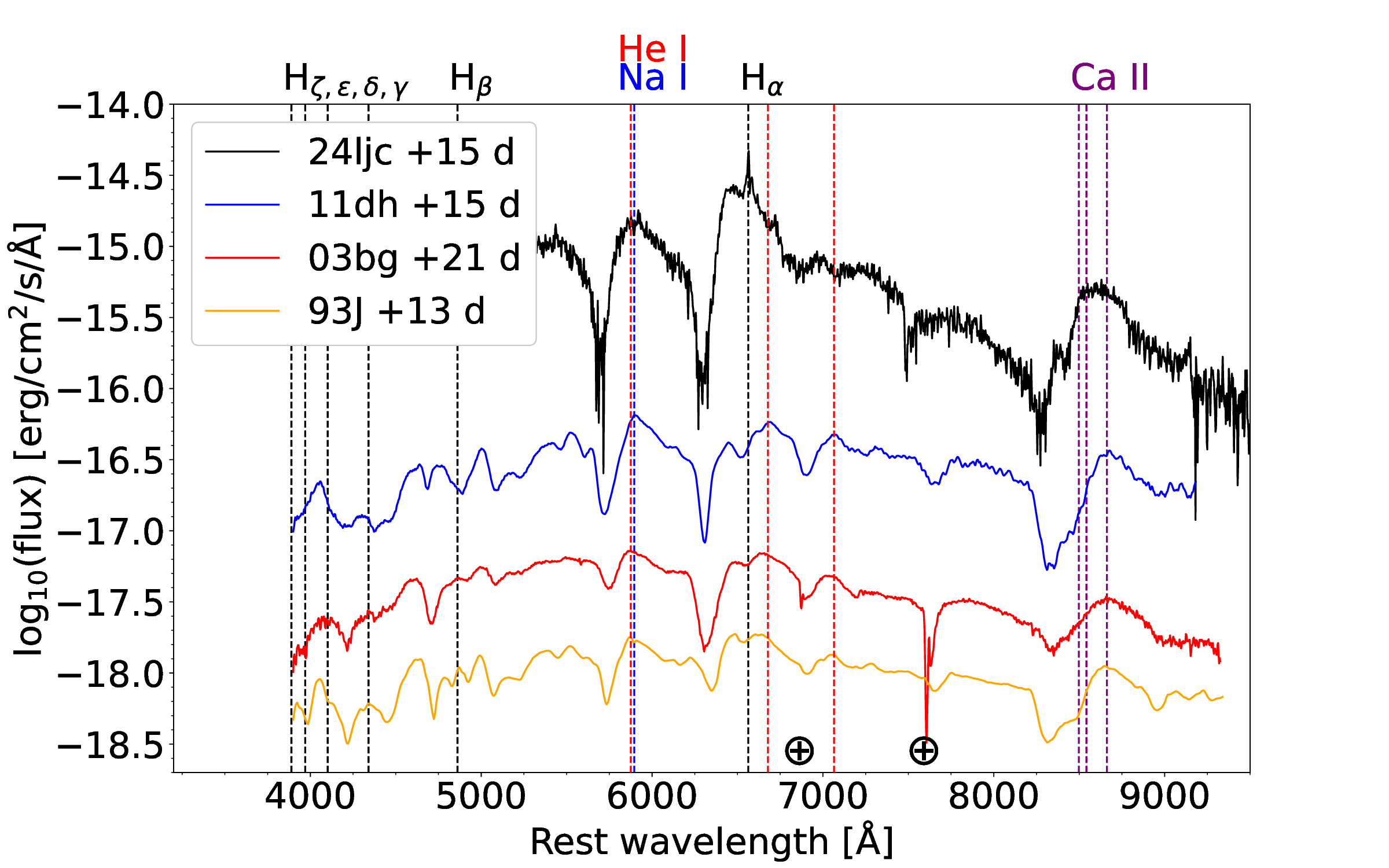}
    \end{minipage}
    \caption{Spectra of SN~2024ljc compared to those of Type IIb SN~1993J \citep{Barbon1995, Matheson2000}, SN~2003bg \citep{Hamuy2003}, and SN~2011dh \citep{Ergon2014}. The spectra have been dereddened and corrected to the wavelength rest frame, and the epochs are measured from the $r$-band maximum. The wavelengths of the most prominent spectral lines are indicated with dashed vertical lines and telluric features with a $\oplus$ symbol. Logarithmic scale is used for flux, and the spectra have been vertically shifted for clarity.}
    \label{fig:24ljc_comp_spec}
\end{figure*}

\twocolumn

\begin{figure}
    \centering
    \includegraphics[width=\linewidth]{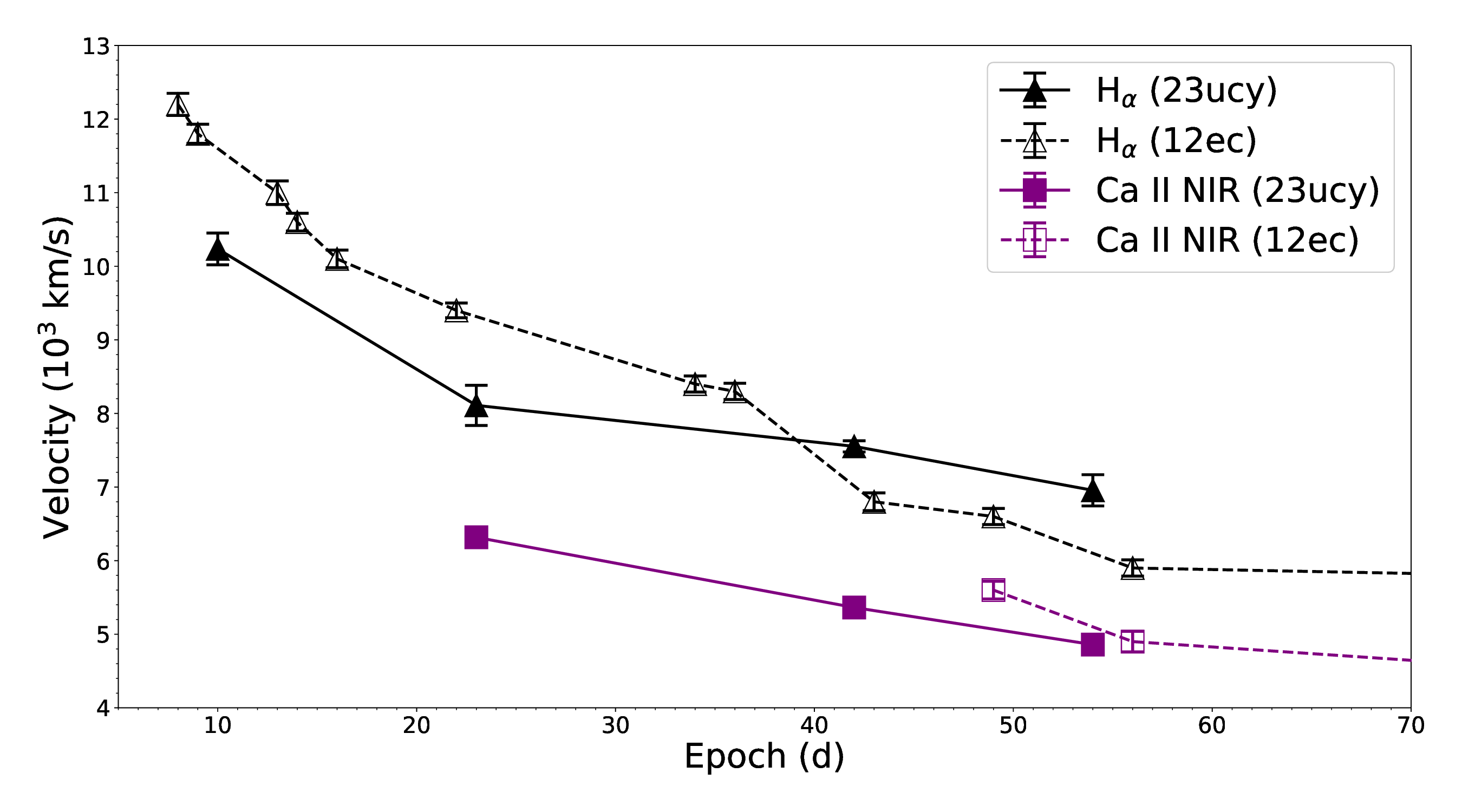}
    \caption{Line velocities for the H$_\alpha$ and Ca~{\sc ii} NIR lines of SN~2023ucy in solid symbols compared to those of SN~2012ec \citep{Barbarino2015} in open symbols.}
    \label{fig:linevel_23ucy}
\end{figure}

\begin{figure}
    \centering
    \includegraphics[width=\linewidth]{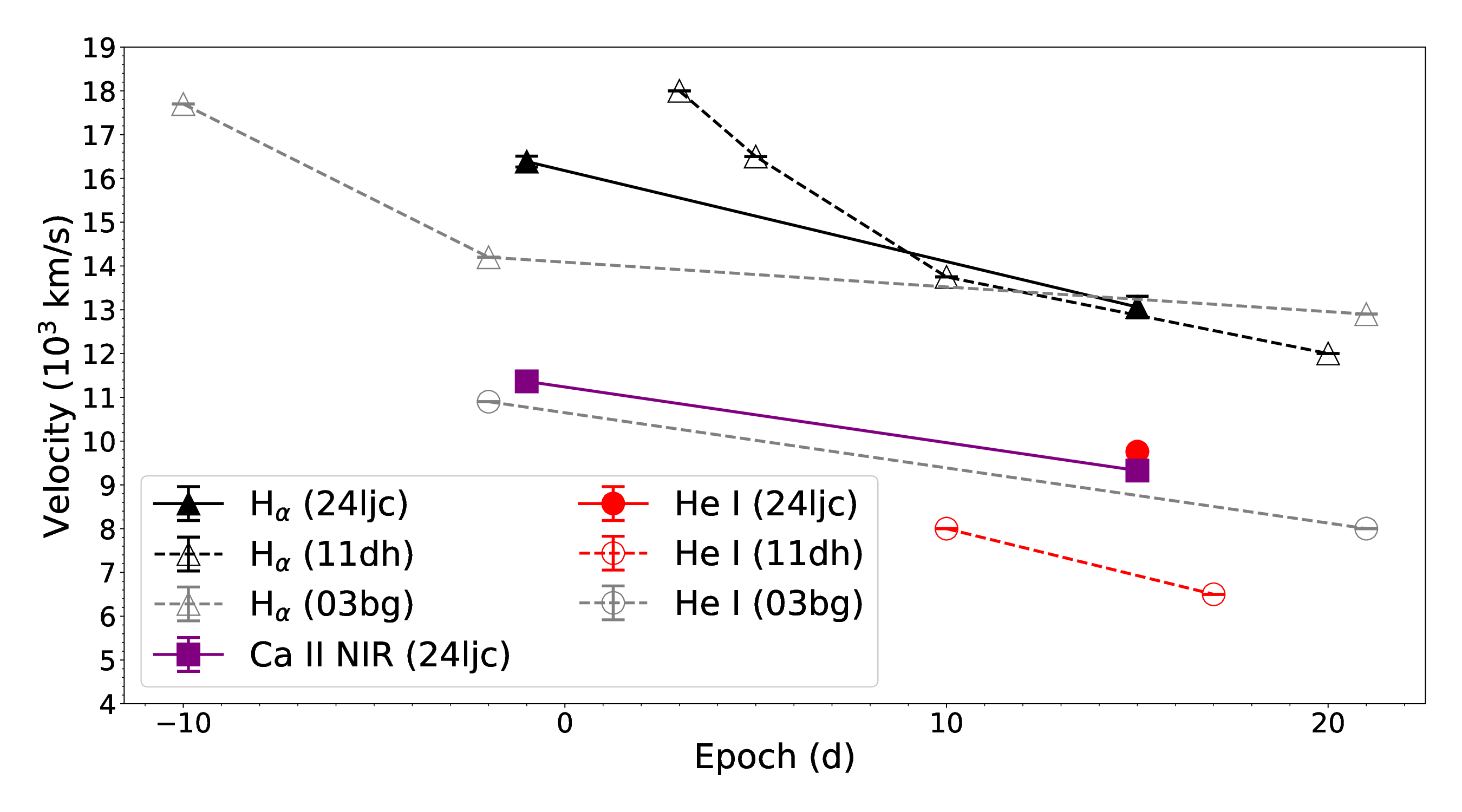}
    \caption{Line velocities for the H$_\alpha$, He~{\sc i} (5876~Å), and Ca~{\sc ii} NIR lines of SN~2024ljc in solid symbols compared to those of SN~2011dh \citep{Ergon2014} and SN~2003bg \citep{Hamuy2009} in open symbols.}
    \label{fig:linevel_24ljc}
\end{figure}

\begin{figure}
    \centering
    \includegraphics[width=\linewidth]{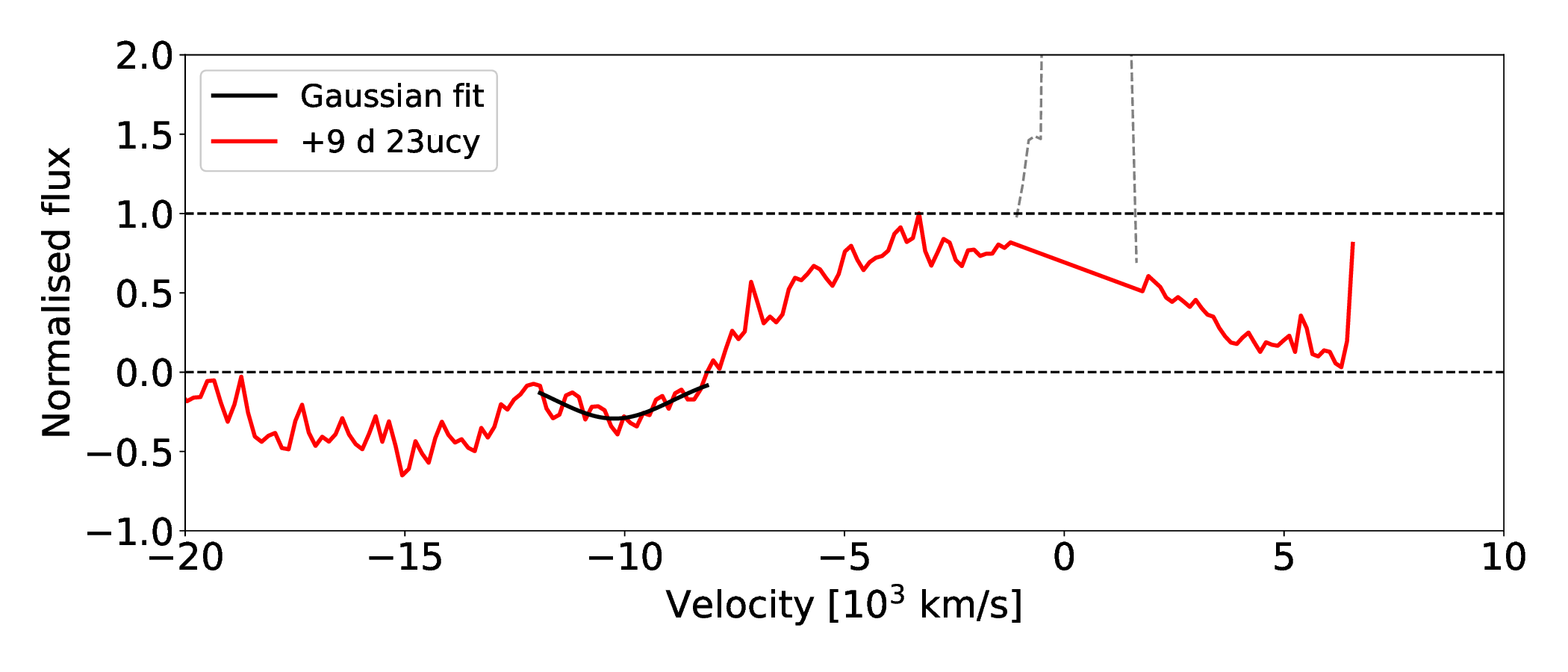}
    \includegraphics[width=\linewidth]{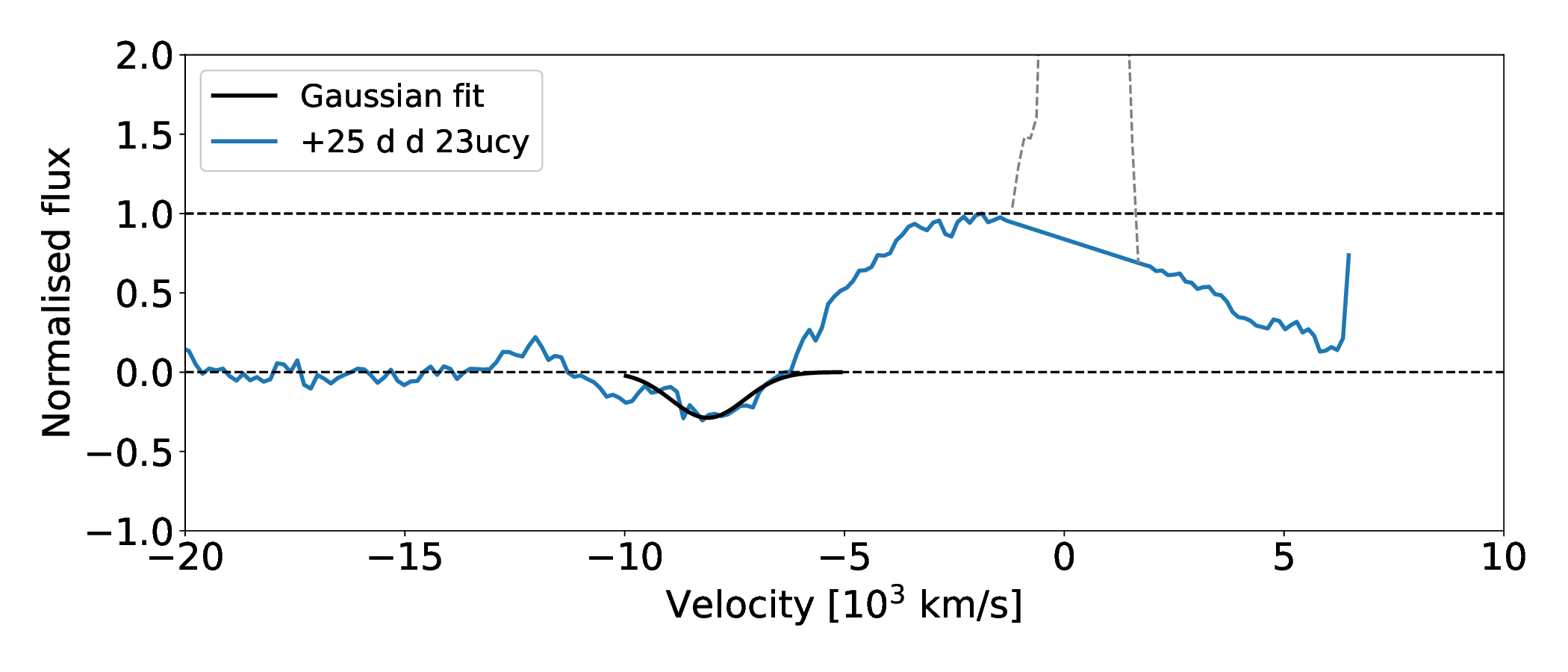}
    \includegraphics[width=\linewidth]{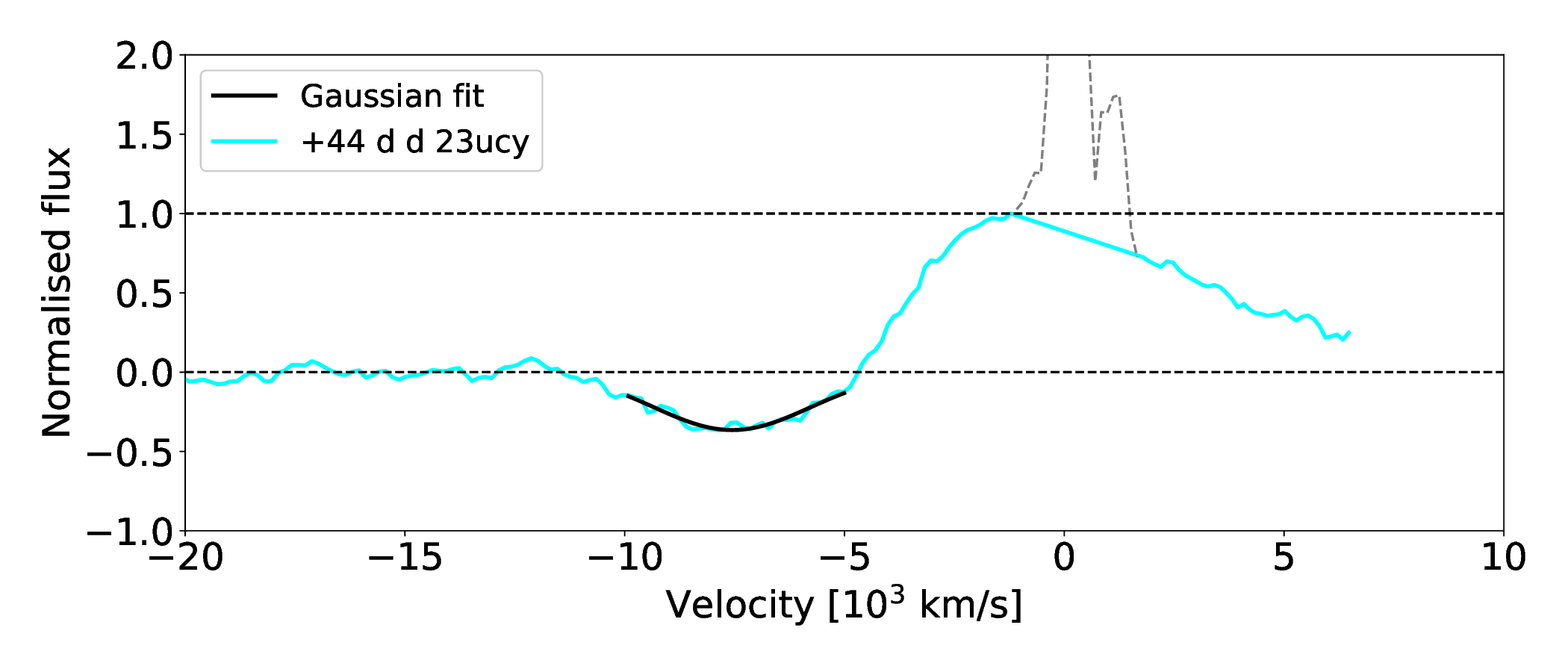}
    \includegraphics[width=\linewidth]{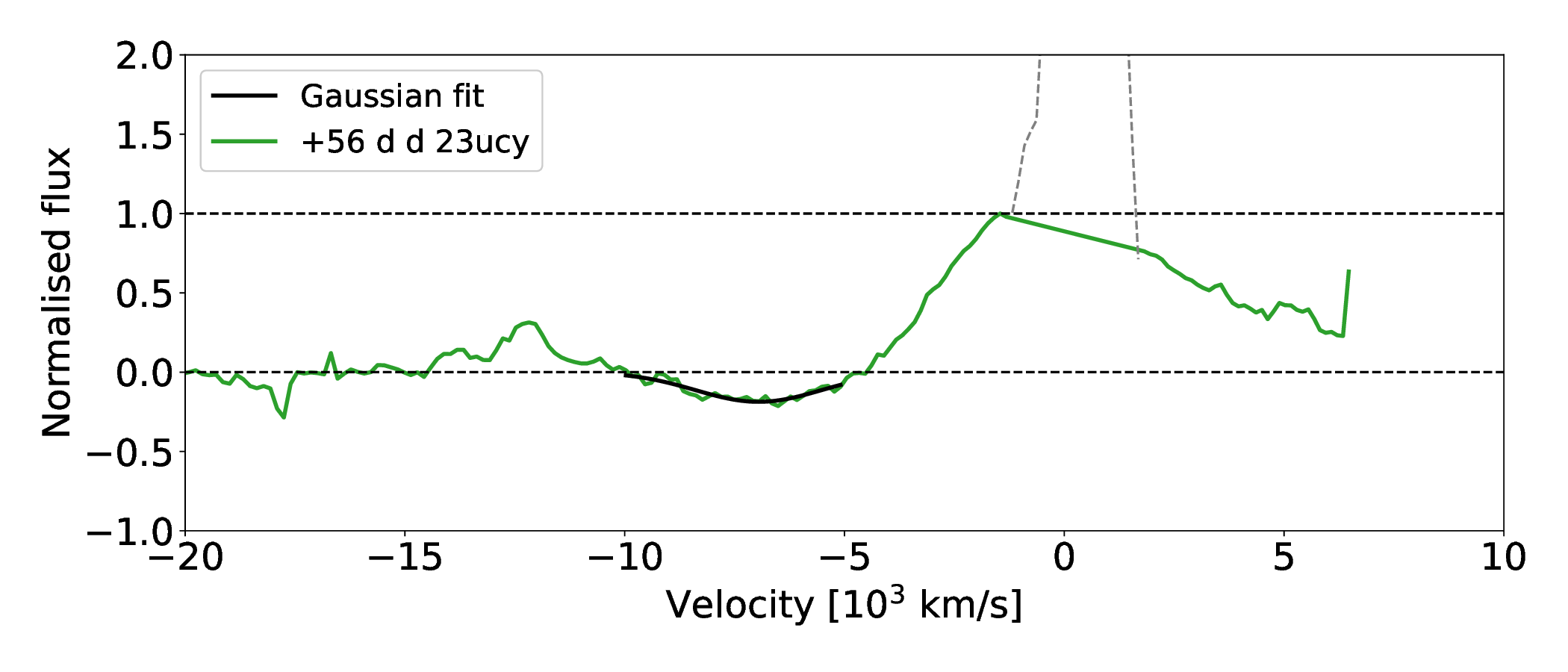}
    \caption{Gaussian fits to the P Cygni absorption feature of the H$_\alpha$ line of SN~2023ucy. The spectra have been continuum-subtracted and normalised. The narrow host lines (grey dashed lines) have been removed from the spectrum.}
    \label{fig:23ucy_pcyg}
\end{figure}

\begin{figure}
    \centering
    \includegraphics[width=\linewidth]{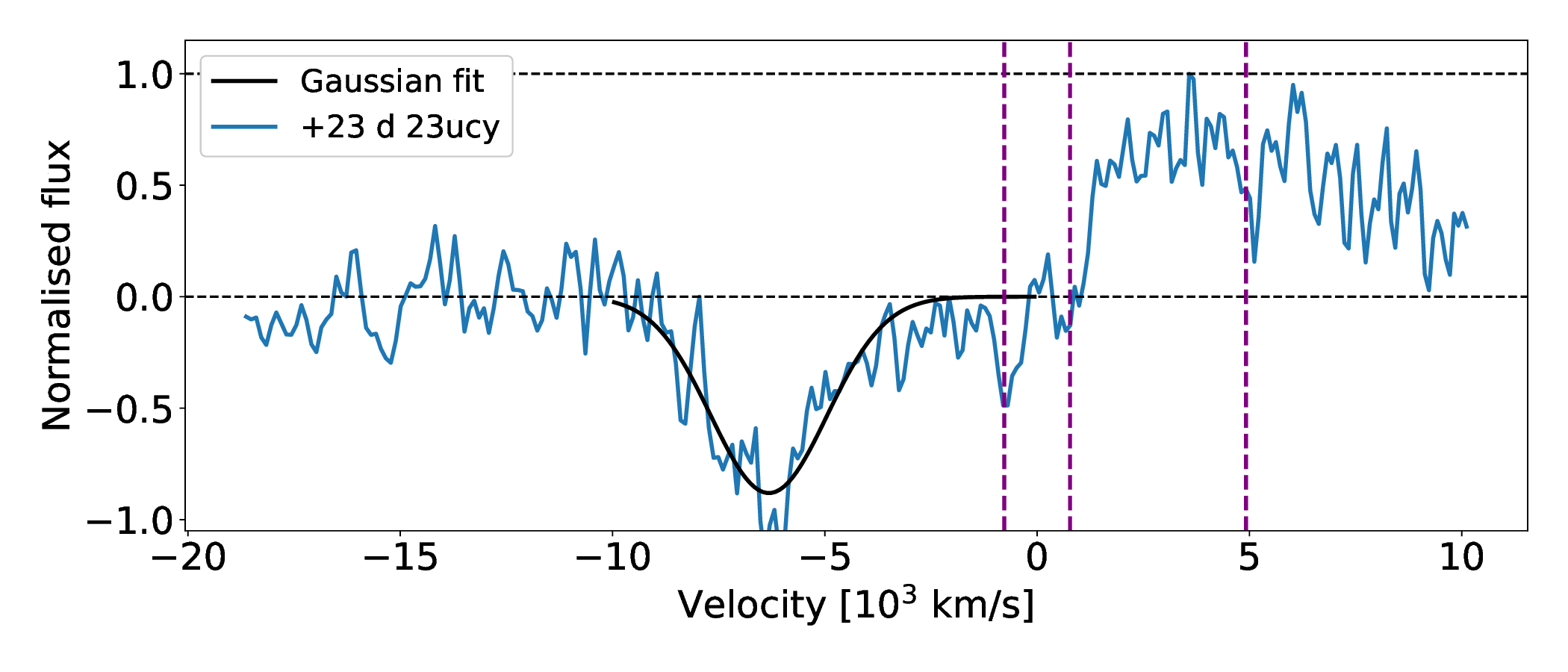}
    \includegraphics[width=\linewidth]{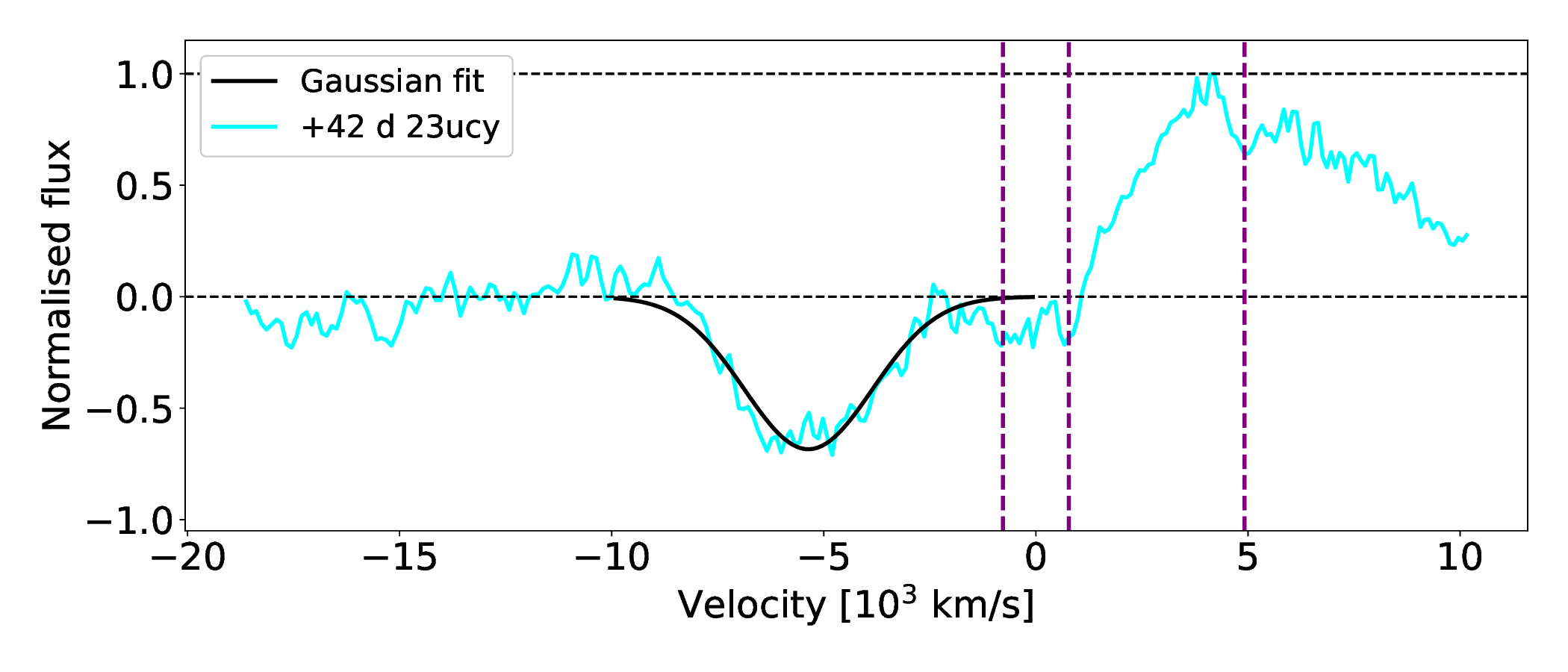}
    \includegraphics[width=\linewidth]{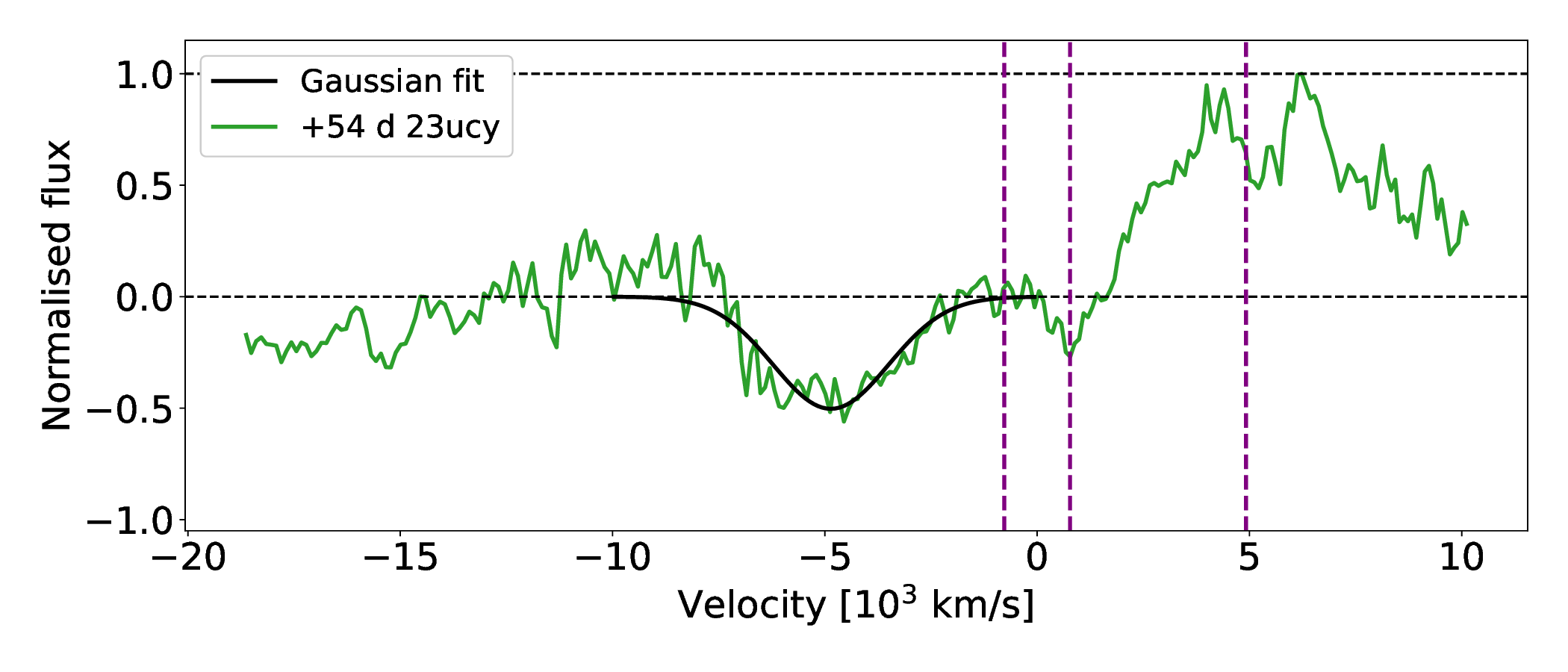}
    \caption{Gaussian fits to the P Cygni absorption component of the Ca~{\sc ii} NIR feature of SN~2023ucy. The spectra have been continuum-subtracted and normalised. The velocities are measured from 8520~Å to match the definition used for SN~2012ec by \cite{Barbarino2015}. The locations of the Ca~{\sc ii} NIR triplet spectral lines are indicated with vertical dashed purple lines.}
    \label{fig:CaNIR_23ucy}
\end{figure}

\begin{figure}
    \centering
    \includegraphics[width=\linewidth]{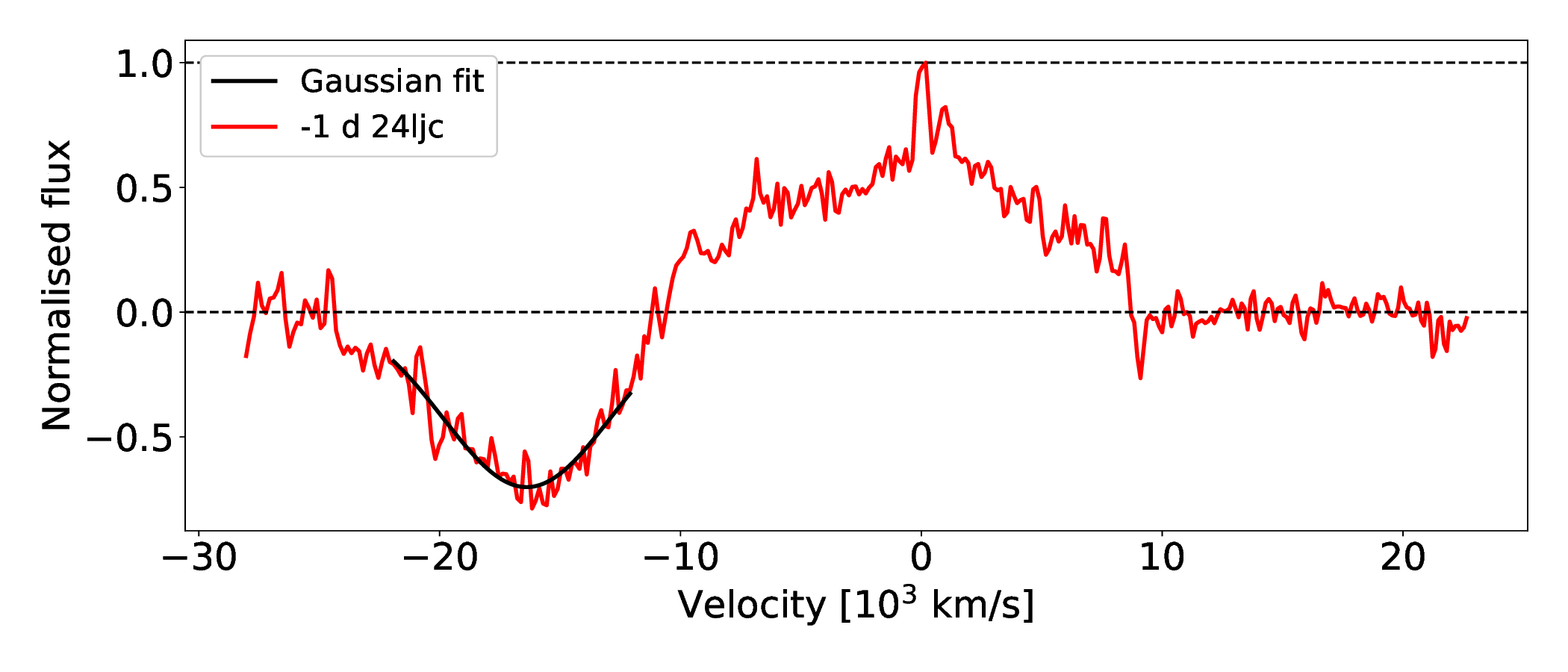}
    \includegraphics[width=\linewidth]{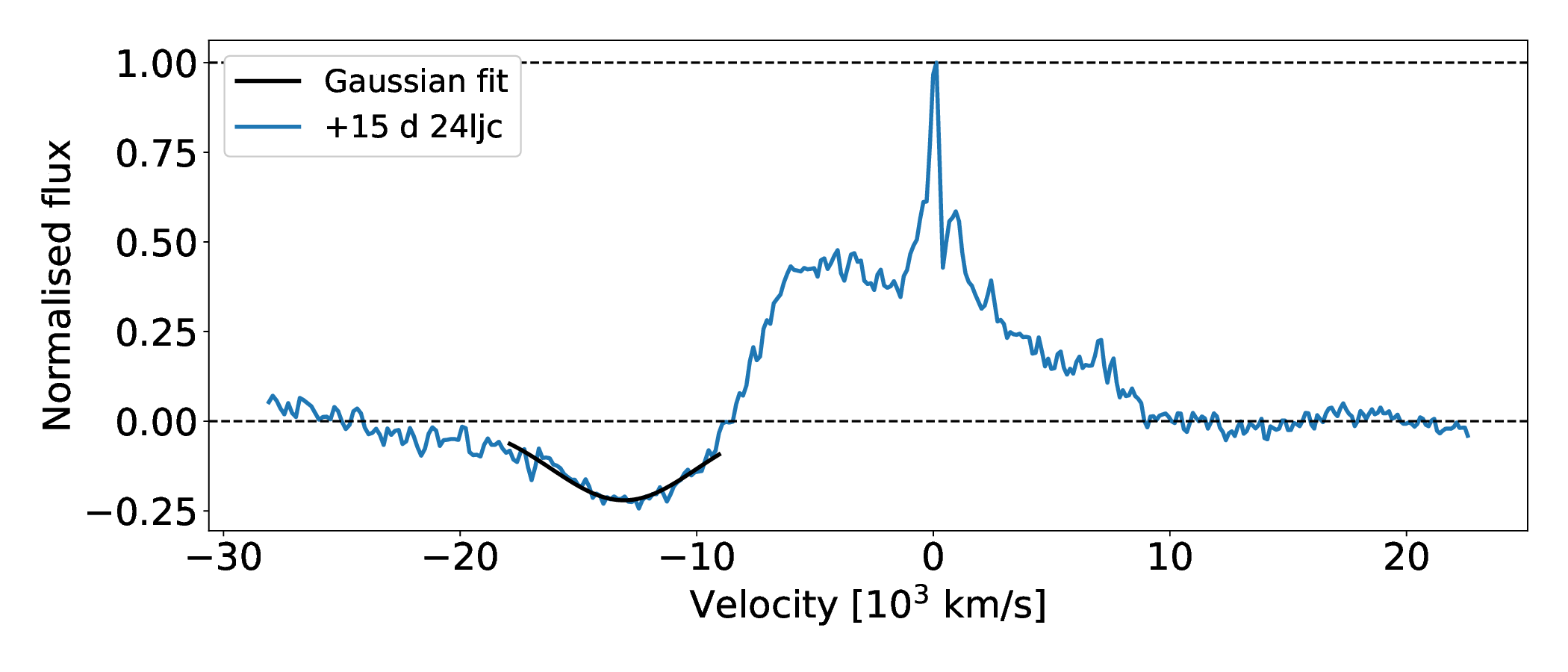}
    \caption{Gaussian fits to the P Cygni absorption feature of the H$_\alpha$ line of SN~2024ljc. The spectra have been continuum-subtracted and normalised.}
    \label{fig:ha_24ljc}
\end{figure}

\begin{figure}
    \centering
    \includegraphics[width=\linewidth]{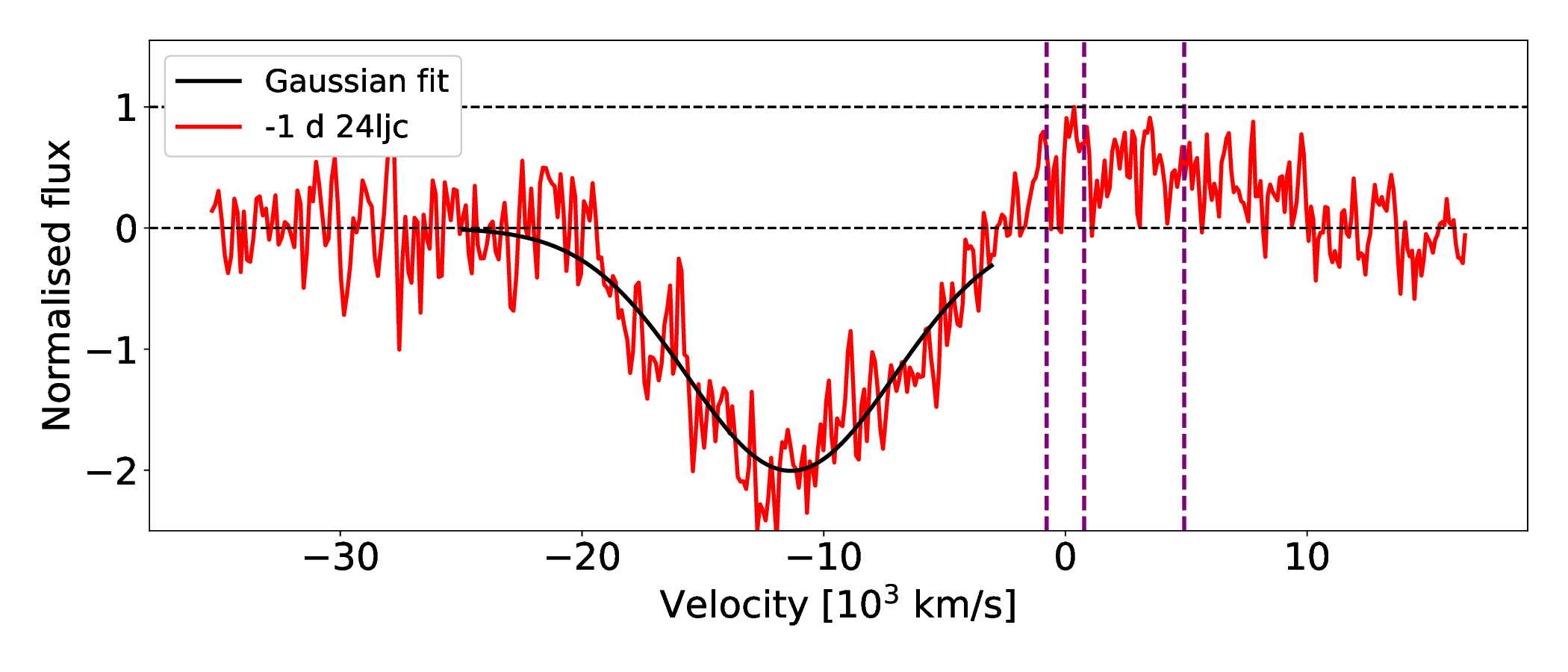}
    \includegraphics[width=\linewidth]{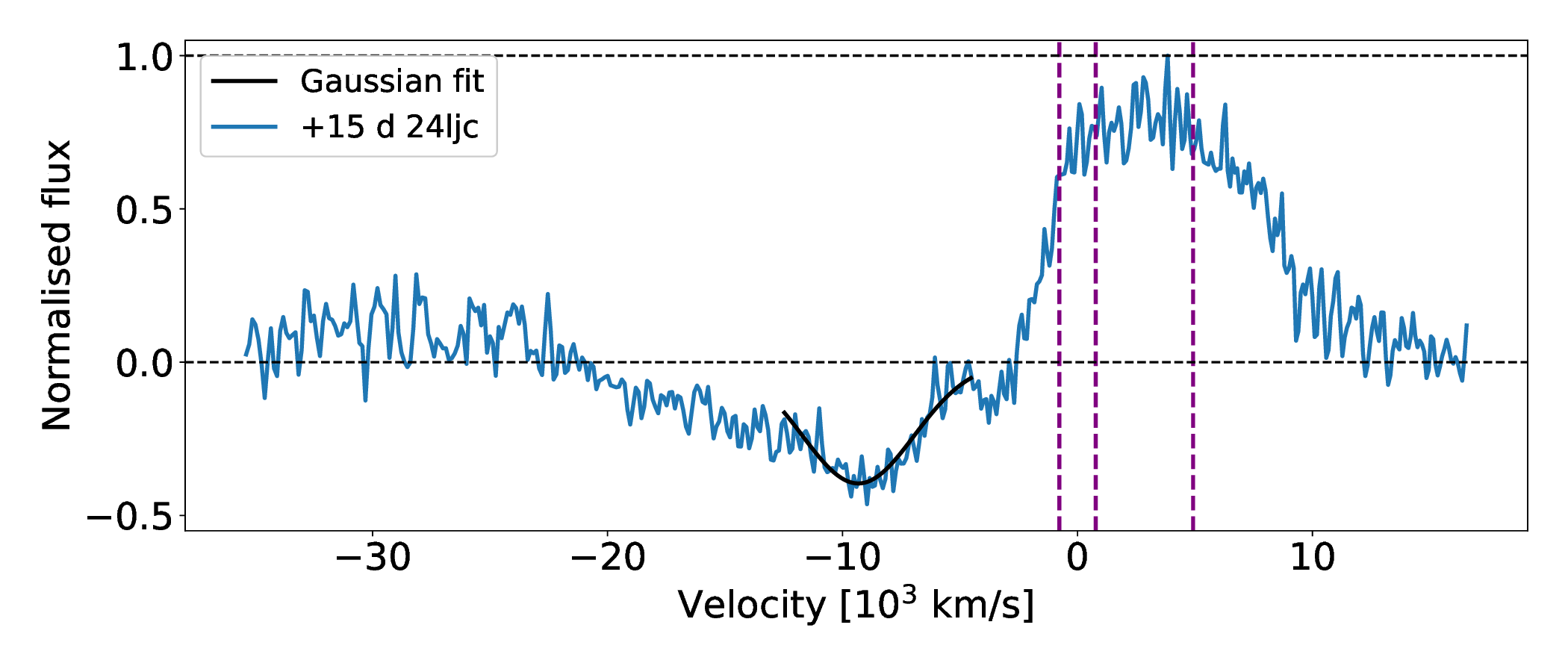}
    \caption{Gaussian fits to the P Cygni absorption component of the Ca~{\sc ii} NIR feature of SN~2024ljc. The spectra have been continuum-subtracted and normalised, and velocities are measured from 8520~Å. The locations of the Ca~{\sc ii} NIR triplet spectral lines are indicated with vertical dashed purple lines.}
    \label{fig:CaNIR_24ljc}
\end{figure}

\end{appendix}

\label{lastpage}

\end{document}